\documentclass[fleqn,usenatbib]{mnras}

\usepackage{newtxtext,newtxmath}
\usepackage[T1]{fontenc}

\usepackage{graphicx}  % Including figure files
\usepackage{amsmath}	% Advanced maths commands
\usepackage{subcaption}
\usepackage{macros}
\title[Disk settling and heating in FIRE MW-mass galaxies]{Disk settling and dynamical heating: histories of Milky Way-mass stellar disks across cosmic time in the FIRE simulations}

\author[F. McCluskey et al.]{
Fiona McCluskey$^{1}$\thanks{E-mail: fmccluskey@ucdavis.edu},
Andrew Wetzel$^{1}$, %\orcidlink{0000-0003-0603-8942}$,
Sarah R.~Loebman$^{2}$, %\orcidlink{0000-0003-3217-5967}$,
Jorge Moreno$^{3}$, %\orcidlink{0000-0002-3430-3232}$,
Claude-Andr\'e Faucher-Gigu\`ere$^{4}$, 
\newauthor
\hspace{0.25mm} Philip F. Hopkins$^{5}$\\%\orcidlink{0000-0002-4900-6628}$ \\
$^{1}$Department of Physics \& Astronomy, University of California, Davis, 1 Shields Ave, Davis, CA 95616, USA\\
$^{2}$Department of Physics, University of California, Merced, 5200 Lake Road, Merced, CA 95343, USA\\
$^{3}$Department of Physics \& Astronomy, Pomona College, Claremont, CA 91711, USA\\
$^{4}$Department of Physics \& Astronomy and CIERA, Northwestern University, 1800 Sherman Ave, Evanston, IL 60201, USA\\
$^{5}$TAPIR, California Institute of Technology, Mailcode 350-17, Pasadena, CA 91125, USA
}
\date{}

\pubyear{2023}

\begin{document}
\label{firstpage}
\pagerange{\pageref{firstpage}--\pageref{lastpage}}
\maketitle

\begin{abstract}
We study the kinematics of stars both at their formation and today within 14 Milky Way (MW)-mass galaxies from the FIRE-2 cosmological zoom-in simulations.
We quantify the relative importance of cosmological disk settling and post-formation dynamical heating.
We identify three eras:
a Pre-Disk Era (typically $\gtrsim8\Gyr$ ago), when stars formed on dispersion-dominated orbits; an Early-Disk Era ($\approx8-4\Gyr$ ago), when stars started to form on rotation-dominated orbits but with high velocity dispersion, $\sigma_{\rm form}$; and a Late-Disk Era ($\lesssim4\Gyr$ ago), when stars formed with low $\sigma_{\rm form}$.
$\sigma_{\rm form}$ increased with time during the Pre-Disk Era, peaking $\approx8\Gyr$ ago, then decreased throughout the Early-Disk Era as the disk settled and remained low throughout the Late-Disk Era.
By contrast, the dispersion measured today, $\sigma_{\rm now}$, increases monotonically with age because of stronger post-formation heating for Pre-Disk stars. 
Importantly, most of $\sigma_{\rm now}$ was in place at formation, not added post-formation, for stars younger than $\approx10\Gyr$.
We compare the evolution of the three velocity components: at all times, $\sigma_{\rm R,form}>\sigma_{\rm \phi,form}>\sigma_{\rm Z,form}$.
Post-formation heating primarily increased $\sigma_{\rm R}$ at ages $\lesssim4\Gyr$ but acted nearly isotropically for older stars.
The kinematics of young stars in FIRE-2 broadly agree with the range observed across the MW, M31, M33, and PHANGS-MUSE galaxies.
The lookback time that the disk began to settle correlates with its dynamical state today: earlier-settling galaxies currently form colder disks.
Including stellar cosmic-ray feedback does not significantly change disk rotational support at fixed stellar mass.
\end{abstract}

%%%%%%%%%%%%%%%%%%%%%%%%%%%%%%%%%%%%%%%%%%%%%%%%%%

%%%%%%%%%%%%%%%%% BODY OF PAPER %%%%%%%%%%%%%%%%%%
\section{Introduction}

The present-day kinematics of stars encode a galaxy's dynamical history. A star's orbit is set by both its initial dynamics, as inherited from the star-forming interstellar medium (ISM) at its formation, and all post-formation dynamical perturbations that the star experienced over its lifetime. Thus, stellar kinematics provide dual insight into the past state of the ISM and the dynamical processes that shaped the galaxy.
However, the extent to which a star's current motion results from its initial state \textit{or} from post-formation perturbations remains poorly understood. The inability to reliably disambiguate these two remains a significant obstacle in connecting the kinematics and morphology of a galaxy today to its formation history \citep[for example][]{Somerville15, Naab17}.

Overcoming this ambiguity requires understanding the kinematics with which stars formed throughout the entire lifetime of a galaxy.
In the solar neighborhood, the youngest stellar populations share the kinematics of the gas disk from which they formed: young stars are on nearly circular orbits with total velocity dispersions, $\sigma_{\rm tot} \approx 25 - 30 \kms$ \citep{Edvardsson93, Casagrande11}.
However, the kinematics of stars that formed at earlier times are likely different than those forming in the Milky Way (MW) today, based on observations of high-redshift galaxies, which provide statistics on how the initial state of stars evolved over cosmic time.
Such observations reveal that MW-like thin disks today are the endpoint of a metamorphosis: galactic disks at higher redshifts ($z \gtrsim 1 - 2$) were thicker, clumpier, and more turbulent than their thin counterparts at lower redshifts \citep[for example][]{schreiber-09, Stott16, Birkin23} \citep[though see][]{Rizzo20}. The high turbulence of MW-mass progenitors probably arose from the combination of stellar feedback, given their higher star formation rates (SFRs) \citep[for example][]{Madau14, FS20}, gravitational instabilities \citep[for example][]{Elmegreen07, Dekel09, Lehnert09}, and higher merger and accretion rates, which lead to higher gas fractions and surface densities \citep[for example][]{Genzel14, Tacconi20}.

Thus, high-redshift observations show that disk galaxies `settled' over cosmic time \citep{Kassin-12}, becoming less turbulent and more rotationally supported since $z \sim 1 - 2$ \citep[for example][]{Wisnioski15, Johnson18}. Specifically, the velocity dispersion, $\sigma$, of gas disks decreased while their rotational velocity, $v_{\rm \phi}$, increased.
Although most works used emission lines from ionized gas to measure high-redshift kinematics, recent works indicate that the dispersions of atomic and molecular gas, while $\approx 15 \kms$ lower, evolved similarly \citep[for example][]{Ubler-19, Girard21}.
Thus, stars, which are born from cold dense gas, formed with larger dispersions at earlier times when turbulence was higher, and later formed with progressively lower dispersions as turbulence decreased and the galaxy's disk settled.

Alongside understanding how stars kinematically formed, it is equally important to understand how they evolved post-formation. Over their lifetimes, stars undergo repeated scattering interactions with  perturbations within the disk, such as giant molecular clouds (GMCs) \citep{Spitzer51, Wielen77, Lacey84}, spiral arms \citep {Barbanis67, Carlberg85, Minchev06}, and bars \citep{Saha10, Grand16}.  Stars are effectively collisionless and dissipationless, and thus these interactions result in enduring increases to their random orbital energy, or `dynamical heating'. In turn, a stellar population's $\sigma$ increases over its lifetime. Additional heating may arise from feedback-driven outflows and potential fluctuations \citep{Badry2016}, the galaxy's global asymmetries -- including torquing from the misalignment of the stellar and gaseous disk \citep{Roskar10} -- cosmic accretion, mergers, and interactions with satellites\citep[for example][]{Quinn93, Brook04, Villalobos08, Belokurov18}. However, the relative impact of each mechanism remains uncertain.

Furthermore, the scattering environment of the disk evolves across cosmic time. At earlier times, a galaxy's gas fraction and gas surface density were higher \citep[for example][]{Shapley11, Swinbank12, Tacconi20}, such that scattering by GMCs was likely more frequent. Prior to the formation of a disk, galaxies generally lacked strong spiral structure and thus spiral-driven heating \citep[for example][]{Margalef22}. On the other hand, certain processes that can drive non-adiabatic rapid heating were more prevalent earlier, including mergers \citep[for example][]{Conselice05} and feedback-driven outflows \citep[for example][]{Genzel14}.
 
One can understand the relative strength of these heating mechanisms by comparing the observed relationship between velocity dispersion and stellar age with predictions from various heating models. The magnitude of each component of $\sigma$ (along $R$, $\phi$, or $Z$) provides insight into the relative strength of different mechanisms: spiral structure increases the in-plane dispersion but is generally inefficient at increasing the vertical, whereas GMCs heat stars more isotropically and can redirect in-plane heating \citep[for example][]{Carlberg87, JenkinsBinney90, Sellwood13}. 

The MW is an ideal test-bed for such comparisons because it offers unparalleled access to the 3D motions, positions, and ages of individual stars.
%, thus inspiring the use of ‘Galactic Archaeology’ to unravel its formation history \citep[for example][] {bhg-16, Helmi21}.
In particular, the observed monotonic increase in velocity dispersion with age for stars in the solar neighborhood provides an important benchmark \citep[for example][]{Stromberg46, Wielen77, Casagrande11}.
\citet{Nordstrom2004} and \citet{Holmberg09} argued that this observed trend is consistent with MW stars forming on `disky' orbits that then underwent dynamical heating by the combination of GMCs and spiral arms. \citet{Mackereth19} and \citet{Ting19} extended this analysis to include stars located outside the solar neighborhood ($R = 3 - 14 \kpc$) and found similar results: the observed relations are consistent with stars younger than $\approx 8 \Gyr$ having formed with a low, constant dispersion and subsequently being heated by both GMCs and spiral arms. However, as before, the observed dispersions of older stars (age $\gtrsim 8 \Gyr$) are higher than those predicted by heating models \citep{Mackereth19}. This discrepancy could arise if old stars were born kinematically hotter, or if their dynamical heating differed from what current models predict. Likely, both factors contribute, but disentangling this requires insight into the MW's early history.

Fortunately, recent MW observations now allow us to characterize this bygone era. Astrometric data from Gaia \citep{Gaia16} along with stellar properties from large spectroscopic surveys, such as APOGEE \citep{APOGEE17}, GALAH \citep{GALAH15, GALAH18}, and LAMOST \citep{LAMOST12}, now provide 6D kinematics and elemental abundances for stars throughout the Galaxy. Most notably, this data revealed that the MW underwent a massive merger -- the Gaia-Sausage-Enceladus (GSE) -- $\approx 8$ to $11 \Gyr$ ago \citep[for example][]{Belokurov18, Helmi18}.
The GSE merger, probably the most major merger of the MW, contributed most of the stars in the MW's stellar halo, and it likely dynamically heated MW stars to more eccentric halo-like orbits, and potentially it triggered intense star-formation and even the formation of the MW's low-$\alpha$ disk \citep[for example][]{Haywood18, Gallart19, Mackereth19a, Bonaca20, Das20}. Recent works suggest that the MW developed this thick -- but continuously-settling -- disk early in its history, with estimates for the onset of coherent rotation ranging from $\approx 10$ to $13 \Gyr$ ago \citep{Belokurov22, Conroy22, Xiang22}, and that the MW's thin-disk began to form later, $\approx 8 - 9 \Gyr$ ago. However, post-formation dynamical heating leads to ambiguity about the nature and timing of these transitions, highlighting the difficulties innate to separating birth kinematics from post-formation dynamical heating, especially for old, metal-poor stars.

Cosmological simulations of the formation of MW-mass galaxies provide important tools for understanding the relative impacts of cosmological disk settling and post-formation dynamical heating \citep[see][for recent reviews]{Naab17, Vogelsberger20}. 
%and thus understand observations of the MW and nearby galaxies today 
Cosmological zoom-in simulations provide the high resolution needed to model a galaxy's internal structure, including the multi-phase ISM that sets the birth kinematics of stars and thus the dynamical structure of the disk, within an accurate cosmological context, allowing for non-equilibrium dynamical perturbations, mergers, mass growth, and so forth.
Many works used such cosmological zoom-in simulations to show that (1) MW-like galaxies form radially `inside-out' and vertically `upside-down', transitioning from a thick, turbulent galaxy at early times to a thin, rotationally dominated disk at later times, and (2) the present-day kinematics of stars result from their kinematics at birth and subsequent dynamical heating, both of which change over cosmic time \citep[for example][]{Bird12, Brook12, Martig14, Grand16, Bonaca17, Ceverino17, Ma-17, Buck20, Agertz21, Bird-21}.

We build upon a series of papers that used FIRE-2 simulations to study the formation of galactic disks.
%Numerous works have used the FIRE simulations to find that MW-like progenitors were turbulent, gas-rich, and morphologically-irregular at early cosmic times ($z \gtrsim 1 - 2$) and hosted highly time-variable `bursty' star formation \citep[for example][]{aa17_cbc, Sparre17, muratov17, el-badry18, Flores21}. However, these bursty and chaotic galaxies transformed into stable disks with steady star formation by $z=0$ \cite[for example][]{Ma-17, Garrison-Kimmel18, Hung19}. 
%Previous works have found that MW-like galaxies in FIRE formed in fairly distinct phases defined by the co-evolution of various galaxy properties. 
%At early times, chaos reigned: gas accreted haphazardly onto the galaxy's inner regions via supersonic filamentary flows, SF occurred in bursts such that feedback drove significant gas outflows, the quasi-isotropic ISM displayed little coherent rotation and was supported by turbulence and bulk flows, and stars were born on eccentric orbits \citep[for example][]{Muratov15, muratov17, aa17_cbc, Sparre17, el-badry18, Flores21, Gurvich22, Yu22}. Galaxies then rapidly transitioned into stable, orderly disks at late times. By $z=0$, gas accretes coherently onto the galaxy via rotating, subsonic flows, SF is steady such that outflows cease, and stars form on nearly circular orbits in a thin-disk \citep[][]{Ma-17, orr20, Yu2021, Hafen22}.
\citet{Stern21} found that virialization of the inner circumgalactic medium (CGM), that is, the onset of a stable hot halo, coincided with the formation of a thin disk, a change from `bursty' to `steady' star formation, and the suppression of gaseous outflows; the presence of a hot halo can allow infalling gas to align coherently with a galaxy's disk \textit{before} it accretes onto the disk \citep{Hafen22}.
%The formation of these halos is a consequence of galactic mass growth: the cooling time of gas in the circumgalactic medium (CGM) increases with galaxy mass\citep[for example][]{Birnboim03}, so that once a galaxy's mass passes a certain threshold ($\approx10^{12} M_\odot$), the gas cooling time exceeds the (previously larger) freefall time and the inner-CGM `virializes'\citep{Stern21}. 
Additional works found that the virialization of the inner CGM and the associated transition from bursty to steady star formation coincided with the transition from `thick-disk' to `thin-disk' formation \citet{Yu2021, Yu22} and the emergence of a rotationally-supported thin gaseous disk from a previously quasi-isotropic ISM \citep{Gurvich22}.
More recently, \citet{Hopkins23} investigated the physical causes of such transitions. While the initial formation of a disk depends on the development of a sufficiently centrally concentrated mass profile, the transition from bursty to smooth star formation generally arises later, when the absolute depth of the gravitational potential (within the radius of star formation) crosses a critical threshold. This threshold is similar to the threshold for virialization of the inner CGM.%, such that the transition to smooth star formation typically accompanies inner-CGM virialization and thin-disk formation.

In this work, we use the FIRE-2 cosmological zoom-in simulations to study cosmological disk settling and dynamical heating across the formation histories of MW-mass galaxies.
We study the kinematics of stars both at birth and today, emphasizing different phases of disk-galaxy formation. We examine how stellar formation kinematics manifest themselves today, that is, how post-formation dynamical heating alters the initial orbits of stars. By doing so, we aim to understand how the current kinematics of the MW and nearby galaxies relate to their formation histories.

\section{Methods}
\label{section:methods}

\subsection{FIRE-2 Simulations}

We analyze 14 cosmological zoom-in simulations of MW/M31-mass galaxies from the Feedback in Realistic Environments (FIRE) project\footnote{
FIRE project web site: \href{http://fire.northwestern.edu}{http://fire.northwestern.edu}
}, which include both dark matter and baryons (gas and stars). The simulations use the GIZMO gravity plus hydrodynamics code \citep{Hopkins15} with the meshless finite-mass (MFM) Godunov method, which provides adaptive spatial resolution while maintaining excellent conservation of energy and (angular) momentum.

We ran these simulations using the FIRE-2 physics model \citep{Hopkins18}.
FIRE-2 models the dense, multi-phase ISM and incorporates metallicity-dependent radiative cooling and heating processes for gas across temperatures $10 - 10^{10}$ K, including free-free, photoionization and recombination, Compton, photoelectric and dust collisional, cosmic ray, molecular, metal-line, and fine structure processes. The simulations self-consistently generate and track 11 elements (H, He, C, N, O, Ne, Mg, Si, S, Ca, Fe), including a model for sub-grid mixing/diffusion via turbulence \citep{Hopkins16, Su17, Escala18}. FIRE-2 also includes photoionization and heating from a spatially uniform, redshift-dependent UV background \citep{Faucher09} that reionizes the Universe at $z \approx 10$.

Crucially for this work, FIRE-2 simulations resolve the phase structure in the ISM, allowing gas to collapse into giant molecular clouds and form stars \citep{Benincasa20, Guszejnov20}. Star formation occurs in locally self-gravitating, Jeans-unstable, dense ($n_{\rm SF} > 1000$ $\rm cm^{-3}$), molecular \citep[following][]{KG11} gas. After reaching these criteria, a gas cell probabilistically converts to a star particle in a local free-fall time. Our simulations, like all simulations, have a finite resolution. As such, our star-formation model is an extrapolation of the collapse below the resolution scale. That said, numerous resolution tests carried out in \citet{Hopkins18} indicate that uncertainties induced by our resolution limit are minor; for example, the GMC mass function and the relation between internal gas velocity dispersion and GMC size for these simulations are both independent of resolution.

Each star particle represents a single stellar population sampled from a \citet{Kroupa01} initial mass function, with mass and metallicity inherited from its progenitor gas cell. Once formed, star particles follow stellar evolution models that tabulate feedback event rates, luminosities, energies, and mass-loss rates from STARBURST99 v7.0 \citep{Leitherer99}. FIRE-2 implements the major channels for stellar feedback, including core-collapse and white-dwarf (Ia) supernovae, stellar winds, radiation pressure, photoionization, and photoelectric heating.

We generated cosmological zoom-in initial conditions at $z \approx 99$ using \textsc{MUSIC} \citep{hahn11}. Each zoom-in region is embedded within a cosmological box of side length $70.4 - 172 \Mpc$. The simulations use a flat $\Lambda$CDM cosmology with parameters broadly consistent with \citet{planck20}: $h = 0.68 - 0.71$, $\Omega_{\Lambda} = 0.69 - 0.734$, $\Omega_{\rm m} = 0.266 - 0.31$, $\Omega_{\rm b} = 0.0455 - 0.048$, $\sigma_{\rm 8} = 0.801 - 0.82$ and $n_{\rm s} = 0.961 - 0.97$.

Our simulation set consists of 14 MW/M31-mass galaxies: 8 isolated galaxies from the \textit{Latte} suite, and 6 galaxies from the \textit{ELVIS on FIRE} suite of Local Group (LG)-like pairs of galaxies. The \textit{Latte} suite, with the exception of m12z, has an initial baryon particle mass of $7070 \Msun$, although stellar mass loss leads to star particles having typical masses $\approx 5000 \Msun$, and dark matter particle mass of $3.5 \times 10^5 \Msun$. m12z has initial baryon and dark matter particle masses of $4200 \Msun$ and $2.1 \times 10^5 \Msun$, respectively. The \textit{ELVIS on FIRE} suite has a mass resolution $\sim 2 \times$ better than the \textit{Latte} suite: Romeo \& Juliet have initial baryon particle masses of $3500 \Msun$, while Romulus \& Remus and Thelma \& Louise have $4000 \Msun$.

Both star and dark-matter particles have fixed gravitational force softenings (comoving at $z > 9$ and physical thereafter) with a Plummer equivalent of $\epsilon_{\rm star} = 4 \pc$ and $\epsilon_{\rm dm} = 40 \pc$. Gas cells have fully adaptive gravitational softenings that match the hydrodynamic kernel smoothing: softening lengths are $\approx 20 - 40 \pc$ at typical ISM densities ($n \sim 1 \rm cm^{-3}$) and $< 1 \pc$ in the densest regions.

Certain aspects of the FIRE-2 ISM model have been subject to scrutiny within the literature. In particular, \citet{Kim23} and \citet{Wibking23} argued that the equilibrium gas cooling in FIRE-2 transitions between the warm and cold phase in the neutral ISM at unphysically low pressures and densities due to our order of too-low free-electron fraction at a given gas density. This implies that the diffuse ISM in FIRE-2 is entirely in the cold neutral phase. However, their idealized tests did not encompass how FIRE-2 simulations actually implement gas cooling; instead, they only tested one aspect of gas cooling in FIRE-2 and only in the presence of a metagalactic UV background, with no local sources. The free-electron fraction and ISM phase structure in actual FIRE-2 simulations agree much better with the observations presented in those works.\footnote{
The FIRE collaboration has made numerous improvements to the gas cooling model in recent years: the default cooling in FIRE-2 since 2020, and in FIRE-3, includes all terms claimed to be missing from earlier FIRE models.
See the GIZMO user guide:
\href{www.tapir.caltech.edu/~phopkins/Site/GIZMO_files/gizmo_documentation.html}{http://www.tapir.caltech.edu/$\sim$phopkins/Site/GIZMO\_files/gizmo\_documentation.html}
}
Furthermore, the specifics of the ISM cooling model are largely inconsequential for the properties that we examine here, because the thermal pressure essentially never dominates over the turbulent pressure in the cold, dense ISM resolved in these simulations.
Indeed, as \citet{Hopkins18} showed, dramatic changes to the assumed ISM cooling rates lead to only minor effects on most galaxy dynamical properties.
Indeed, we tested re-simulating one of our simulations (m12i) over the last $500 \Myr$ using the newer FIRE-3 model \citep{Hopkins23}, which includes numerous updates to gas cooling, and this results in nearly identical velocity dispersions for cold gas and stars.

For our main results (Sections~\ref{subsec:stellardynamics}, \ref{subsec:component comparsion}, \ref{subsec:formation vs post-formation}, and \ref{subsec:heating rate}), we present properties averaged across our galaxy sample, not including the lower-mass galaxies m12r and m12z or Juliet from the \textit{ELVIS on FIRE} suite. However, in Sections~\ref{subsec:galaxy mass trends}, \ref{subsec:observation} and \ref{subsec:disk settling time}, where we explore dependence on galaxy mass, we include all 14 galaxies. m12r and m12z have stellar masses of 1.8 and $1.5 \times 10^{10} \Msun$, lower than the MW's and M31's stellar masses of $\approx 5 \times 10^{10}$ and $\approx 10^{11} \Msun$, respectively. Furthermore, m12r and m12z have stellar kinematics that are weakly or never dominated by disk-like motion. Juliet has a stellar mass closer to the MW's, at $3.3 \times 10^{10} \Msun$, and disk-dominated kinematics at $z = 0$, but we exclude it, because a rapid gas accretion event late in its history leads to anomalous properties that skew the sample-averaged properties towards trends unique to Juliet. We will examine these cases in more detail in future work.

\subsection{Measuring Stellar Dynamics Today and at Formation}
\label{subsec:selection}

Motivated primarily by measurements of the MW, we select stars that are in each galaxy's `solar annulus': $R = 6 - 10 \kpc$ and $|Z| < 3 \kpc$ today.  In future work, we will examine trends with radius, to understand the drivers of post-formation dynamical heating and their application to MW observations in greater detail; briefly, we find that the dispersion today generally decreases with $R$ today, but that neither the dispersion at formation nor the dispersion today depend on $R$ at $R \gtrsim 1 - 2 \kpc$.
The effect of the vertical selection on our results is minimal, as we show in Appendix~\ref{a:vertical_selection}.

For all of our results besides those in Section~\ref{fig:local_vs_global}, we show `disk-wide' (or more aptly, `annulus-wide') kinematics. That is, we compute median velocity and velocity dispersions using all the stars (in a given age bin) within the `solar annulus'. However, in Section~\ref{fig:local_vs_global}, we show `local' kinematics, meaning we compute the median velocity and dispersion within smaller patches (of diameter 150 or 500 pc) spanning an annulus centered at 8 kpc, and then average across all patches.

A key advantage of simulations is their ability to study properties both today and when stars formed. In both cases, we select stars on the basis of their \textit{current} coordinates. However, when determining properties at formation, we only include stars that formed in situ, which we define following \citet{Bellardini22} as forming within a total distance of 30 kpc comoving from the galactic center. We do not apply this in-situ selection when measuring properties today, to more directly compare with observations.
Across our simulations, ex-situ stars comprise only $\sim 4\%$ of all stars currently within the disk ($R < 20 \kpc$ and $|Z| < 3 \kpc$), but they can make up a non-trivial percentage within our oldest age bins: we show the effect of our in-situ selection on velocity dispersion versus age in Appendix~\ref{a:insitu_selection}.

We compute stellar positions and velocities relative to the galaxy's center. We determine the orientation of the galaxy at each snapshot (independently) by computing the rotation tensor and axis ratios of the principal axes of the stellar disk, which we define via the moment of inertia tensor of the 25\% youngest star particles inside a radius enclosing 90\% of the total stellar mass within 10 kpc physical. We compute properties at formation using the galaxy's orientation at that time, not at $z = 0$.

Our simulations store 600 snapshots from $z = 99$ to 0, with a time between snapshots of $20 - 25 \Myr$, and the past 20 Myr (immediately before today) includes 10 snapshots spaced by 2 Myr.
We measure a star's `formation' properties at the first snapshot after it formed, so a star particle on average experienced $\approx 10 \Myr$ of evolution when we record its properties.
This limits to some extent our ability to measure the earliest dynamical evolution, especially because, in our simulations, stars are born clustered in GMCs \citep[for example][]{Benincasa20, Guszejnov20}, where they can be subject to various gravitational interactions on short timescales.
Thus, `formation' properties in our analysis necessarily refer to the combination of birth properties and early dynamical evolution.

When computing the velocities of a stellar population, we select star particles using 250 Myr bins of age, which corresponds to the dynamical/orbital time at the solar annulus today.
When plotting velocities, we use the mass-weighted median velocity of all star particles within a given age range; we find nearly identical results using the mass-weighted mean instead. To ensure that outlier stars do not skew our results, we compute the velocity dispersion by finding half of the difference between the mass-weighted 16th and 84th-percentile velocities of all stars in a given age range, which is equivalent to the standard deviation for normal distribution. We compute these quantities (in each age bin) individually for each galaxy and then show the mean and standard deviation across our 11 galaxies.

\begin{table*}
\centering
\begin{tabular}{|p{0.65 cm}|c|c|c|c|c|c|c|c|c|c|}
\hline
Name & $M_{\rm star}^{\rm 90}$ & $v_{\rm \phi}^{\rm disk}$ & $\sigma_{\rm Z}^{\rm disk}$ & $\sigma_{\rm tot}^{\rm disk}$ & $\sigma_{\rm Z}^{\rm local}$ & $\sigma_{\rm tot}^{\rm local}$ & $t_{\rm lb} \left[ \left(\frac{v_{\rm \phi}}{\sigma_{\rm tot}} \right)_{\rm form}>1\right]$ & $t_{\rm lb} \left[ \left(\frac{v_{\rm \phi}}{\sigma_{\rm tot}} \right)_{\rm now} > 1 \right]$ & $t_{\rm lb} \left[ \left( \frac{v_{\rm \phi}}{\sigma_{\rm Z}} \right)_{\rm now} > \sqrt{3} \right]$ & $t_{\rm lb}^{\rm burst}$ \\
& $\mathrm{[10^{10} M_\odot]}$ & [km/s] & [km/s] & [km/s]  & [km/s] & [km/s] & [Gyr] & [Gyr] & [Gyr] & [Gyr] \\
\hline
m12m & 10.0 & 288 & 16.7 & 50.1 & 13.3 & 33.1 & 9.21 & 6.90 & 7.68 & 3.81 \\
Romulus & 8.0 & 265 & 20.4 & 55.9 & 16.6 & 41.8 & 7.42 & 7.16 & 7.16 & 4.90 \\
m12b & 7.3 & 272 & 17.9 & 56.6 & 13.1 & 37.5 & 7.42 & 7.42 & 7.42 & 6.32 \\
m12f & 6.9 & 256 & 19.9 & 74.9 & 13.5 & 31.6 & 7.42 & 6.39 & 6.39 & 5.01 \\
Thelma & 6.3 & 230 & 21.2 & 62.8 & 19.4 & 43.9 & 4.35 & 4.35 & 4.25 & 2.57 \\
Romeo & 5.9 & 249 & 14.1 & 40.3 & 12.2 & 31.3 & 11.0 & 10.2 & 10.2 & 6.52 \\
m12i & 5.3 & 238 & 20.1 & 52.4 & 12.8 & 27.5 & 6.65 & 6.39 & 6.39 & 3.14 \\
m12c  & 5.1 & 240  & 23.1 & 62.8 & 19.5 & 38.4 & 6.49& 6.14 & 6.39 & 3.70 \\
m12w & 4.8 & 163 & 16.6 & 83.9 & 5.5 & 29.8 & 4.09 & 4.6 & 5.89 & 0 \\
Remus & 4.0 & 216 & 13.9 & 40.3 & 10.7 & 25.0 & 7.93 & 7.93 & 7.93 & 5.88 \\
Juliet & 3.3 & 211 & 16.6 & 45.6 & 15.0 & 32.6 & 4.35 & 4.61 & 4.61 & 4.40 \\
Louise & 2.3 & 177 & 14.6 & 37.5 & 11.0 & 26.1 & 7.16 & 6.9 & 7.16 & 5.56 \\
m12z & 1.8 & 92.6 & 21.4 & 67.6 & 19.0 & 39.0 & 0.51 & - &- & - \\
m12r & 1.5 & 127 & 17.4 & 47.1 & 7.3 & 15.4 & 5.9 & 0.26 & 2.3 & - \\
\hline
sample mean & 6.0 & 236 & 18.0 & 56.1 & 13.5 & 32.4 & 7.19 & 6.76 & 6.98 & 4.31 \\
\hline
\end{tabular}
\vspace{-1 mm}
\caption{
\textbf{Stellar properties at $z = 0$ of the FIRE-2 simulations of MW/M31-mass galaxies that we analyze}, in order of decreasing stellar mass. The first column lists the galaxy name: `m12' indicates an isolated galaxy; otherwise the galaxy is in a Local Group-like pair. $M_{\rm star}^{\rm 90}$ is the stellar mass within $R_{\rm star}^{\rm 90}$, the radius enclosing 90\% of the stellar mass within 20 kpc. $v_{\rm \phi}^{\rm disk}$, $\sigma_{\rm Z}^{\rm disk}$, and $\sigma_{\rm tot}^{\rm disk}$ are the \textit{disk-wide} median azimuthal velocity, vertical velocity dispersion, and total velocity dispersion, respectively, of stars younger than 250 Myr in a cylindrical radius $R = 6 - 10 \kpc$ and $|Z| < 3 \kpc$ from the midplane. $\sigma_{\rm Z}^{\rm local}$ and $\sigma_{\rm tot}^{\rm local}$ are the \textit{local} vertical and total velocity dispersion of stars younger than 100 Myr measured in 500 pc patches centered at $R = 8 \kpc$. The first 3 lookback times are the onset of disk formation, defined as the oldest age/lookback time that stars permanently passed above a disk-wide threshold: $v_{\rm \phi} / \sigma_{\rm tot} > 1$ at formation, $v_{\rm \phi} / \sigma_{\rm tot} > 1$ measured at present, and $v_{\rm \phi} / \sigma_{\rm Z} > 1.8$ measured at present. The last column lists the lookback time when the galaxy transitioned from bursty to steady star formation from \citet{Yu2021}. The last row shows the average quantities for our sample (excepting m12r, m12z, and Juliet, see text).
\vspace{-2 mm}
}
\label{table:table1}
\end{table*}

\section{Results}

\subsection{Stellar Kinematics versus Age}
\label{subsec:stellardynamics} 

\subsubsection{Defining the Eras of Disk Evolution}
\label{subsubsec:disk eras}

Motivated by the kinematic trends that we explore and by previous analyses of these galaxies \citep[for example][]{Garrison-Kimmel18, Ma-17, FG18, Yu2021, Gurvich22}, we divide our galaxies' histories into three eras, which we term the `Pre-Disk', `Early-Disk', and `Late-Disk' Eras.
Each of our galaxies began in the Pre-Disk Era, when the gas lacked coherent rotation, such that stars formed on nearly isotropic orbits. Then, each galaxy formed a stable, coherent disk and transitioned into the Early-Disk Era. Specifically, the start of the Early-Disk Era marks the time when all subsequent generations of stars that formed with rotationally-dominated kinematics, with  $(v_{\rm \phi} > \sigma_{\rm tot})_{\rm form}$. Table~\ref{table:table1} lists the lookback time, $t_{\rm lb}$, that each galaxy transitioned into the Early-Disk Era, $t_{\rm lb}\left[ \left(v_{\rm \phi} / \sigma_{\rm tot} \right)_{\rm form} > 1 \right]$.

These disks were kinematically hot, with high $\sigma$, at the start of the Early-Disk Era, but they became increasingly settled, with increasingly cold and rotationally-dominated kinematics, throughout it. As a result, $(v_{\rm \phi} / \sigma_{\rm tot})_{\rm form}$ grew most rapidly throughout this era. Eventually, the growth of $(v_{\rm \phi} / \sigma_{\rm tot})_{\rm form}$ slowed and the galaxies transitioned into the Late-Disk Era, when the kinematics at formation did not evolve much as stars formed on highly rotational orbits in a dynamically cold disk. On average, galaxies in our sample transitioned from the Pre-Disk to the Early-Disk Era $\approx 8 \Gyr$ ago and from the Early-Disk to the Late-Disk Era $\approx 4 \Gyr$ ago, though with significant galaxy-to-galaxy scatter in these ages.

\citet{Hopkins23} argued that two aspects of the galaxy's gravitational potential drive the transitions between these eras.
The \textit{shape} of the potential, specifically the formation of a sufficiently centrally-concentrated mass profile, drives the initial formation of a disk, that is, the Pre-Disk to Early-Disk transition.
%A sufficiently centralized potential provides the galaxy with a dynamic center, allowing for the conservation of angular momentum,  stabilizes the disk against strong non-axisymmetric modes, such as spiral arms or bars, and promotes orbit-mixing, facilitating the coherence of angular momentum.
Separately, the \textit{absolute depth} of the potential (within the radius of star formation) drives the transition from bursty to smooth star formation, which roughly coincides with the transition from the Early-Disk to Late-Disk era, although this connection may be less direct.
%A sufficiently deep potential `traps' and/or confines outflows near the disk, enabling short recycling times such that star formation bursts `blur' together in time.
Similarly, our Pre-Disk and Early-Disk Eras generally correspond to the `bursty star formation' phase, in \citet{Stern21}, \citet{Gurvich22} and \cite{Yu2021}, while our Late-Disk Era corresponds to their `steady star formation' phase.

\subsubsection{Median velocity versus age}
\label{subsubsec:median velocity}

We start by analyzing the general dynamic evolution of our simulated galaxies. Figure~\ref{fig:fig1} (top) shows the sample-averaged azimuthal, radial, vertical, and total median velocities at formation (dashed lines) and at present (solid lines) versus stellar age. We label the Late-, Early-, and Pre-Disk Eras on the top of each column, and we mark the start of the Early and Late-Disk Eras with shaded bars. The time that one era ended and another began is not exact, in part because Figure~\ref{fig:fig1} shows sample-averaged quantities, so the shaded bars represent the approximate transition time and not an instantaneous boundary.

Figure~\ref{fig:fig1} (top left) shows the median azimuthal (rotational) velocity, $v_{\rm \phi}$, versus age.
For most of the Pre-Disk Era, stars formed with $v_{\rm \phi, form} \approx 0 - 20 \kms$: in the early galaxy, stars did not coherently rotate. Eventually, $v_{\rm \phi, form}$ began to increase as the disk began to settle, and the galaxy transitioned into the Early-Disk Era. The slight growth of $v_{\rm \phi, form}$ towards the end of the Pre-Disk Era arises because we average over multiple galaxies, so earlier-transitioning galaxies bias the sample's mean: within individual galaxies, the rise of $v_{\rm \phi, form}$ is generally (but not always) sharp and coincident with the transition to the Early-Disk Era (see Figure~\ref{fig:casestudies}). $v_{\rm \phi, form}$ strongly rose throughout the Early-Disk Era, but its growth weakened once the galaxy transitioned into the Late-Disk Era ($\approx 4 \Gyr$ ago), likely because the total mass within our radial range changed little during this period \citep{Garrison-Kimmel18}.
$v_{\rm \phi, now}$ primarily follows the same trends with age as $v_{\rm \phi, form}$. However, stars that formed in the Pre-Disk Era have significantly larger $v_{\rm \phi}$ now than at formation, because they were torqued onto more coherent rotational orbits as the disk subsequently formed and settled. Likely, gas-rich mergers that seeded the disk's orientation also torqued up these existing stars \citep[for example][]{Santistevan21}.
Bellardini et al., in preparation, show that this increase in $v_{\rm \phi}$ corresponds to an increase in specific angular momentum for these pre-existing stars, that is, they were indeed torqued. In contrast, stars that formed in the Late-Disk Era on nearly circular orbits now have slightly smaller $v_{\rm \phi}$ now than at formation because of dynamical heating/scattering.

Figure~\ref{fig:fig1} (top row, column 2 and column 3) shows the median radial velocity, $v_{\rm R}$, and the median vertical velocity, $v_{\rm Z}$, versus stellar age. Previous work analyzing FIRE simulations found that bursty star formation in the Pre-Disk galaxy drove significant gaseous outflows \citep[for example][]{Muratov15, aa17_cbc, Feldman17, Sparre17, Pandya21, Stern21}. These outflows sometimes remained star-forming and produced stars that inherited their progenitor's outflowing dynamics \citep{Badry2016, Yu2021}. We find complementary results: stars that were born in the Pre-Disk Era had positive (outward) $v_{\rm R, form}$, indicating the presence of star-forming outflows, and $v_{\rm Z, form}$ that rapidly fluctuated around 0. Once the early disk formed, $v_{\rm R, form}$ decreased to 0, in part because feedback-driven outflows declined and eventually ceased: throughout the rest of the Early-Disk and Late-Disk Eras, stars had $v_{\rm R, form} \approx v_{\rm Z, form} \approx 0$.

At present, both $v_{\rm R, now}$ and $v_{\rm Z,now} \approx 0$ for stars that formed in \textit{all} eras, meaning that old stars' subsequent dynamical evolution overrode any preference for outward/upward/downward motion. 
This difference also reflects our selection of  stars that remain within the disk today.
Our galaxies largely formed radially `inside-out' \citep[for example][]{Garrison-Kimmel18}, such that old in-situ stars typically formed inside of our current radial range of $6 - 10 \kpc$, and as \citet{Badry2016} and \citet{Ma-17} discussed for the FIRE simulations, old in-situ stars typically reside at larger radii today, compared to when they formed, because bursts of feedback-driven outflows generated large fluctuations in the gravitational potential at early times that moved these stars outward. We note that this change in radius for old stars is distinct from `radial migration', in that the vast majority of these stars have moved coherently outward, while radial migration entails a roughly equal number of inward and outward moving stars. 
As such, old stars in our sample are more likely than young stars to be near their apocenters, and as a result, have $v_{\rm R, now}$ closer to 0.
Bellardini et al., in preparation will examine this radial redistribution in detail.

Figure~\ref{fig:fig1} (top right) shows the median total velocity, $v_{\rm tot}$, versus age. The growth of the galaxy's mass drove much of the evolution of $v_{\rm tot}$, because the underlying gravitational potential determines its virial velocity. In turn, the evolution of $v_{\rm tot, form}$ indicates that the galaxy's mass increased over time, most rapidly in the Pre-Disk Era and only minimally in the Late-Disk Era.
The evolution of $v_{\rm tot, form}$ largely resembles that of $v_{\rm \phi, form}$, with the exception of the Pre-Disk Era, when $v_{\rm tot, form}$ grew steadily, while $v_{\rm \phi, form}$ remained near 0. This is not surprising: $v_{\rm tot}$ depends only on the magnitude of each component, while $v_{\rm \phi}$ also depends on the sign. In the Pre-Disk Era, stars rotated, just not coherently, such that the median $v_{\rm \phi, form} \approx 0$ while $\sqrt{v_{\rm \phi, form}^2} > 0$. However, once the galaxy transitioned into the Early-Disk Era, stars had larger and larger fractions of their (increasing) $v_{\rm tot,form}$ in rotation.
At present, $v_{\rm tot}$ has almost no dependence on age, with all stars near $v_{\rm tot,now} \approx v_{\rm virial} \approx 200 - 240 \kms$, as a result of virialization and dynamical heating.

\subsubsection{Velocity dispersion versus age}
\label{subsubsec:velocity dispersion}

\begin{figure*}
\includegraphics[width = \textwidth]{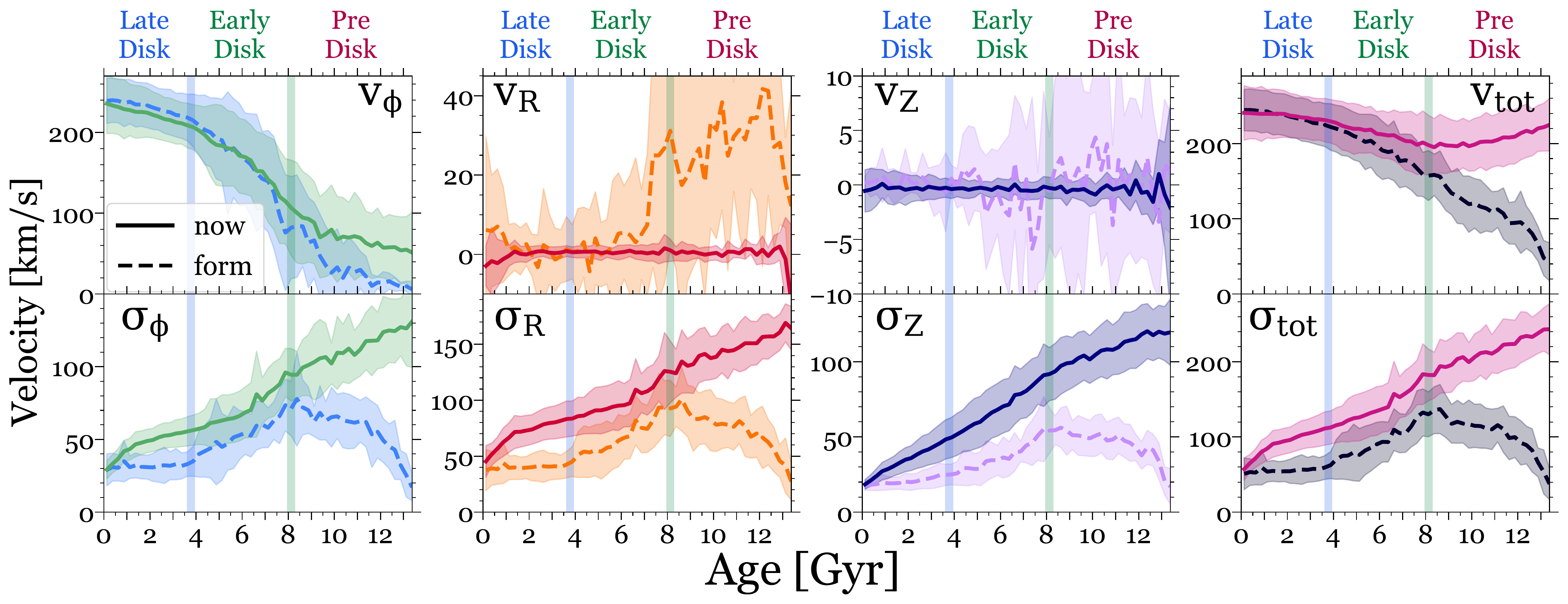}
\vspace{-6 mm}
\caption{
\textbf{Median velocities and velocity dispersions for stars versus age}, within cylindrical $R = 6 - 10 \kpc$ and $|Z| < 3 \kpc$ at $z = 0$.
Age bins have a width of 250 Myr.
The lines show the average and the shaded regions show the standard deviation across 11 \ac{MW}-mass galaxies, with velocities measured at $z = 0$ (solid) and at the time of formation (dashed, including only in-situ stars). The shaded vertical bars show when our galaxies transitioned from the Pre-Disk to the Early-Disk Era ($\approx 8$ Gyr ago) and the Early-Disk to the Late-Disk Era ($\approx 4$ Gyr ago), on average: we define these eras in Section~\ref{subsubsec:disk eras}.
\textbf{Top row}: Median velocity versus age.
For most of the Pre-Disk Era, stars formed with little net rotation, but $v_{\rm \phi, form}$ rapidly increased in the Early-Disk Era when the disk initially formed, and it increased at a lower rate in the Late-Disk Era when the disk further settled.
Stars that formed in the Pre-Disk Era have $v_{\rm \phi, now} > v_{\rm \phi, form}$ from dynamical torquing after formation, while stars that formed in the Late-Disk Era now have slightly smaller $v_{\rm \phi, now}$ than at formation from post-formation dynamical heating.
$v_{\rm R}$ and $v_{\rm Z}$ fluctuated at formation in the Pre- and Early-Disk Eras, because of frequent mergers, rapid accretion, bursty star formation, and gaseous outflows, but today they are 0 across all ages.
$v_{\rm tot, form}$ steadily increased at all times, but $v_{\rm tot, now}$ is nearly independent of age, because all stars have been dynamically heated to orbit near the virial velocity.
\textbf{Bottom row}: Velocity dispersion versus age.
All 3 components and the total show the same trends with age.
$\sigma_{\rm form}$ increased throughout the Pre-Disk Era as the galaxy grew, peaking $\approx 8 \Gyr$ ago, but decreased throughout the Early-Disk Era and remained constant throughout the Late-Disk Era.
By contrast, $\sigma_{\rm now}$ monotonically increases with stellar age because of the dynamical heating of stars after their formation.
Thus, \textit{the kinematics of stars today do not simply reflect their kinematics at formation.}
\vspace{-2 mm}
}
\label{fig:fig1}
\end{figure*}

Figure~\ref{fig:fig1} (bottom) shows the azimuthal, radial, vertical, and total stellar velocity dispersion, $\sigma$, versus age, at both the time of the star's formation and at present. All components exhibit similar trends with age, albeit with different normalizations. Here we focus on general trends with age; we compare the 3 components in Section~\ref{subsec:component comparsion}.

$\sigma_{\rm form}$ reflects the turbulence in star-forming gas and shows distinct behavior in each kinematic era. The oldest stars formed with the lowest $\sigma$, reflecting their low $v_{\rm tot,form}$ in the low-mass galaxy progenitor at early times. As the galaxy grew during the Pre-Disk Era, $\sigma_{\rm form}$ grew, reflecting the deepening of the gravitational potential in addition to an increase in absolute gas turbulence, from accretion, mergers, bursty and more vigorous star formation, and outflows \citep[for example][]{Ma-17, el-badry18, Hung19, Flores21}.

$\sigma_{\rm form}$ reached a maximum during the transition to the Early-Disk Era. 
This peak in $\sigma_{\rm form}$ reflects the competing effects of the galaxy continuing to become more massive (increasing $v_{\rm tot,form}$) but also starting to settle into a disk, decreasing the relative contribution to $\sigma_{\rm form}$.
$\sigma_{\rm form}$ steadily decreased throughout the Early-Disk Era as the disk settled.
This steady decrease in turbulence could occur only after the initial onset of the disk, when it could self-regulate to maintain a turbulent Toomre $Q \sim 1$ in star-forming gas \citep[for example][]{Hopkins12, FG13, Wisnioski15, Krumholz18, Gurvich20, Orr20}. As such, equilibrium models imply that a gas disk maintains $\sigma / v_{\rm \phi} \approx M_{\rm gas}(< r) / M_{\rm tot}(< r) \approx f_{\rm gas}$, so $\sigma$ reduced as $f_{\rm gas}$ did.

Finally, upon the transition to the Late-Disk Era, $\sigma_{\rm form}$ reached a floor and remained nearly constant, reflecting the plateauing gas fraction and correspondingly steady and low turbulence and SFR during this era \citep{Ma-17, Yu2021, Garrison-Kimmel18, Hung19, Gurvich22}.

While $\sigma_{\rm form}$ shows distinct trends in each era, $\sigma_{\rm now}$ shows markedly different behavior, monotonically increasing with age across all eras.
\textit{Thus, the additional dispersion from post-formation dynamical heating, in combination with $\sigma_{\rm form}$, produces the observed continuous increase in dispersion with stellar age; neither process alone explains it.
This also means that $\sigma_{\rm now}$ does not directly and simply reflect the formation history of the galaxy.}
This difference is most evident for stars that formed in the Pre-Disk Era, which have the highest values of $\sigma_{\rm now}$ today but formed with the smallest values of $\sigma_{\rm form}$.
By contrast, stars forming in the Early-Disk Era exhibit an age evolution that follows a similar slope at formation and at present, modulo a normalization offset.
Finally, stars that formed in the Late-Disk Era have $\sigma_{\rm now}$ increasing with age their while $\sigma_{\rm form}$ is constant.
We examine the age dependence of post-formation dynamical heating in detail in Section~\ref{subsec:formation vs post-formation}.

In conclusion, the age dependence of $\sigma_{\rm form}$ differed in the Pre-Disk, Early-Disk, and Late-Disk Eras, reflecting the distinct kinematics of these eras. At present, $\sigma_{\rm now}$ increases monotonically with age, with a relation to $\sigma_{\rm form}$ that differs for each era.

\subsubsection{Rotational support versus age}
\label{subsubsec:rotational_support}

\begin{figure}
\includegraphics[width = \columnwidth]{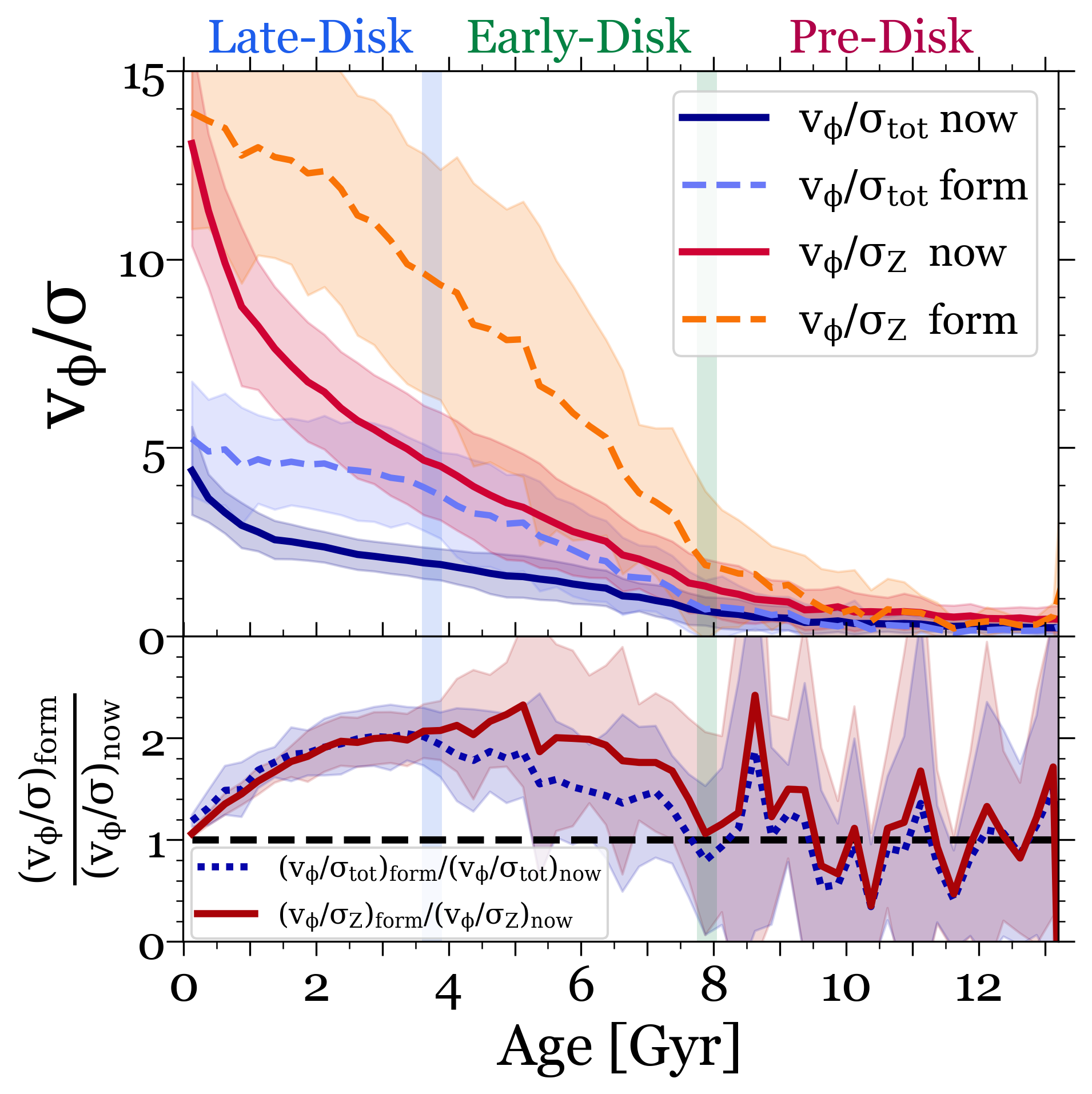}
\vspace{-6 mm}
\caption{
\textbf{Top}: The ratio of median $v_{\rm \phi}$ to $\sigma$, that is, how rotationally supported or `disky' the stars are, versus age.
We average across 11 galaxies, and the shaded region shows the galaxy-to-galaxy standard deviation.
The solid purple line shows $(v_{\rm \phi} / \sigma_{\rm tot})_{\rm now}$, while the red shows $(v_{\rm \phi} / \sigma_{\rm Z})_{\rm form}$. The dashed light-purple and orange lines show the same but measured when the star formed.
In the Pre-Disk Era, stars formed with $v_{\rm \phi} / \sigma \approx 0$, but once the disk began to form, $(v_{\rm \phi} / \sigma_{\rm Z})_{\rm form}$ rapidly increased until today, so vertical settling continued until today.
By contrast, $(v_{\rm \phi} / \sigma_{\rm tot})_{\rm form}$ also initially increased rapidly during the Early-Disk Era, but it remained roughly constant in the Late-Disk Era.
\textbf{Bottom}: $(v_{\rm \phi} / \sigma)_{\rm form} / (v_{\rm \phi} / \sigma)_{\rm now}$ for $\sigma_{\rm tot}$ (dotted navy) and $\sigma_{\rm Z}$ (solid red), which measures how `disky' stars were when they formed relative to today. Both ratios have similar values and exhibit similar trends with age. Stars that formed in the Early- and Late-Disk Eras have ratios above 1 (and near 2 for most ages), because post-formation dynamical heating increased their dispersions. Pre-Disk stars have $(v_{\rm \phi} / \sigma)_{\rm now} \approx  (v_{\rm \phi} / \sigma)_{\rm form} \approx 0$, with fluctuations from mergers/accretion and feedback.
\vspace{-6 mm}
}
\label{fig:fig2}
\end{figure}

The ratio $v_{\rm \phi} / \sigma$ is a dimensionless measure of the degree of rotational support, that is, how `disky' or `dynamically cold' a disk is. Figure~\ref{fig:fig2} (top) shows the ratio of median $v_{\rm \phi}$ to $\sigma_{\rm Z}$ and $\sigma_{\rm tot}$, versus stellar age, averaged across our 11 galaxies. We do not show the ratios with $\sigma_{\rm R}$ or $\sigma_{\rm \phi}$, because they exhibit qualitatively similar trends as $\sigma_{\rm tot}$, albeit with higher normalizations.

In the Pre-Disk era, $(v_{\rm \phi} / \sigma_{\rm tot})_{\rm form} < 1$ (by definition). When $v_{\rm \phi, form}$ permanently exceeded $\sigma_{\rm tot, form}$ ($\approx 8 \Gyr$ ago on average), a galaxy transitioned into the Early-Disk Era, when successively younger populations were born with increasing $(v_{\rm \phi} / \sigma_{\rm tot})_{\rm form}$ and $(v_{\rm \phi} / \sigma_{\rm Z})_{\rm form}$.
%That is, stars formed dynamically-colder and thinner over time during the Early-Disk Era.
After the galaxy transitioned into the Late-Disk Era ($\approx 4 \Gyr$ ago on average), $(v_{\rm \phi} / \sigma_{\rm tot})_{\rm form}$ grew only slightly, 
%Such behavior reflects $\sigma_{\rm tot, form}$ also being at a (generally steady) minimum and $v_{\rm \phi, form}$ being at a (slightly growing) maximum.
so stars during this era formed with the same degree of total rotational support.
By contrast, $(v_{\rm \phi} / \sigma_{\rm Z})_{\rm form}$ continued to increase throughout the Late-Disk Era, with only slightly slower growth than in the Early-Disk Era,
%Thus, even though stars formed with a broadly constant ratio of rotation to total random motion in the Late-Disk Era, they formed with an increasing ratio of rotation to \textit{vertical} random motion. 
%meaning that populations formed kinematically `thinner'.
reflecting the more significant reduction in $\sigma_{\rm Z, form}$ over time.
Thus, we find more significant \textit{vertical} `settling' during the Late-Disk Era.
%A stellar population's post-formation evolution depends on the dynamical era when it formed.
%Stars that formed $\gtrsim$ 8 Gyr ago in the Pre-Disk Era and the start of the Early Disk Era with $v_{\rm \phi} / \sigma$ < 1 maintain their low values today. %Even though these stars may now have larger \vphi because they were torqued onto more rotational orbits, post-formation dynamical heating has imparted an even greater amount of dispersion. 
%Stars that formed after the disk began to settle have larger values of $v_{\rm \phi} / \sigma$ at formation than at present due to post-formation dynamical heating. 
%These ratios rise most rapidly between 0 and 1 Gyr, suggesting that the strongest decrease in a stellar population's rotational support occurs within their first Gyr.

Relative to formation, stars have lower $v_{\rm \phi} / \sigma$ as measured today. To better understand how the post-formation change in $v_{\rm \phi} / \sigma$ depends on age, Figure~\ref{fig:fig2} (bottom) shows the ratio $(v_{\rm \phi} / \sigma)_{\rm form} / (v_{\rm \phi} / \sigma)_{\rm now}$ versus age. This ratio indicates how `disky' a stellar population formed relative to how disky it is today, from the combined effect of post-formation dynamical heating and coherent torquing.

Younger stars, which formed in the Late-Disk Era, became increasingly less disky with age, as post-formation dynamical heating increased their dispersion. Stars older than $\approx 2 \Gyr$ exhibit ratios near 2, meaning they are currently half as disky as when they formed.
For stars that formed in the Early-Disk Era, this trend \textit{reverses}, as the post-formation change in diskiness \textit{lessens} with age, and the oldest stars in this era are now equally as disky as when they formed.
This arises because $(v_{\rm \phi} / \sigma)_{\rm form}$ decreased with age \textit{faster} than the post-formation change in $v_{\rm \phi} / \sigma$ increased.
Now, the ratio with $\sigma_{\rm tot}$ exhibits nearly identical behavior as the ratio with $\sigma_{\rm Z}$, indicating that both `3D diskiness' and 'vertical diskiness' evolved after formation identically relative to their values at formation.
%In the Early-Disk Era, both ratios exhibit similar qualitative behavior, but  the ratio with $\sigma_{\rm tot}$ is now less than that with $\sigma_{\rm Z}$. This slight difference indicates that $\sigma_{\rm Z, now}$ has a higher fractional contribution from post-formation heating than $\sigma_{\rm tot, now}$, which we discuss in further detail in Section~\ref{subsec:formation vs post-formation}.

Stars that formed in the Pre-Disk Era have values of $(v_{\rm \phi} / \sigma)_{\rm form} / (v_{\rm \phi} / \sigma)_{\rm now}$ that fluctuate around 1. Some stars have higher $v_{\rm \phi} / \sigma$ now than when they formed, from post-formation torquing from events like mergers. Also, the division of intrinsically small velocity values amplifies the fluctuations.
%from the division of intrinsically small values, because old stars have values of $(v_{\rm \phi} / \sigma)_{\rm form}$ and $(v_{\rm \phi} / \sigma)_{\rm now} < 1$. Therefore, the exact behavior of $(v_{\rm \phi} / \sigma)_{\rm form} / (v_{\rm \phi} / \sigma)_{\rm now}$ for stars that formed in the Pre-Disk Era is quantitatively limited.

\subsection{Case Studies}
\label{subsec:case studies}

\begin{figure*}
\centering
\includegraphics[width = 0.5 \textwidth]{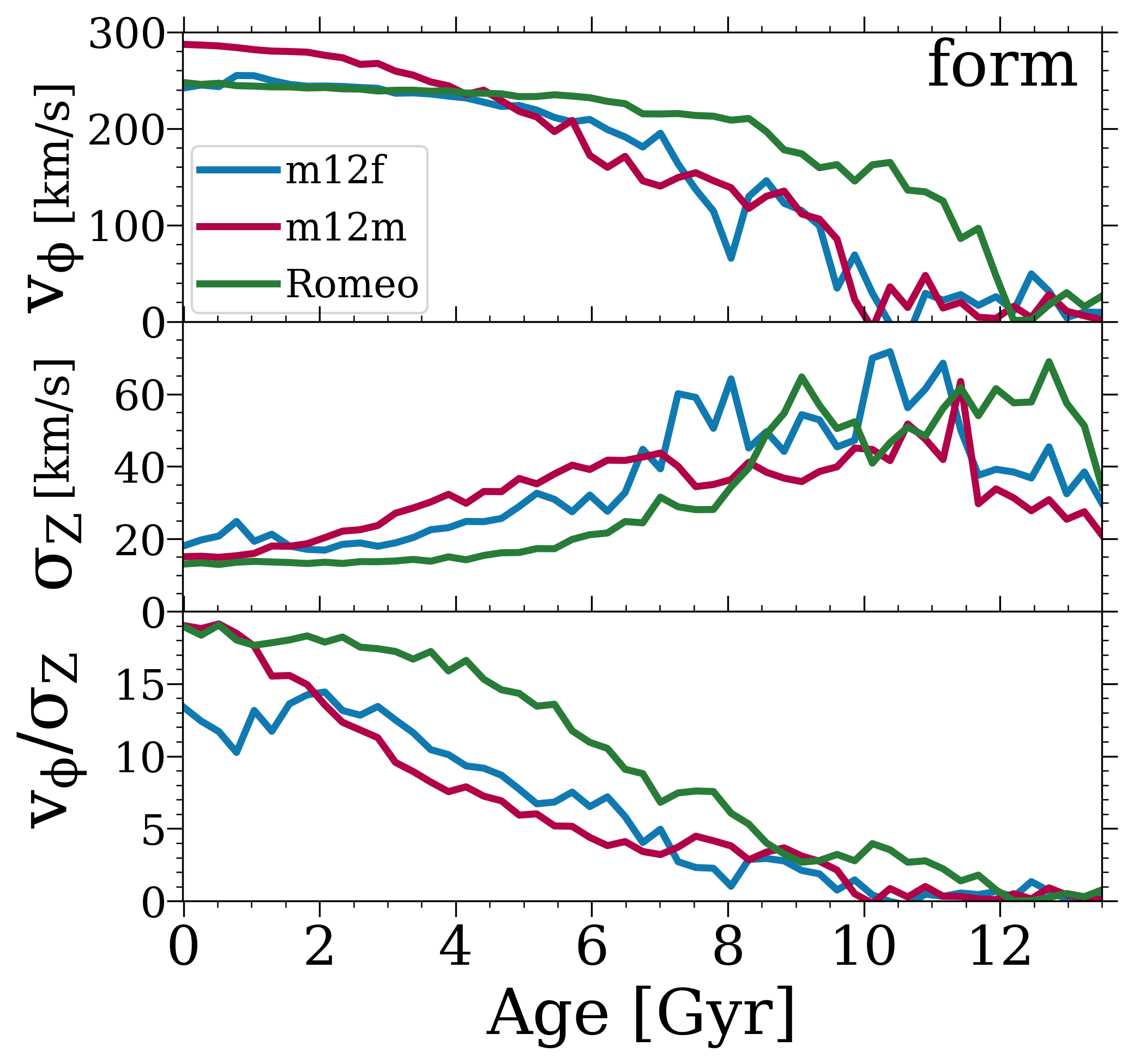}\hfill 
\includegraphics[width = 0.5 \textwidth]{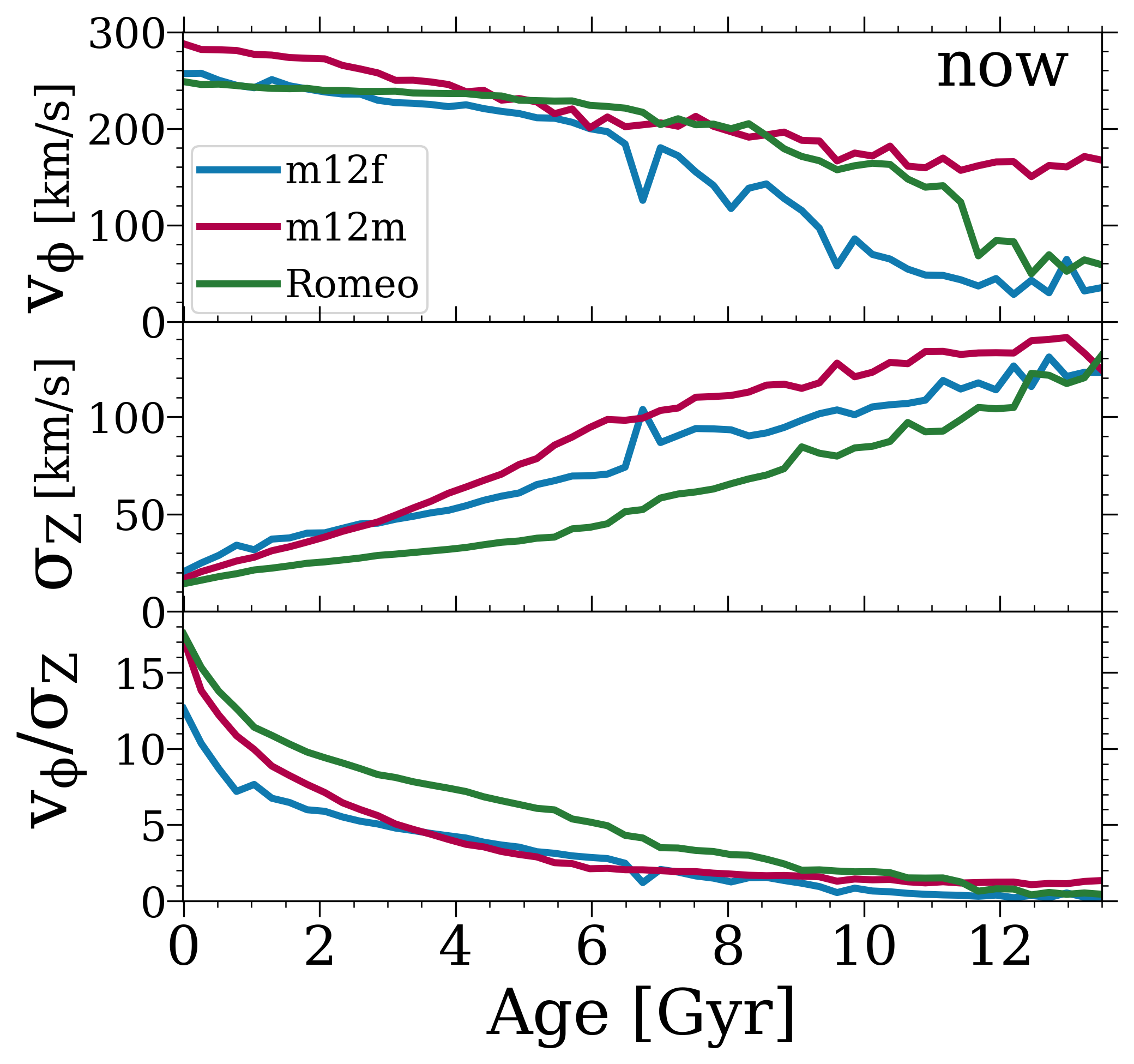}\hfill
\vspace{-2 mm}
\caption{
\textbf{Median azimuthal velocity and velocity dispersion of stars versus age for 3 galaxies}: m12f (blue), m12m (pink), and Romeo (green). Left shows velocities at the time of formation, while right shows velocities today.
\textbf{Top}: Median azimuthal velocity, $v_{\rm \phi}$.
Romeo is the earliest settling galaxy in our sample; it exhibited rapid growth in $v_{\rm \phi}$ at formation $\approx 2 \Gyr$ before either m12m or m12f. However, by $z = 0$ the later-settling m12m exhibits higher $v_{\rm \phi}$ at almost all ages, including its oldest stars.
\textit{Thus, kinematics at a given age as measured today do not necessarily indicate those at formation}.
\textbf{Middle}: Vertical velocity dispersion, $\sigma_{\rm Z}$. At formation, all of the galaxies show large fluctuating velocity dispersions for old stars, which peaked at different times, but this peak does not reflect itself in the current velocity dispersion versus age, at right.
m12f underwent a major merger $\approx 7 \Gyr$ ago, corresponding to a spike in the velocity dispersion in both panels.
\textbf{Bottom:} Ratio of median azimuthal velocity to vertical velocity dispersion, as a measure of rotational support.
Romeo is the most rotationally supported galaxy at all ages, measured both today and at formation, though m12m catches up for young stars today.
Despite differing histories, all 3 galaxies have similar velocity dispersions for young stars today, although they are more separated in $v_{\rm \phi} / \sigma_{\rm Z}$.
\vspace{-4 mm}
}
\label{fig:casestudies}
\end{figure*}

While Figures~\ref{fig:fig1} and \ref{fig:fig2} show trends averaged across 11 galaxies, Figure~\ref{fig:casestudies} shows these trends for 3 individual galaxies: m12f (blue), m12m (pink), and Romeo (green).

Romeo is one of our closest analogues to the MW today, residing in a LG-like pair and having a stellar mass of $5.9 \times 10^{10} \Msun$, similar to the MW.
Furthermore, recent observational analysis suggests that the MW's disk began to form $\approx 12 \Gyr$ ago \citep{Belokurov22, Conroy22, Xiang22}, and as we quantify further below, of all our galaxies, Romeo's disk began to form the earliest and thus nearest in time to this estimate for the MW, likely because of its forming in LG-like environment \citep{Garrison-Kimmel18, Santistevan20}.
Figure~\ref{fig:casestudies} (left) shows that Romeo's $v_{\rm \phi, form}$ rapidly increased and its $(v_{\rm \phi} / \sigma_{\rm Z})_{\rm form}$ began to rise $\approx 12$ Gyr ago and it transitioned into its Early-Disk Era $\approx 11 \Gyr$ ago.
Although Romeo transitioned the earliest, its evolution largely follows the same trends as most of our other galaxies.

m12f highlights the effect of major mergers on stellar kinematics.
Its $v_{\rm \phi, form}$ and $(v_{\rm \phi} / \sigma_{\rm Z})_{\rm form}$ began to increase $\approx 10 \Gyr$ ago, during its transition to the Early-Disk Era. However, both decreased $\approx 8$ Gyr ago, when a galaxy flew by and torqued its disk. After this, $v_{\rm \phi, form}$ and $(v_{\rm \phi} / \sigma_{\rm Z})_{\rm form}$ increased again. m12f then experienced a major merger $\approx 6.9$ Gyr ago. Although the dispersion was high during the merger, afterwards $\sigma_{\rm Z,form}$ sharply decreased while $(v_{\rm \phi} / \sigma)_{\rm form}$ increased. Thus, a major merger led to a more rotationally-supported disk and may have triggered its transition to the Late-Disk Era. This agrees with previous work that found that gas-rich mergers can trigger the settling of galactic disks, particularly if they occur when the potential is sufficiently deep to prevent outflows after bursts of merger-triggered star formation \citep[for example][]{Hopkins09, Moreno21, Santistevan21, McElroy22, He23}.

At present, the high dispersion during the merger and the sharp decline after remain discernible. Although the merger more dramatically impacts the in-plane components of the dispersion (notably, \sigmaphi at present spikes from $60 \kms$ to $150 \kms$ back down to $55 \kms$ within 1 Gyr during the merger), for consistency, we display only the vertical component.
Thus, the present-day kinematics of stars, specifically `spikes' in the velocity dispersion versus age, can reveal past merger events, as we will explore in future work. After its disk began to settle, m12f underwent a smaller, prograde merger $\approx 1.5$ Gyr ago that triggered a significant burst of star formation \citep{Yu2021}, increased $\sigma_{\rm form}$, and decreased $(v_{\rm \phi} / \sigma)_{\rm form}$. Thus, certain mergers occurring in the Pre- or Early-Disk Eras increased the galaxy's rotational support, while mergers occurring in the Late-Disk Era decreased the disk's rotational support.

We next highlight m12m, whose evolution deviates from our other galaxies. m12m has the largest stellar mass at $z = 0$ in our sample, $\approx 10^{11} \Msun$, more similar to M31.
Furthermore, m12m underwent numerous early mergers, including two particularly massive mergers $\approx 10$ and 9 Gyr ago, but it had a quiescent merger history over the past 8 Gyr.

m12m has our second earliest transition into the Early-Disk Era: its $(v_{\rm \phi} / \sigma_{\rm tot})_{\rm form}$ permanently surpassed unity $\approx 9$ Gyr ago. Despite this, m12m has yet at $z = 0$ to transition fully into its Late-Disk Era: its $v_{\rm \phi, form}$ continues to increase (albeit only slightly), all but its vertical component of $\sigma_{\rm form}$ have yet to become constant, and $(v_{\rm \phi} / \sigma_{\rm tot})_{\rm form}$ continues to strongly increase. This may be because m12m began to develop a bar $\approx 4.8$ Gyr ago, leading to a bar-buckling event that formed a significant X-shaped bulge composed primarily of younger stars (age $< 4.8 \Gyr$) \citep{Debattista19}.
m12m's kinematics are also unique at present; most notably, its $v_{\rm \phi, now}$ exhibits no significant decrease for old stars; instead, all ages have $v_{\rm \phi, now} \gtrsim 160 \kms$, which is almost $3 \times$ faster than our sample's average, likely as a result of its multiple major mergers. Of our 3 case studies, m12m generally had the lowest $\sigma_{\rm Z, form}$ for stars older than $\approx 8$ Gyr, but at present m12m has the highest $\sigma_{\rm Z, now}$ at almost all ages, because its stars have been more dynamically heated.

Comparing these case studies provides useful context and caution for interpreting population-wide trends.
First, the dispersion of young stars does not necessarily reflect the dispersion of older stars: across the three galaxies, the difference in $\sigma_{\rm Z, now}$ is only a few $\kms$ for the youngest stars (age $< 250$ Myr), but jumps to $14 \kms$ at 1 Gyr old and $57 \kms$ at 6 Gyr old. In fact, the rate at which $\sigma_{\rm now}$ increases with age varies between galaxies.
On a related note, how $\sigma_{\rm form}$ compares between galaxies can be different than how $\sigma_{\rm now}$ does. That is, stars of a certain age may now have larger/smaller dispersions than those in another galaxy despite forming with smaller/larger dispersions. For example, Romeo generally had the highest values of $\sigma_{\rm Z, form}$ for stars older than 6 Gyr while m12m had the lowest. However, this ordering flips at present: today, $\sigma_{\rm Z}$ in m12m is higher than in Romeo. Thus, the dispersion with which stars formed does not set their present dispersion, because the amount and age dependence of post-formation heating can differ significantly between galaxies.

On the contrary, $v_{\rm \phi} / \sigma_{\rm Z}$ provides a similar comparison between galaxies at both formation and present-day. $(v_{\rm \phi} / \sigma_{\rm Z})_{\rm now}$ maintains its relative ordering between galaxies from formation and even largely preserves the trends with age exhibited by each individual galaxy: Romeo still has the highest values for stars younger than $\approx 10$ Gyr, m12f and m12m still have similar values for intermediate-aged stars, and m12f still has the lowest values for young stars. Although the differences between each galaxy's $v_{\rm \phi} / \sigma_{\rm Z}$ are less pronounced today than at formation, \textit{present day observations of $(v_{\rm \phi} / \sigma)_{\rm now}$ (versus age) give a better representation of a galaxy's formation history than $\sigma_{\rm now}$ and $v_{\rm \phi, now}$ alone.}

\subsection{Comparing Velocity Components}
\label{subsec:component comparsion}

\begin{figure}
\includegraphics[width = \columnwidth]{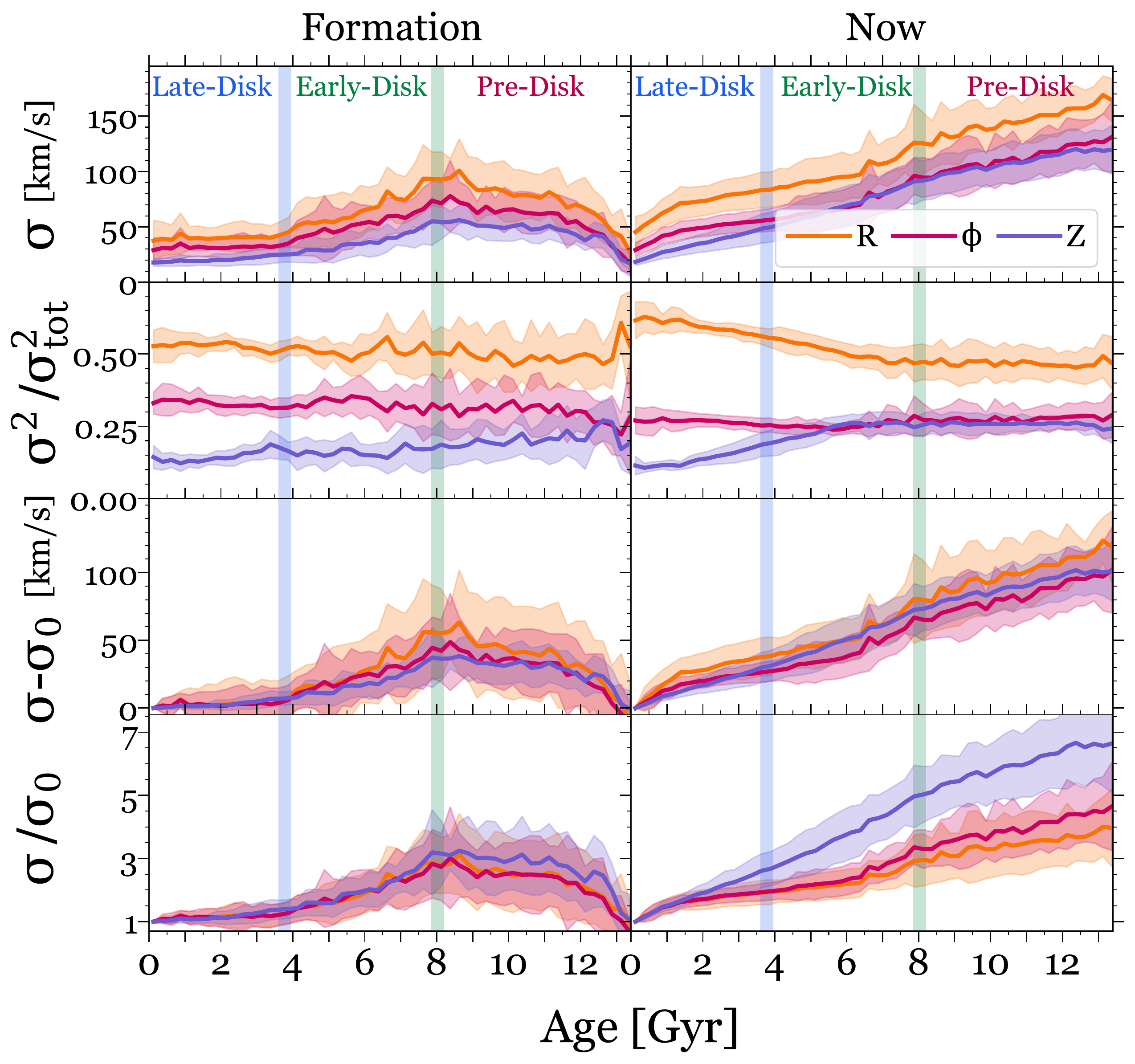}
\vspace{-6 mm}
\caption{
\textbf{Comparing the evolution of the 3 components of velocity}: lines show the radial (orange), azimuthal (pink), and vertical (purple) velocity dispersion at formation (left) and today (right) versus age.
We average across 11 galaxies, and the shaded regions show the galaxy-to-galaxy standard deviation.
\textbf{Top row}: Velocity dispersion, $\sigma$.
At formation, all components exhibit constant ordering, with $\sigma_{\rm R}$ > $\sigma_{\rm \phi}$ > $\sigma_{\rm Z}$, and similar dependence on age.
Today, $\sigma_{\rm R}$ dominates at all ages, while $\sigma_{\rm \phi}$ and $\sigma_{\rm Z}$ are near identical for older stars.
However, $\sigma_{\rm Z}$ falls below $\sigma_{\rm \phi}$ for stars that formed in the Late-Disk Era.
\textbf{Second row}: Ratio of each component's $\sigma^2$ relative to the total at that age, as a metric of the relative kinetic energy in each component.
At both formation and today, $\sigma_{\rm R}^2$ dominates the kinetic energy.
At formation, kinetic energy partition is nearly independent of age, while today, it is nearly independent for ages > 6 Gyr, but because of post-formation heating, the importance of $\sigma_{\rm Z}^2$ decreases while $\sigma_{\rm R}^2$ increases for younger and younger stars.
\textbf{Third row}: Absolute change in the velocity dispersion relative to $\sigma_{\rm 0}$, the dispersion of the youngest stars (age < 250 Myr at $z = 0$) for each component.
At formation and today, all components are nearly self-similar, indicating that each component changes with age by about the same absolute amount (although at present, the radial component is slightly higher, because of its stronger post-formation heating).
\textbf{Bottom row}: Velocity dispersion normalized by its value for the youngest stars, which quantifies the fractional change in $\sigma$ over time.
At formation, all components display similar values and age dependence, such that intermediate-age stars formed with 3 times higher $\sigma$ than the youngest or oldest stars did.
As measured today, $\sigma_{\rm Z}$ exhibits the strongest fractional dependence on age, despite it showing the weakest absolute dependence on age in the top panels, as a result of post-formation heating.
\vspace{-2 mm}
}
\label{fig:comps}
\end{figure}

All three spatial components of velocity dispersion share similar trends with age, but we now compare them in detail. Figure~\ref{fig:comps} shows the radial, azimuthal, and vertical velocity dispersions, averaged across the 11 galaxies, at formation (left) and at present (right).

Figure~\ref{fig:comps} (top row) compares the dispersions. The relative order of the components at formation is constant with age: $\sigma_{\rm R, form}$ > $\sigma_{\rm \phi, form}$ > $\sigma_{\rm Z, form}$. At present, $\sigma_{\rm R, now}$ dominates at all ages. However, $\sigma_{\rm Z, now}$ now matches $\sigma_{\rm \phi, now}$ for stars that formed in the Early- and Pre-Disk Eras. Thus, the similarity between $\sigma_{\rm Z, now}$ and $\sigma_{\rm \phi, now}$ for intermediate-aged and old stars arises from vertical post-formation dynamical heating, not from formation. Measurements of $\sigma_{\rm now}$ in the Solar neighborhood find $\sigma_{\rm R}$ > $\sigma_{\rm \phi}$ > $\sigma_{\rm Z}$ \citep[for example][]{Wielen77, Dehnen98, Nordstrom2004}.
While our results agree for stars that formed in the Late-Disk Era, we find that $\sigma_{\rm \phi, now} \approx \sigma_{\rm Z, now}$ for stars that formed in the Early and Pre-Disk Eras, potentially indicating that post-formation heating of $\sigma_{\rm Z}$ is greater for older stars in our galaxies than in the MW.

Figure~\ref{fig:comps} (second row) shows the ratio of each component's $\sigma^2$ to the total $\sigma^2$ at that age, which compares each component's contribution to the overall kinetic energy in dispersion. Because $\sigma_\textrm{total}^2 \equiv \sigma_{\rm R}^2 + \sigma_{\rm \phi}^2 + \sigma_{\rm Z}^2$, the sum of this ratio in all three components is 1.
Remarkably, the ordering and relative value of each component at formation are independent of age. That is, $\sigma_{\rm form}$ was distributed across its 3 components the same way across the galaxy's entire formation history, with $\sigma_{\rm R, form}^2$, $\sigma_{\rm \phi,form}^2$, and $\sigma_{\rm Z, form}^2$ accounting for $\approx 51\%$, 32\%, and 17\% of $\sigma_{\rm tot, form}^2$, on average. This roughly constant partitioning across cosmic time is striking, given that it persists across the eras of disk formation, including the transition from dispersion-dominated to rotation-dominated orbits, from bursty to steady star formation, and the abatement of galactic outflows.
Given this consistent partitioning at formation during both early and late times, we posit that the drivers of (star-forming) gas dispersion common in the early galaxy, such as star-burst driven outflows, and those common in the later galaxy, such as spiral arms, have similar impacts on how the energy in dispersion at formation is partitioned between components. However, we we relegate testing this hypothesis to future work.
%This constancy may indicate that drivers of radial heating more common in the early galaxy, such as outflows and mergers, have a similar relative impact on $\sigma_{\rm R, form}^2$ as the drivers of radial heating in the later galaxy, such as spiral arms; however, we relegate testing this hypothesis to future work.

Currently, $\sigma_{\rm R, now}^2$ dominates in all ages, as Figure~\ref{fig:comps} (top) shows. However, unlike at formation, the relative contributions of $\sigma_{\rm R, now}^2$ and $\sigma_{\rm Z, now}^2$ are constant with age only for stars older than $\approx 6$ Gyr, which formed in the Pre-Disk Era and the beginning of the Early-Disk Era. $\sigma_{\rm R, now}^2$ still contributes $\approx 50\%$ of the random kinetic energy of these old stars, while $\sigma_{\rm \phi, now}^2$ and $\sigma_{\rm Z, now}^2$ both contribute $\approx 25\%$.

For stars younger than 6 Gyr, the relative contribution of $\sigma_{\rm R, now}^2$ increases for younger stars, even surpassing its contribution at formation, because young stars have been heated most strongly along their radial component, as we show below. In contrast, the relative contribution of $\sigma_{\rm Z, now}^2$ decreases for younger stars, more directly reflecting the low $\sigma_{\rm Z, form}$ of these young stars. Interestingly, stars of all ages have approximately 25\% of their random kinetic energy from $\sigma_{\rm \phi, now}^2$, slightly less than when they formed.

We define $\sigma_{\rm 0}$ as the dispersion of our youngest stars (age < 250 Myr at $z = 0$). To examine the degree of self-similarity in the evolution of the different velocity components, Figure~\ref{fig:comps} (third row) shows $\sigma - \sigma_{\rm 0}$, the \textit{absolute change} in the velocity dispersion with age.
All velocity components have more similar evolution in $\sigma - \sigma_{\rm 0}$ than in $\sigma$. This is most self-similar at formation, but it also holds to a lesser degree at present. In particular, $\sigma_{\rm R, form}$ exhibits a greater absolute change only for intermediate-aged stars ($\approx 6 - 10$ Gyr), whereas $\sigma_{\rm R, now}$ has a greater absolute change at all ages, because $\sigma_{\rm R, now}$ had the largest amount of dispersion added by post-formation dynamical heating for all but intermediate-aged stars, as we discuss below.
%At formation, the change is near 0 and is constant over the past 5 Gyr%: stars formed with similar dispersions as they do today.
%\textit{Relative to stars forming today, all three components of the dispersion at formation have changed by the same absolute amount.}
%At intermediate times, the dispersion was higher, as expected, but unlike in the case for the raw dispersion, all components have similar values, with the radial only slightly dominating $\approx$  7-9 Gyr ago. Meaning that although the radial component of the raw dispersion at formation dominates, it always `leads' by the same amount, independent of the galaxy's age or dynamical state.

Figure~\ref{fig:comps} (bottom row) shows $\sigma / \sigma_{\rm 0}$, the \textit{fractional change} in the velocity dispersion with age, relative to today. At formation, this exhibits significant self-similarity, with all components showing near identical evolution for stars that formed in the Early- and Late-Disk Eras. In particular, at the onset of the Early-Disk Era, stars formed with $\approx 3 \times$ the dispersion as young stars today for all velocity components.
However, the striking self-similarity at formation does \textit{not} hold at present. Although all components evolve relatively similarly for ages $0 - 3$ Gyr, the vertical component increasingly dominates for older stars, with the oldest stars having vertical dispersions $\approx 7 \times \sigma_{\rm Z, 0}$. Because $\sigma_{\rm Z}$ had similar fractional evolution at formation, its distinctive behavior at present must arise from a greater fractional amount of post-formation heating for stars that formed in the Early- and Pre-Disk Eras, which we explore next.

\subsection{Disk Settling versus Post-Formation Heating}
\label{subsec:formation vs post-formation}

%As we have shown, the present kinematics of stars can be significantly different from their kinematics at birth, because stars are dynamically heated over their lifetimes. Furthermore, the dispersion that stars formed with has also changed across cosmic time. Notably, $\sigma_{\rm form}$ decreased throughout the Early-Disk Era, reaching a minimum throughout the Late-Disk Era. We term this decrease in $\sigma_{\rm form}$ `disk settling'.

As we have shown, both disk settling and dynamical heating shape the relation between $\sigma_{\rm now}$ and stellar age.
We now quantify the relative importance of these two processes versus age.

\begin{figure}
\includegraphics[width = \columnwidth]{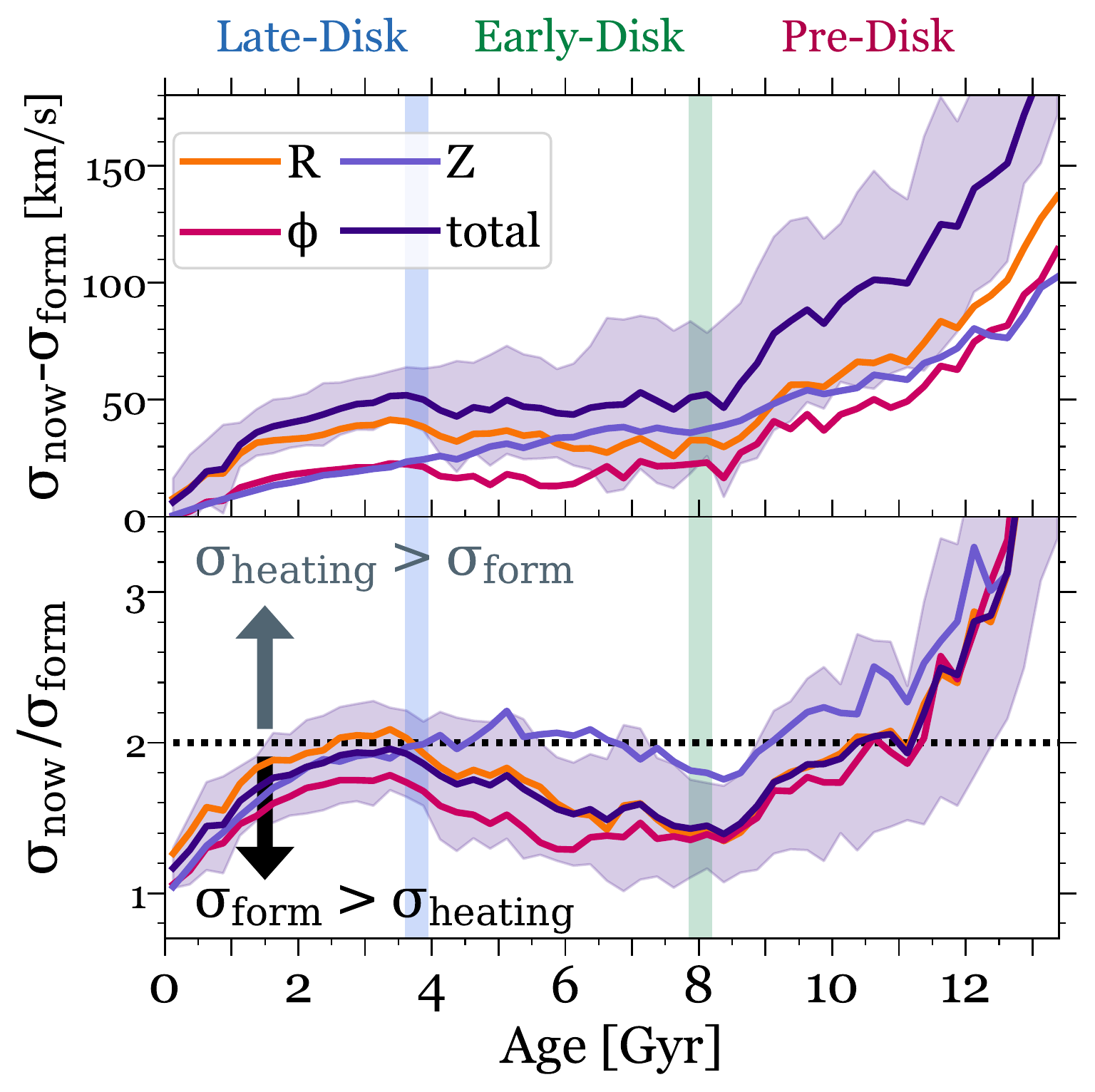}
\vspace{-6 mm}
\caption{
\textbf{Comparison between the velocity dispersion today versus at formation, which quantifies the amount of post-formation dynamical heating, versus age}, for different components of velocity.
We average across 11 galaxies, and the shaded region shows the galaxy-to-galaxy standard deviation.
\textbf{Top}: Absolute change in dispersion from formation to today.
The amount of heating increases with age. However, for all components except $\sigma_{\rm Z}$, the heating is constant for ages $\approx 3 - 8 \Gyr$, thus, intermediate-age stars were heated by a similar amount.
\textbf{Bottom}: Ratio of dispersion today relative to that at formation, which shows an `S'-shaped dependence on age. The dotted horizontal line at 2 divides stars whose current dispersion is primarily from formation (below) versus post-formation (above).
The combination of small dispersions at birth and rapid post-formation heating leads to a steadily increasing ratio with age up to $\approx 2 \Gyr$.
Intermediate-age stars have a dispersion today mostly from that at formation, because they formed when the intrinsic dispersion within the galaxy was highest.
The oldest stars have the largest contribution from post-formation heating.
\textit{For stars younger than $10 \Gyr$, most of their dispersion was in place at formation}, though $\sigma_{\rm Z}$ shows nearly equal contributions from formation and subsequent heating at ages $4 - 8 \Gyr$.
}
\label{fig:form_comp}
\end{figure}

Figure~\ref{fig:form_comp} shows the absolute and fractional impact of post-formation dynamical heating on the present-day radial, azimuthal, vertical, and total velocity dispersion of stars versus age. Most notably, both absolute and fractional post-formation dynamical heating follow distinct trends with age for stars that formed in each of our three kinematic eras.

Figure~\ref{fig:form_comp} (top) shows the difference between dispersion at present and formation, $\sigma_{\rm now} - \sigma_{\rm form}$, versus age, which directly conveys the amount of dispersion added post-formation. We focus primarily on the total dispersion, noting that the radial and azimuthal components show similar behavior. However, we discuss the vertical dispersion separately below, because it shows qualitatively different behavior.

For stars that formed in the Late-Disk Era, the amount of dispersion added post-formation increases with age, but at a declining rate, consistent with the idea that post-formation dynamical heating from scattering processes causes the observed increase in $\sigma_{\rm now}$ with age \citep[for example][]{Spitzer51}.
However, stars that formed in the Early-Disk Era have been dynamically heated by the same total amount, $\approx 50 \kms$, regardless of age. Considering the monotonic increase of $\sigma_{\rm tot}$ in Figure~\ref{fig:fig1}, one may not intuitively expect this plateau. For example, stars that formed at the start of the Early-Disk Era ($\approx 8 \Gyr$ ago) currently have $\sigma_{\rm tot, now}$ that is almost $80 \kms$ higher than stars born at the end of the Early-Disk Era ($\approx 4 \Gyr$ ago), but have $5 \kms$ \textit{less} $\sigma_{\rm tot}$ added post-formation. Thus, \textit{the monotonic increase of $\sigma_{\rm tot, now}$ with age for stars that formed in the Early-Disk Era ($\approx 4 - 8 \Gyr$ ago) arose not from the dependence of post-formation heating on age, but rather, from the amount of disk settling in $\sigma_{\rm form}$ versus age}.

%The radial and azimuthal components display trends with age similar to those of the total. These plateaus may indicate that the primary drivers of in-plane post-formation heating cannot significantly heat stars above a certain dispersion threshold.

%In the Early- and Late-Disk Eras, the processes that heated stars in the early galaxy dwindled as the merger rate declined, the galaxy's potential deepened, and star formation steadied, among numerous other interconnected factors.
Once a disk started to form, scattering with structures within the now well-defined disk plane likely drove most post-formation heating, though interactions with satellite galaxies likely also contributed.
% meaning post-formation heating was most rapid for stars immediately after they formed, when they were near the disk's midplane and/or near GMCs and spiral arm overdensities.
As stars were dynamically heated, they experienced larger radial and vertical oscillations. Larger vertical oscillations decrease the amount of time that stars spend near the midplane, and increase their vertical velocities when passing through the midplane. In turn, the ability of stars to be heated along the in-plane direction decreases as vertical oscillations increase. Similarly, stars with large radial eccentricities encounter spiral structure at different orbital phases; the effects of these varied encounters tend to average out, such that spiral-driven heating becomes progressively weaker for increasingly eccentric orbits \citep{bt-08}.
Thus, stars that formed with low dispersions in the Late-Disk Era were initially efficiently heated. However, this efficiency correspondingly decreased as their dispersions increased over their lifetimes, which explains their near-exponential behavior for $\sigma_{\rm tot, now} - \sigma_{\rm tot, form}$, as shown in the top row of Figure~\ref{fig:form_comp}. 
Similarly, because stars that formed in the Early-Disk and Pre-Disk Eras had large dispersions at birth, they likely underwent little post-formation heating over the course of the Late-Disk and Early-Disk Era (though old stars likely experienced significant heating within the Pre-Disk Era ).

On the other hand, post-formation \textit{vertical} heating shows different trends than in-plane heating across the Late and Early-Disk Eras. The amount of $\sigma_{\rm Z}$ added post-formation never plateaus but instead monotonically increases with age by $\approx 7 \kms$ per Gyr. Therefore, \textit{post-formation vertical heating is likely a steady, continuous process}, implying a different dynamical origin than in-plane post-formation heating.
We will pursue a detailed analysis of the origin of in-plane and vertical heating in future work.

For stars that formed in the Pre-Disk Era, the amount of post-formation heating increases steadily with age for all components, including vertical. Older stars have more dispersion added post-formation, not only because they have had more time to be heated, but also because the early galaxy experienced the most frequent mergers, starbursts, and potential fluctuations that increased the dispersion of existing stars \citep[for example][]{El-Badry18-anc}. Likely, this phase accounted for most of the heating of stars that formed in the Pre-Disk Era.
In later eras, more localized, temporally-stable processes (that is, scattering by non-axisymmetric structures within the disk) primarily heated the stars.

Figure~\ref{fig:form_comp} (bottom) shows the ratio of dispersion at present to dispersion at formation, $\sigma_{\rm now} / \sigma_{\rm form}$, versus age, which quantifies the fraction of the present dispersion that existed at formation versus arose from post-formation heating. Values of $\sigma_{\rm now} / \sigma_{\rm form} > 2$ indicate that post-formation heating contributes more to the current dispersion than the dispersion from formation, while values $< 2$ indicate that most of the dispersion was in place at the time of formation.

The evolution of the total dispersion largely matches that of the radial and azimuthal components.
$\sigma_{\rm now} / \sigma_{\rm form}$ has an `S'-shaped dependence on age, somewhat different than in the top panel. Post-formation heating increases rapidly with age for stars that formed in the Late-Disk Era. This era's oldest stars have current (radial, vertical, and total) dispersions with approximately equal contributions from heating and formation. Furthermore, stars that formed $\approx 4 \Gyr$ ago, during the transition to the Late-Disk Era, experienced the largest fractional increase from post-formation heating of all stars younger than 10 Gyr (for all but the vertical dispersion). These stars formed in a dynamically-settled thin disk with low dispersion and had more time to experience dynamical heating than stars that formed afterward.

However, strikingly, this dependence on age \textit{reverses} in the Early-Disk Era, when $\sigma_{\rm now} / \sigma_{\rm form}$ decreases with age.
As Figure~\ref{fig:form_comp} (top row) showed, the absolute amount of dispersion added post-formation is constant throughout this era. However, as  Figure~\ref{fig:fig1} showed, $\sigma_{\rm form}$ increases with age throughout this era. Thus, the fractional contribution of $\sigma_{\rm form}$ at birth became increasingly important with age in this era. Besides stars younger than $\approx 1 \Gyr$, the oldest stars from the Early-Disk Era have the lowest dependence on post-formation heating, with only 40\% of their current $\sigma_{\rm tot}$ from post-formation heating, corresponding with their maximum $\sigma_{\rm form}$ and comparatively low amount of post-formation heating.
In essence, stars were not able to be dynamically heated efficiently during this Early-Disk Era, because they already were born with high dispersions.

For stars that formed in the Pre-Disk Era, $\sigma_{\rm now} / \sigma_{\rm form}$ increases rapidly with age, because $\sigma_{\rm form}$ decreases with age (given the galaxy's lower mass) while the amount of post-formation heating (and thus $\sigma_{\rm now}$) increases with age (from mergers, starbursts, and strong outflows).
Thus, \textit{the current total dispersion of stars is primarily from post-formation heating for stars older than $\approx 11$ Gyr, and primarily inherited at formation for stars younger than 11 Gyr}.

The evolution of the vertical component largely resembles that of the other components for stars that formed in the Late-Disk Era. However, while $\sigma_{\rm now} / \sigma_{\rm form}$ of the in-plane components noticeably decrease with age for stars that formed in the Early-Disk Era, the vertical component remains relatively constant near 2. In fact, $\sigma_{\rm Z,now} / \sigma_{\rm Z, form} \approx 2$ for stars with ages between $2 - 10 \Gyr$, meaning that the current vertical dispersions of stars with ages $2 - 10$ Gyr have equal contributions from post-formation heating and formation.
Thus, \textit{for most stars, the fractional impacts of vertical post-formation heating and vertical disk settling are about the same}.
For stars younger than 2 Gyr, $\sigma_{\rm Z,form}$ is more important, while for stars older than 10 Gyr, vertical post-formation heating is.

In summary, the monotonic rise of the current velocity dispersion with stellar age arises from the \textit{combined} effects of \textit{both} cosmological disk settling and the dynamical processes that heat stars after they form.
In understanding the origin of the current \textit{shape} of the relation between the total velocity dispersion and age, we conclude that: \textit{(1) post-formation heating dominates the dispersion for stars that formed in the Late-Disk Era} (age $\lesssim 3.5 \Gyr$), when the dispersion at formation was flat with age, (2) \textit{disk settling dominates for stars that formed in the Early-Disk Era} (age $\approx 4 - 8 \Gyr)$, when post-formation heating was flat with age, and (3) \textit{post-formation heating strongly dominates for stars that formed in the Pre-Disk Era} (age $\gtrsim 10 \Gyr$).
Overall, post-formation dynamical heating only dominates the current dispersion for stars that formed $\approx 3$ Gyr before the disk started to form, or, equivalently, \textit{the stellar velocity dispersion at present is primary inherited at formation and not from post-formation dynamical heating} for stars younger than $\approx 10 \Gyr$.

Our results agree with \citet{Yu22}, who also analyzed these FIRE-2 galaxies and found that most stars currently on `thin-disk-like', `thick-disk-like', or `spheroid-like' orbits were born that way, that is, the type of orbit a star is currently on primarily reflects its formation orbit, not its post-formation evolution.

\subsection{Effective Rate of Post-formation Heating}
\label{subsec:heating rate}

Stars that formed in the Pre-Disk Era have the most dispersion added post-formation, as Figure~\ref{fig:form_comp} (top) shows. One might assume (naively) that the amount of heating simply scales with age, because older stars have had more time to be heated. However, Figure~\ref{fig:form_comp} (top) showed that the amount of post-formation heating does not necessarily increase with age across all eras. For example, stars $8 \Gyr$ old have experienced the same absolute amount of post-formation heating as stars $4 \Gyr$ old. 
To provide more insight, we examine the `effective' rate that stars today experienced post-formation heating, on average, via $(\sigma_{\rm now} - \sigma_{\rm form}) / \rm age$. This quantity is \textit{not} the actual instantaneous rate of increase in dispersion. Rather, it represents the time-averaged, overall rate at which the dispersion increased since formation. Thus, one can think of this effective rate as the amount of post-formation heating normalized to the stellar populations' age.

\begin{figure}
\includegraphics[width = \columnwidth]{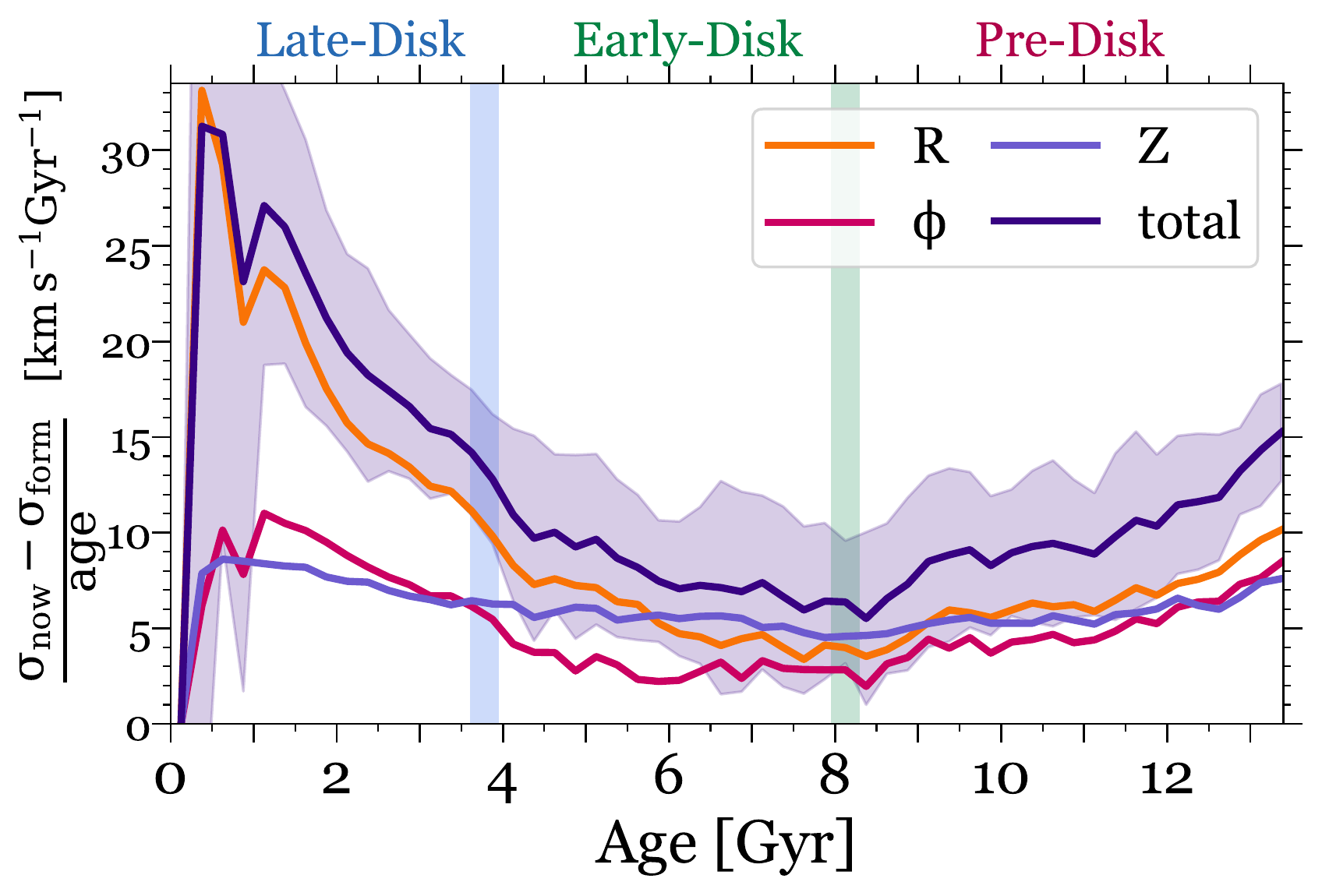}
\vspace{-6 mm}
\caption{
\textbf{The effective rate of dynamical heating since formation, $(\sigma_{\rm now} - \sigma_{\rm form}) / \rm age$, versus stellar age}, for each component of velocity.
We average across 11 galaxies, and the shaded region shows the galaxy-to-galaxy standard deviation.
The youngest stars have the highest effective rate of heating, implying that the rate of dynamical heating was most rapid immediately after stars formed and became more gradual over time. Increases in $\sigma_{\rm R}$ strongly dominate for stars that formed in the Late-Disk Era, but all components exhibit broadly similar effective rates for star that formed in the Early- and Pre-Disk Eras.
\vspace{-4 mm}
}
\label{fig:heating_rate}
\end{figure}

Figure~\ref{fig:heating_rate} shows this effective rate versus stellar age, for radial, vertical, azimuthal, and total dispersions. First, we focus on the total dispersion: $(\sigma_{\rm tot, now} - \sigma_{\rm tot, form})/ \rm age$.
The youngest stars experienced the largest effective rate of increase, which rapidly decreases with age for stars that formed during the Late-Disk Era.
For stars that formed in the Early-Disk Era, the effective rate was low, reaching a minimum near the transition to the Pre-Disk Era.
This strong age dependence implies that, for stars that formed in a disk, most dynamical heating occurred shortly after birth, and their heating rate slowed as they became progressively dynamically hotter (excepting the vertical component). By contrast, stars that formed in the Pre-Disk Era had effective rates of heating that \textit{increase} with age, likely because more dramatic, galaxy-wide, non-equilibrium perturbations like mergers and accretion, starbursts, and galactic winds, were more prominent. The number of these perturbation events that stars in the Pre-Disk era experienced increases with stellar age, such that the oldest stars have experienced the most cumulative heating.

Furthermore, for stars that formed in the Late-Disk Era, the increase in $\sigma_{\rm R}$ strongly dominates over the other components, so radial heating dominates the total heating. For example, the youngest stars have effective rates near 10 km/s/Gyr for $\sigma_{\rm \phi}$ and $\sigma_{\rm Z}$, but above 30 km/s/Gyr for $\sigma_{\rm R}$, which may indicate that heating by spiral arms and/or bars dominates most early heating for stars forming in the Late-Disk Era. However, the effective radial rate decreases with age while the other components remain more constant. Thus, stars that formed in the Early- and Pre-Disk Eras have nearly identical rates for all components of $\sigma$, indicating that \textit{heating was more isotropic in the early galaxy}.
%The near flatness of the effective increase in $\sigma_{\rm Z}$ at all ages is particularly striking.
Thus, \textit{post-formation heating operated differently over cosmic time}. Any model of post-formation heating thus should treat early cosmic times (prior to the initial formation of the disk) and later cosmic times (after the formation of the disk) separately.
%However, despite the effective rate of heating increasing with age for stars that formed in the Pre-Disk Era, the youngest stars (that formed in the Late-Disk Era) still have the largest effective rates.

\subsection{Disk Kinematics versus Galaxy Mass}
\label{subsec:galaxy mass trends}

So far, we showed trends averaged across our 11 galaxies that are similar to the MW in stellar mass.
Now, we analyze all 14 galaxies in our sample individually, including the lower-mass m12r, m12z, and Juliet to expand the mass range (see Table~\ref{table:table1} for details).

\begin{figure}
\includegraphics[width = \columnwidth]{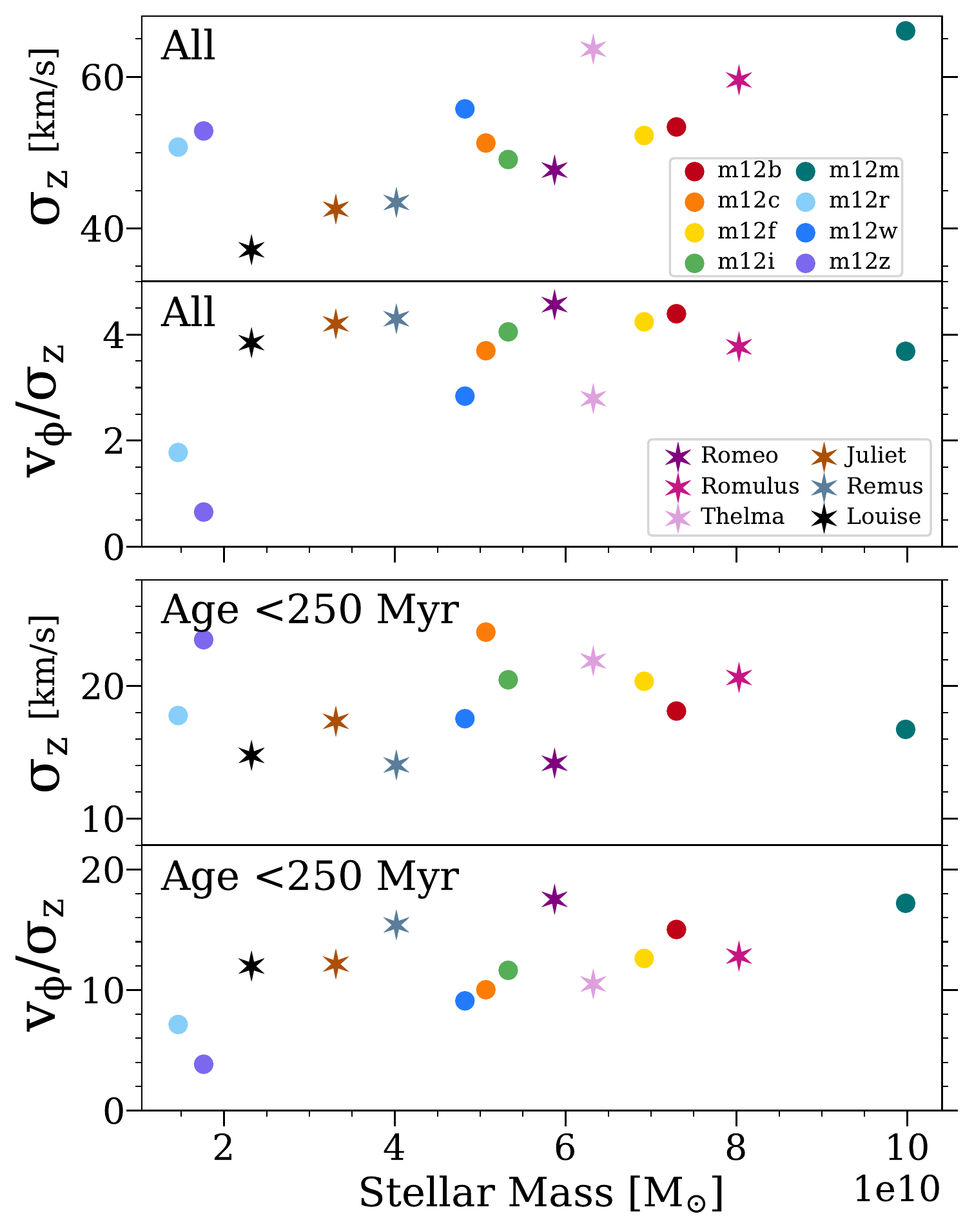}
\vspace{-5 mm}
\caption{
\textbf{Vertical velocity dispersion, $\sigma_{\rm Z}$, and $v_{\rm \phi} / \sigma_{\rm Z}$ today versus stellar mass} for 14 galaxies.
We show the isolated (m12) galaxies via circles and LG-like paired galaxies via stars.
\textbf{Top panels}: stars of all ages in the galaxy. The dispersion for our disk-dominated galaxies increases with mass.
All galaxies have approximately constant $v_{\rm \phi} / \sigma_{\rm Z}$ regardless of mass, excepting m12r and m12z.
\textbf{Bottom panels}: stars with ages $< 250 \Myr$. Here, $\sigma_{\rm Z}$ does not correlate with mass, though $v_{\rm \phi} / \sigma_{\rm Z}$ is slightly higher in at higher mass.
Overall, galaxies in LG-like pairs tend to have lower dispersions and more rotational support at a given mass; however, this difference is more evident for all stars than for young stars.}
\label{fig:stellar_mass}
\end{figure}

Figure~\ref{fig:stellar_mass} shows how stellar kinematics depend on the current stellar mass (within $R_{\rm 90}$) of a galaxy.
We focus on $\sigma$ and $v_{\rm \phi} / \sigma$, as before.
Although we do not show it, \vphi increases strongly with stellar mass, 
%having Pearson correlation coefficients of 0.93 and 0.94 and low p-values of 1.0 and 1.5 $\times 10^{-6}$ for all and young (age < 250 Myr) stars, respectively, 
which follows from the deeper gravitational potential of more massive galaxies \citep[see also][]{Hopkins23}.
We show the isolated galaxies from the \textit{Latte} suite as circles and the galaxies in LG-like pairs from the \textit{ELVIS on FIRE} suite as stars.
%As above, we calculate velocities and dispersions using only stars within $R = 6 - 10 \kpc$ and $|Z| < 3 \kpc$, but we measure stellar masses using all stars within $R_{\rm 90}$ of each galaxy.

Figure~\ref{fig:stellar_mass} (top panels) shows $\sigma_{\rm Z}$ (top row) and $v_{\rm \phi} / \sigma_{\rm Z}$ (second row), for stars of all ages.
If we exclude the two lowest-mass galaxies, m12r and m12z, whose disks have yet to or have just begun to settle, $\sigma_{\rm Z}$ increases with stellar mass. This trend is remarkably strong among LG-like galaxies, with the exception of Thelma, which underwent the most major mergers and has a higher $\sigma_{\rm Z}$ than galaxies of similar or higher mass. Unlike $\sigma_{\rm Z}$, $v_{\rm \phi} / \sigma_{\rm Z}$ of all stars does not depend on mass at $> 2 \times 10^{10} \Msun$. Although we do not show it, we performed an identical analysis for $\sigma_{\rm R}$ and $\sigma_{\rm \phi}$ and found similar trends.

On average, galaxies in LG-like pairs have lower dispersions and more rotational support than isolated galaxies. These colder kinematics likely relates to their more extended disk sizes at present: LG-like galaxies have $R^{*}_{90} \approx 1 \kpc$ larger than isolated galaxies for all stars and $\approx 3 \kpc$ larger among stars younger than 250 Myr \citep{Garrison-Kimmel18, Bellardini22}. These larger sizes likely relate to these galaxies transitioning to the Early-Disk Era earlier than our isolated galaxies, on average, as we show below.

Figure~\ref{fig:stellar_mass} (middle panels) shows the same but for stars younger than 250 Myr, which reflects the current dynamical state of the ISM and star formation, less affected by the galaxy's integrated merger and heating histories. Now, $\sigma$ does not depend on stellar mass, while $v_{\rm \phi} / \sigma$ increases modestly with stellar mass, which implies that more massive galaxies have more vertically settled gas disks.

Figure~\ref{fig:stellar_mass} (bottom panels) shows the same but for $\sigma_{\rm tot}$ of young stars, which exhibits broadly similar trends as each $\sigma$ component: $\sigma_{\rm tot}$ does not depend much on stellar mass while $v_{\rm \phi} / \sigma_{\rm tot}$ increases moderately, although weaker than for $v_{\rm \phi} / \sigma_{\rm Z}$.
Thus, a galaxy's stellar mass has little effect on the current velocity dispersion of young stars in the solar annulus and at most a modest effect on $v_{\rm \phi} / \sigma$, for the masses we explore.

\subsection{Comparisons with observations}
\label{subsec:observation}

Thus far, we have focused on our galaxies' \textit{disk-wide} kinematics, that is, the median velocities and dispersions for all stars (in a given age bin) within our simulated `solar annulus' of $R = 6 - 10 \kpc$ and $|Z| < 3 \kpc$ at $z = 0$. However, observational measurements of nearby external galaxies typically measure kinematics in smaller-scale `patches' and then average over these spatial patches to discuss trends with age within the galaxy. Similarly, most published dispersion values for the MW are for stars within one `patch' -- the solar neighborhood. We now examine how the type of measurement -- `disk-wide' or `local' -- impacts our measured velocity dispersions.

We measure a \textit{local} velocity dispersion within annular bins of side length 150 pc, motivated by the typical ALMA-resolution scale for nearby galaxies, and then we compute the average across the annulus at a given radius. For more context of our results on velocity dispersions of young stars, we also determine the local dispersions of cold and dense gas, $T < 300 \K$ and $n > 10 \rm cm^{-3}$, as a proxy for molecular gas, following \citet{Orr18}.

%We have seen that young stars in our galaxies have values of $\sigma_{\rm now}$ that (1) are lower when measured locally versus disk-wide and (2) are in agreement with observation
%For more context of our results on velocity dispersions of young stars, we also compare against observations of  cold (molecular) gas. As a proxy for cold molecular gas, we follow \citet{Orr18} and select gas with $T < 300 \K$ and $n > 10 \rm cm^{-3}$.

\begin{figure}
\includegraphics[width = \columnwidth]{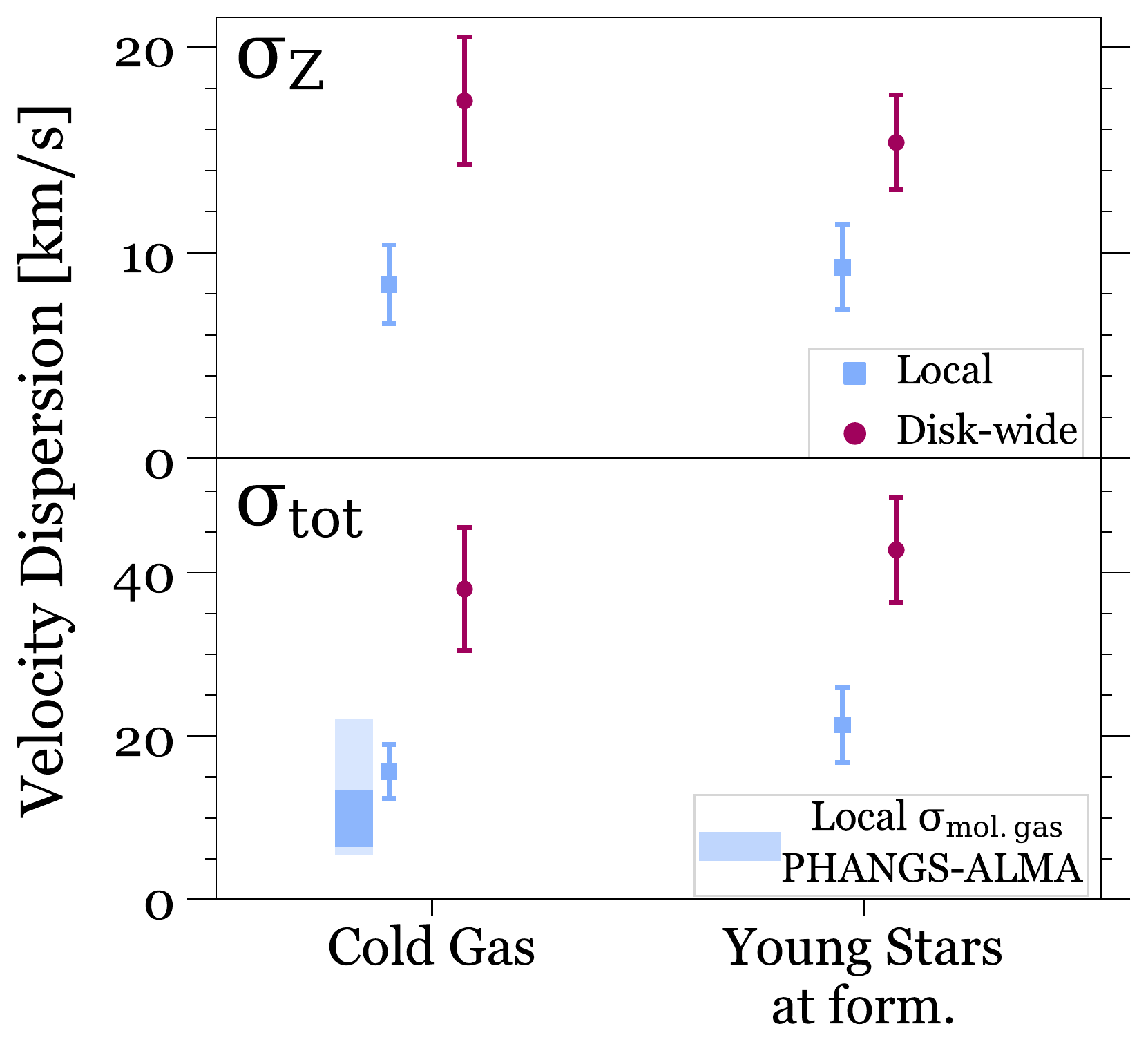}
\vspace{-5 mm}
\caption{
\textbf{Comparison of velocity dispersion, for a disk-wide versus local measurement, for cold dense gas and young stars}: specifically, $\sigma_{\rm Z}$ (top) and $\sigma_{\rm tot}$ (bottom) of cold ($T < 300 \K$) and dense ($n > 10 \rm cm^{-3}$) gas at present (left) and young (age < 250 Myr) stars at formation (right). Points show the mean across all 14 galaxies, while the error bars show the standard deviation. Magenta circles show the dispersion of gas or stars across $R = 6 - 10 \kpc$ and $|Z| < 3 \kpc$, identical to our previous analyses. Blue squares show the local velocity dispersion, within 150 pc patches, with cold gas patches centered at $\approx$ 4 kpc and young star patches centered at 8 kpc (see text for further explanation). The shaded region shows the total dispersion of molecular gas in 19 PHANGS-ALMA galaxies: the light blue shows the sample's full range while the darker blue shows the sample's 16th-84th percentile range.
\vspace{-4 mm}
}
\label{fig:local_vs_global}
\end{figure}

Figure~\ref{fig:local_vs_global} shows the average vertical (top) and total (bottom) dispersion of cold dense gas at present (left) and young (age < 250 Myr) stars at formation (right), averaged our 14 galaxies, for both \textit{local} dispersions (light blue) and \textit{disk-wide} dispersions (magenta). Our disk-wide dispersions use the same geometric selection as all of our previous analyses, while our local dispersions are the annulus-averaged dispersion in patches of side length 150 pc centered at $R = 8 \kpc$. However, for the local dispersion of cold gas, we instead average between the dispersion within annuli centered at $R \approx 3.5$ and $R \approx 4.9 \kpc$ (corresponding to 1/3 and 2/3 of our observational radial range), to match the radial range of the PHANGS observations we compare against (see below). Velocity dispersions in cold gas at $R = 8 \kpc$ are $\approx 72\%$ of those at $R \approx 4.2 \kpc$ in our simulations.
%, such that all below discussion, with the exception of our observational comparison, uses the (non-plotted) values at $R = 8 \kpc$. 

As the results for $\sigma_{\rm Z}$ (top) show, FIRE-2 galaxies have 1D local velocity dispersions in cold gas of $\sigma_{\rm 1D} \approx 8.4 \kms$ at $R \approx 4.2 \kpc$ (as shown, and $\approx 6 \kms$ at $R \approx 8 \kpc$ not shown) -- much lower than the typical disk-wide dispersions in our previous results. We compare against ALMA-measured dispersions of molecular gas CO(2–1) on $\approx 100 \pc$ scales in PHANGS galaxies from \citet{Sun20}, where we restrict the observed sample to the 19 galaxies that also have resolved stellar kinematics in \citet{Pessa23}. We include only gas observations within $R = 4.2 \pm 2 \kpc$, and we measure the local dispersion of cold gas in our FIRE-2 galaxies at concordant radii ($\approx 3.5$ and $4.9 \kpc$).
%Each PHANGS-ALMA galaxy has $\approx 1300$ independent measurements of $\sigma_{\rm LOS}$ with 150 pc resolution across this radial range.
We measure the mean $\sigma_{\rm LOS}$ for each PHANGS galaxy and assume isotropy to convert to a 3D $\sigma_{\rm tot}$.
Figure~\ref{fig:local_vs_global} (bottom left) shows the PHANGS-ALMA sample's full range of mean $\sigma_{\rm tot}$ as a shaded light-blue region and $1 \sigma$ scatter in darker blue. Although our FIRE-2 galaxies have local velocity dispersions in cold gas somewhat higher than typical in PHANGS, all of our FIRE-2 galaxies fall within the observed range.
%This discrepancy is not completely unexpected, as PHANGS-ALMA measurements have 150 pc resolution while we find dispersions within 500 pc patches. The linewidth-size relation, that is, the increase in $\sigma_{\rm LOS}$ of gas within GMCs with increasing projected (mass-weighted) cloud radius, predicts that $\sigma_{\rm LOS}$ decreases as measurement resolution decreases. In turn, our larger `aperture' may account for our slightly larger dispersions.
While a rigorous comparison would require synthetic CO observations, our comparison here indicates that velocity dispersions in cold gas are broadly consistent with those of molecular gas in nearby galaxies.

Most importantly, Figure~\ref{fig:local_vs_global} shows that the dispersion of cold gas in our simulations is significantly larger ($\approx 3$ times) if we measure it disk-wide rather than locally. Similarly, stars at formation have disk-wide $\sigma_{\rm Z}$ and $\sigma_{\rm tot}$ that are $\approx 2$ times larger than their local values at $R = 8 \kpc$. These values broadly agree with those from \citet{Orr20} ($\approx 15 - 30$ km/s), who examined the spatially-resolved dispersion of neutral gas in 7 of our isolated (m12) simulations.
This motivates the need to use local velocity dispersions to compare against spatially-resolved observations of nearby galaxies.
%Thus, while the disk-wide dispersions of gas and stars at formation both have significant contributions from disk-wide azimuthal and radial variations, these contributions are stronger for gas. 
%Most importantly, Figure~\ref{fig:local_vs_global} shows that local dispersions are significantly lower than disk-wide dispersions. Specifically, the vertical and total dispersions of cold gas are $\approx$ 2.8 and 3.3 times higher when measured disk-wide versus locally. This increase is slightly weaker for young stars at formation, with disk-wide vertical and total dispersions being 1.7 and 2 times higher, respectively.

Finally, the \textit{local} $\sigma_{\rm Z}$ and $\sigma_{\rm tot}$ for young stars at formation are 1.5 and 1.9 times larger, respectively, than cold gas (at $R \approx 8 \kpc$). On the other hand, young stars at formation have a \textit{disk-wide} $\sigma_{\rm tot}$ that is about the same as cold dense gas, while the \textit{disk-wide} $\sigma_{\rm Z}$ is lower for stars at formation than cold dense gas.
The higher local dispersion for stars than cold gas is expected, given that we measure them typically $\approx 12 \Myr$ after formation.
%, and this disperison may increase more rapidly in the disk plane.
However, the weaker (and even negative) change from cold gas to stars at formation for the disk-wide dispersion suggests that rapid dynamical heating has less of an effect on the galaxy-wide dispersion, which may be more influenced by spiral arms, warps, etc.
We defer a more detailed investigation to future work.
%That said, the fact this increase for stars is smaller for the disk-wide dispersion than the local may indicate that their early heating has less of an effect on the galaxy-wide dispersion, or that disk-wide kinematic asymmetries -- especially vertical asymmetries -- are stronger than local asymmetries for gas (even cold, dense gas) than for stars. That said, we relegate the testing of these hypotheses to future work.

Figure~\ref{fig:stellar_mass} showed that the \textit{disk-wide} vertical velocity dispersion of young stars in FIRE-2 galaxies is $\approx 15 - 25 \kms$ higher than the MW's $\approx 10 \kms$ \citep{Nordstrom2004, Casagrande11, Anders23}. However, there are key differences in the spatial scales over which we measured our dispersions in Figure~\ref{fig:stellar_mass} versus the scales used for MW observations.
Given the non-trivial differences between the disk-wide and local velocity dispersions shown in Figure~\ref{fig:local_vs_global}, we now revisit our previous comparison between galaxy kinematics and stellar mass from Figure~\ref{fig:stellar_mass}, and now include observed values from the MW and nearby galaxies. This more detailed comparison between stellar kinematics in our simulations, in the MW, and in nearby galaxies is especially important, because detailed MW observations lay the bedrock for much of our understanding of disk formation and dynamical heating. However, the MW is only one galaxy, and it remains unclear how representative the MW is of similarly-massive galaxies. 
%Such a discrepancy may indicate that our simulated galaxies are too kinematically hot compared to observations.
%Therefore, here we explore a more detailed comparison between gas and stellar kinematics in our simulations, in the MW, and in nearby galaxies.

For these comparisons, we compute the average \textit{local} velocity dispersion of our galaxies, as opposed to the \textit{disk-wide} dispersion of our previous results.
Specifically, motivated by the typical measurement scales of the observations that we compare against here, we measure the `local' velocity dispersion within annular bins centered at 8 kpc and then compute the average across the annulus. 
%That is, instead of calculating the dispersion of all stars within each galaxy's entire`solar annulus' (the `disk-wide dispersion'), 
%This local dispersion better encapsulates how observed dispersions are measured in practice. 
%: all of the observed dispersions measured in external galaxies
%We find that stars have local present-day velocity dispersions that are significantly lower than their disk-wide counterparts. For example, stars younger than 100 Myr at R=8 kpc have local dispersions that equal $\approx$ 57, 55, 77, and 63\% of their disk-wide $\sigma_{\rm R}$, $\sigma_{\rm \phi}$, $\sigma_{\rm Z}$, and $\sigma_{\rm tot}$, respectively, corresponding to respective differences of $\sim$ 14, 11, 3, and 16 km/s. Therefore, $\approx$ 37\% of these stars' $\sigma_{\rm tot}$ arises from spatial variations within the disk.

\begin{figure}
\includegraphics[width = \columnwidth]{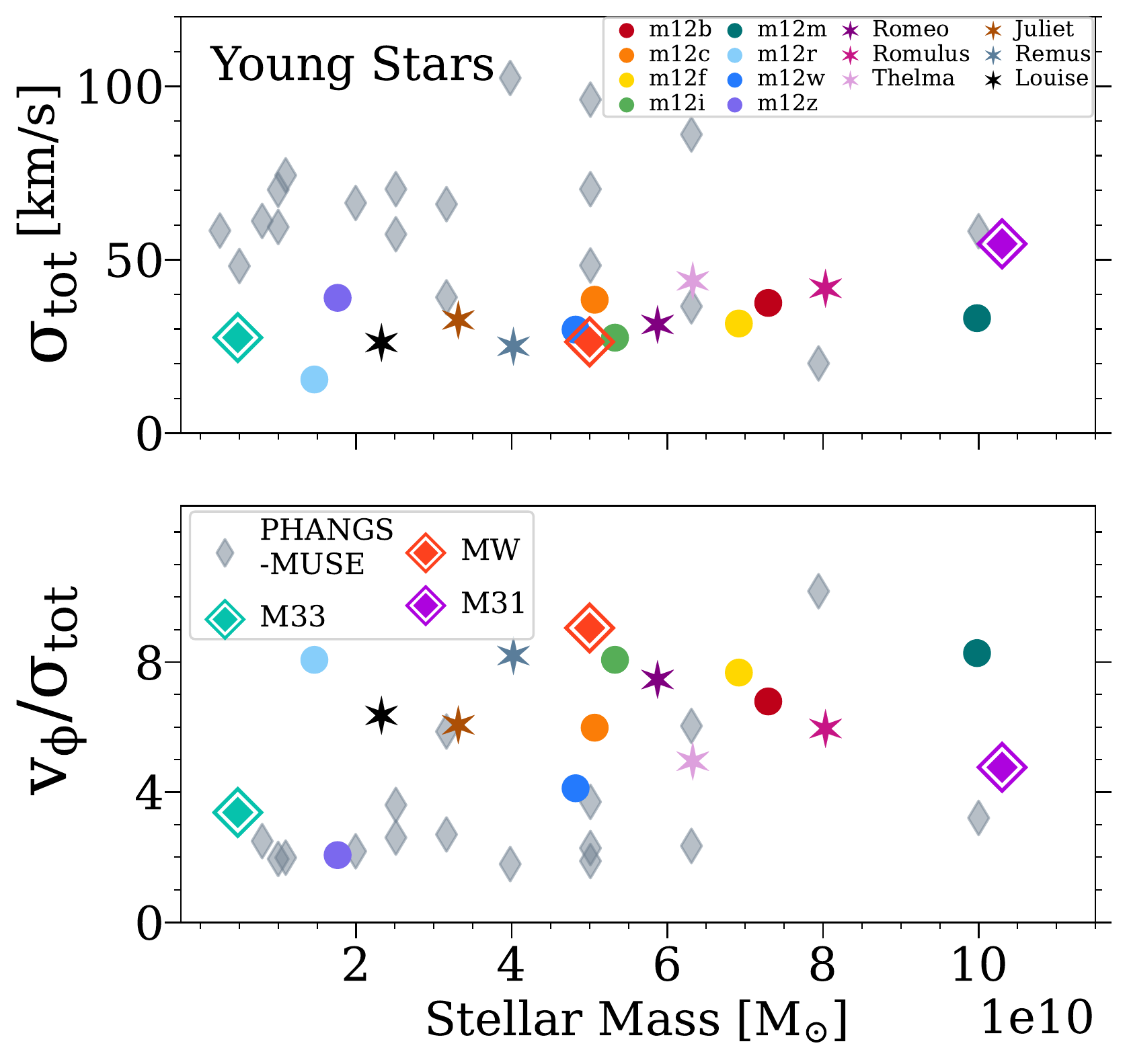}
\vspace{-5 mm}
\caption{
\textbf{Total velocity dispersion, $\sigma_{\rm tot}$, and median $v_{\rm \phi} / \sigma_{\rm tot}$, today for young (age < 100 Myr) stars versus stellar mass} for 14 galaxies.
As before, we show the isolated (m12) galaxies via circles and LG-like paired galaxies via stars.
We also show observations of young stars in the Local Group: the MW \citep{Casagrande11}, M31 \citep{Dorman2015}, and M33 \citep{Quirk22}, and in 19 nearby star-forming disk galaxies from the PHANGS-MUSE survey \citep{Pessa23}.
Young stars in FIRE-2 galaxies have kinematics that broadly agree with most observed galaxies; their $v_{\rm \phi} / \sigma_{\rm tot}$ are slightly lower than the MW, but two galaxies are within 10\% of the MW.
%Young stars in FIRE-2 galaxies have velocity dispersions that are similar to M31 but moderately larger than the MW and M33; however, their $v_{\rm \phi} / \sigma_{\rm tot}$ values agree with both M31 and M33.
Young stars in the MW are kinematically colder than those in M31, M33, and all but one of the 19 galaxies in PHANGS-MUSE.
\vspace{-4 mm}
}
\label{fig:stellar_mass_obs}
\end{figure}

Figure~\ref{fig:stellar_mass_obs} shows the \textit{local} $\sigma_{\rm tot}$ (top) and $v_{\rm \phi} / \sigma_{\rm tot}$ (bottom) for young (age < 100 Myr) stars versus stellar mass. 
We also show the observed velocity dispersion of young stars from \citet{Casagrande11} for the MW, \citet{Dorman2015} for M31 (Andromeda), \citet{Quirk22} for M33 (Triangulum), and \citet{Pessa23} for 19 galaxies from PHANGS-MUSE.
%A more consistent comparison would include generating mock observations for our simulated galaxies and accounting for differences in galaxy inclinations, radial and vertical selection functions, smoothing circle/aperture size, and age and kinematic uncertainties in measurements of observed galaxies.
We show velocity dispersions for only young stars; in McCluskey et al. (in prep.), we will pursue a more rigorous comparison of the full relation between velocity dispersion and age using mock observations.
%That said, we now discuss a few caveats. 
%However, here we simply aim to compare the observed galaxies against each other and explore whether our simulated galaxies have realistic kinematics for young stars.

An important caveat is that each of these observational works measured stellar populations and velocity dispersions somewhat differently.
In particular, while measurements of M31, M33, and the MW used individual stars, PHANGS-MUSE measurements used the integrated spectra from regions of size $\approx 100 \pc$, which may include some contribution from older stars, leading to larger velocity dispersions.
Nonetheless, the larger sample from PHANGS-MUSE provides important statistical context for both our simulations and the Local Group.
M31 and M33 observations measure resolved stars in multiple apertures of radius $\approx 760$ and 350 pc, respectively, while the MW measurement includes stars within 60 pc of the Sun, so this represents only a single region of the MW.
As a compromise to matching all of these works, and to limit numerical noise in our simulations, we use an intermediate aperture of diameter 250 pc (unlike in Figure~\ref{fig:local_vs_global}, where we used smaller 150 pc patches to better match ALMA's resolution). Using a larger region in our simulations increases the measured dispersion, but this effect is minimal for regions $\lesssim 1 \kpc$.

Moreover, we show the youngest stellar population from each work, but their age binning varies: the average age is 100 Myr for the MW, 30 Myr for M31, 80 Myr for M33, and 100 Myr for PHANGS-MUSE.
Therefore, for this comparison only, we measure stars in our simulations today younger than 100 Myr.
This reasonably matches the MW and PHANGS-MUSE, but these younger stars in M31 and M33 may have undergone somewhat less post-formation heating than in the MW, PHANGS-MUSE, or our simulations.

Finally, full 3D kinematics are available only for the MW: measurements of M31, M33, and the PHANGS-MUSE sample are limited to line-of-sight dispersions, $\sigma_{\rm LOS}$. To estimate a total 3D dispersion, we assume isotropy, such that $\sigma_{\rm tot}^2 = 3 \sigma_{\rm LOS}^2$, given the wide variety of observed inclination angles across these observations.
This is a simplification, given that our simulated galaxies show $\sigma$ that is not isotropic, with $\sigma_{\rm R}$ > $\sigma_{\rm \phi}$ > $\sigma_{\rm Z}$.
%The smaller the inclination angle, the larger the contribution from the vertical velocity component. 
As such, our assumption may underestimate the total dispersion of galaxies observed with smaller inclinations (more face-on), including the PHANGS-MUSE sample, which generally has inclinations $\approx 25 - 55^{\circ}$, while our assumption likely overestimates the total dispersion of highly-inclined galaxies, such as M31 at $77^{\circ}$.

Our 14 FIRE-2 galaxies have an average \textit{local} $\sigma_{\rm tot, now} \approx 32 \kms$ for young stars, with isolated and LG-like galaxies exhibiting similar values, on average. While our average $\sigma_{\rm tot, now}$ is slightly higher than the MW and M33's values of $\approx$ 26.3 and 27.5 km/s, it is considerably lower than M31's value of $\approx$ 55 km/s. However, the local total velocity dispersion of young stars shows a slight increase with galaxy mass, unlike the disk-wide vertical dispersion in Figure~\ref{fig:stellar_mass}. In turn, the high dispersion of M31 is likely (at least in part) because of its higher stellar mass.  We emphasize that MW's value of $\sigma_{\rm tot, now}$ is lower than all but 1 of the 21 galaxies in this PHANGS-MUSE sample. That said, 5 of our galaxies exhibit dispersions within 5 km/s of the MW's value, while 9 lie within 10 km/s, and 3, the fairly low-mass m12r, Remus, and Louise, have even lower dispersions than the MW at $\approx$ 15.5, 25.0, and 26.1 km/s, respectively.

\citealt{sanderson-20} (Figure 2) also compared three FIRE-2 galaxies---m12i, m12f, m12m---against observations of the MW and M31; but for M31, they compared the \textit{disk-wide} total (3D) dispersion in FIRE-2 against just the \textit{local} line-of-sight (1D) dispersion of M31.
Therefore, our comparison is more accurate and shows better agreement between FIRE-2 and the MW and M31.
%Figure~\ref{fig:stellar_mass_obs} shows that young stars in FIRE-2 have local $\sigma_{\rm tot}$ within the range of observations. 

Figure~\ref{fig:stellar_mass_obs} (bottom) compares $v_{\rm \phi} / \sigma_{\rm tot}$, which is a better, dimensionless metric of `diskiness' and the degree of rotational support. For each of the PHANGS-MUSE galaxies, we use their maximum $v_{\rm \phi}$ in \citet{Lang20}, which they determined using PHANGS-ALMA measurements of CO (2–1) emission. We note that \citet{Lang20} do not include $v_{\rm \phi}$ values for 3 of the PHANGS galaxies; as such, we do not show these 3 galaxies in the bottom panel. 
As above, we emphasize that this observational compilation shows that \textit{young stars in the MW are kinematically colder, in terms of both $\sigma_{\rm tot}$ and $v_{\rm \phi} / \sigma_{\rm tot}$, than those in all but 1 of 21 nearby galaxies}, including M31 and M33. Specifically, young stars in the MW have $v_{\rm \phi} / \sigma_{\rm tot} \approx$ 9, while M31, M33, and the average PHANGS galaxy have values of $\approx$ 4.8, 3.4, and 3.4. In turn, this initial comparison indicates that the MW is a kinematic outlier compared to similar-mass nearby galaxies; however, we will test this hypothesis more rigorously in McCluskey et al. (in prep.). 

Secondly, young stars in FIRE-2 galaxies show a range of $v_{\rm \phi} / \sigma_{\rm tot}$ that is \textit{consistent} with the range of the observed sample: all FIRE-2 galaxies reside within the observed range. The average $v_{\rm \phi} / \sigma_{\rm tot}$ of FIRE-2 is $\approx$ 6.4, with LG-like and isolated galaxies having similar values.
% of $\approx$ 6.4 and 6.5, respectively. 
If anything, the typical FIRE-2 galaxy is kinematically colder than M31, M33, and the typical galaxy in PHANGS-MUSE.
Although none of the FIRE-2 galaxies have $v_{\rm \phi} / \sigma_{\rm tot}$ quite as high as the MW's value ($\approx$ 9), two galaxies, Remus and m12m, are within 10\% of the MW's value, at 8.3 and 8.2, respectively.
%and 7 (that is, half of our FIRE-2 galaxies) are within 75\%.
Thus, this tentative comparison indicates that \textit{young stars in FIRE-2 have statistically realistic/representative values of  $\sigma_{\rm tot}$ and $v_{\rm \phi} / \sigma_{\rm tot}$}, provided we measure both on spatial scales comparable to those of observations.
%, in agreement with many nearby galaxies.
%, but are not as kinematically cold as young stars in the MW.

\subsection{Time of disk onset}
\label{subsec:disk settling time}

Galaxies transitioned from their Pre-Disk to their Early-Disk Era at their `time of disk onset', which marks the \textit{start} of the \textit{continuous} process of disk settling. Most of our galaxies transitioned $4 - 8 \Gyr$ ago, although Romeo transitioned particularly early, $\approx 11 \Gyr$ ago. We now examine the time of disk onset across our sample, including how it relates to a galaxy's stellar kinematics today. We also reiterate that unlike in the previous subsection, all values are now \textit{disk-wide} quantities, just as in the first 6 subsections of our Results. 

We defined the transition to the Early-Disk Era as when all stellar populations that formed after this time had $(v_{\rm \phi} / \sigma_{\rm tot})_{\rm form} > 1$.
%We chose 1 as our threshold value because this represents when rotation first exceeded random motion.
%, although systems with $(v_{\rm \phi}/\sigma_{\rm tot})_{\rm form} \approx 1$ do not resemble the thin disks observed locally.  
%We found that using a larger/smaller threshold value simply pushed the lookback time that the galaxy transitioned later/earlier, but had little effect on the order that galaxies within our sample transitioned. 
%We previously used $v_{\rm \phi} / \sigma_{\rm tot}$ at \textit{formation} because it reflects the dynamical state of the galaxy's star-forming gas disk at that snapshot. 
Of course, $v_{\rm \phi} / \sigma$ measured today can differ from its formation value. To examine how transition times derived using present-day kinematics compare to our fiducial formation-based times, we also measure the age when $(v_{\rm \phi} / \sigma_{\rm tot})_{\rm now}$ permanently exceeds 1.
Furthermore, we show that $\sigma_{\rm Z}$ evolved somewhat differently post-formation than the other components, so we additionally test this by also finding the oldest age after which $(v_{\rm \phi} / \sigma_{\rm Z})_{\rm now}$ permanently exceeds $\sqrt{3}$, under the simplifying (and in detail incorrect) approximation of isotropy.
%, such that $\sigma_{\rm Z}$ contributes $1 / \sqrt{3}$ of the total dispersion.
For all of these metrics, we tested a range of threshold values, and while this shifts the exact lookback time that a galaxy transitioned, it has little effect on the relative ordering within our sample, which is our primary metric of interest. We also tested transition times determined using only $v_{\rm \phi}$ (at both formation and present-day) and found that they broadly agree with our $v_{\rm \phi} / \sigma$ derived times, differing by $\lesssim 1\Gyr$ for thresholds above $\approx$ 100 km/s. However, we limit our analysis to $v_{\rm \phi} / \sigma$-based metrics, because, unlike $v_{\rm \phi}$ alone, $v_{\rm \phi} / \sigma$ is dimensionless, largely independent of galaxy stellar mass in this mass range, and has a logical threshold ($v_{\rm \phi}>\sigma$). 
%This arises from the continuous nature of disk formation coupled with our requirement that the threshold value is \textit{permanently} met. If $v_{\rm \phi} / \sigma$ permanently exceeded some initial threshold, it will soon grow to exceed larger thresholds as well, generally before a galaxy that has yet to permanently exceed the initial threshold will.

In summary, we use three thresholds to measure the time of disk onset that defines the transition from the Pre-Disk to Early-Disk Era:
\begin{itemize}
\item $v_{\rm \phi} / \sigma_{\rm tot} > 1$ at formation
\item $v_{\rm \phi} / \sigma_{\rm tot} > 1$ as measured today
\item $v_{\rm \phi} / \sigma_{\rm Z} > \sqrt{3} \approx 1.73$ as measured today
\end{itemize}
Table~\ref{table:table1} lists the lookback times/ages of disk onset for each galaxy.

These three metrics give times within $\approx 0.5 - 1 \Gyr$ of each other for most galaxies. Times based on $(v_{\rm \phi} / \sigma_{\rm Z})_{\rm now}$ and $(v_{\rm \phi} / \sigma_{\rm tot})_{\rm now}$ agree best with each other. Thus, the present vertical and total kinematics of stars provide similar insights into the disk's formation. However, lookback times from the kinematics at formation are generally earlier, because dynamical heating tends to decrease $v_{\rm \phi} / \sigma$ of stars post-formation.
A few galaxies yield substantially different times for different metrics. Notably, formation kinematics indicate that m12r transitioned 5.9 Gyr ago, while current kinematics indicate that it settled only 0.26 Gyr ago, because m12r experienced a merger with an LMC-mass galaxy $\approx 0.5 \Gyr$ ago, which significantly heated existing stars.
This lends caution to using present-day kinematics to infer the exact timing of events in a galaxy's formation history.

As Figure~\ref{fig:fig2} shows, once our galaxies transitioned to the Early-Disk Era, their stars formed on progressively cooler orbits as the disk continuously settled.
If, after the time of disk onset, $(v_{\rm \phi} / \sigma)_{\rm form}$ grew at a broadly similar rate in each galaxy, then galaxies that had earlier times of disk onset would have higher $(v_{\rm \phi} / \sigma)_{\rm form}$ today, simply from having more time to settle.

\begin{figure}
\centering
\includegraphics[width = \columnwidth]{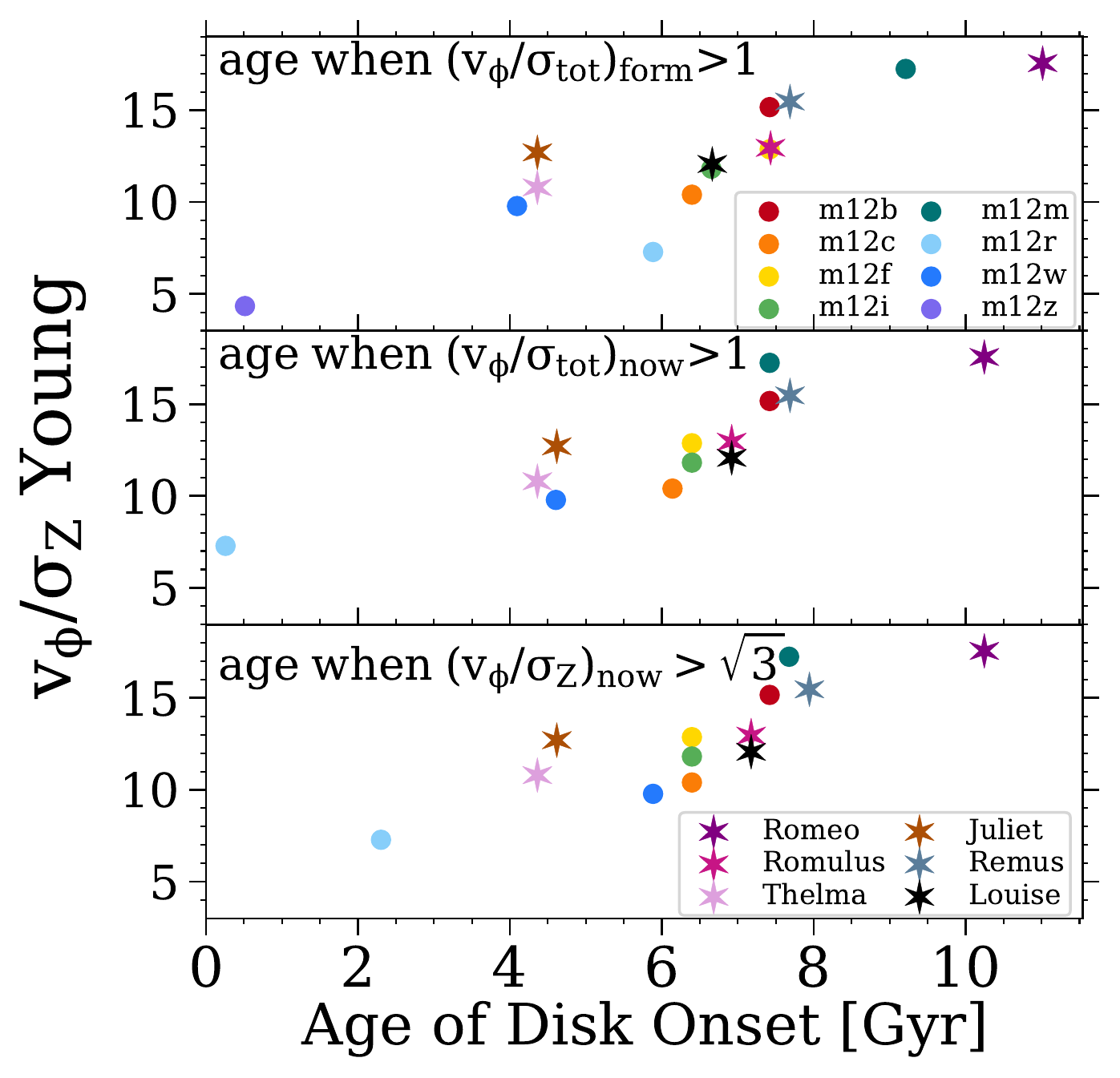}
\vspace{-7 mm}
\caption{
\textbf{Degree of rotational support, $v_{\rm \phi} / \sigma_{\rm Z}$, for young stars (age $< 250 \Myr$) versus the lookback time of disk onset}.
Circles show isolated (m12) galaxies, and stars show galaxies in LG-like pairs.
We examine 3 metrics to determine the time that each galaxy's disk started to settle, which yield similar but occasionally different times and ordering between galaxies.
\textbf{Top}: disk onset time defined as age when $v_{\rm \phi} / \sigma_{\rm total}$ at formation permanently exceeded 1.
\textbf{Middle}: disk onset time defined age as when $v_{\rm \phi} / \sigma_{\rm total}$ as measured today permanently exceeded 1.
\textbf{Bottom}: disk onset time defined age as when $v_{\rm \phi} / \sigma_{\rm Z}$ as measured today permanently exceeded 1.8.
Overall, using kinematics at formation versus today or measuring different velocity components all lead to similar times of disk onset.
\textit{All 3 metrics show that disks that started to settle earlier currently form stars on more rotationally supported orbits}, which implies that the MW, which is unusually dynamically cold today, started to settle unusually early.
}
\label{fig:disktime}
\end{figure}

Figure~\ref{fig:disktime} tests this idea, showing the time of disk onset versus $v_{\rm \phi} / \sigma_{\rm Z}$ of stars younger than 250 Myr for our 14 galaxies. The three rows show our three thresholds in $v_{\rm \phi} / \sigma$ for the onset of disk settling, which all show similar trends. We show isolated galaxies as circles and galaxies in LG-like pairs as stars, as in Figure~\ref{fig:stellar_mass}. Figure~\ref{fig:disktime} shows that the time of disk onset was $\approx 4.5 - 8 \Gyr$ ago ($z \approx 0.4 - 1$) for the majority of our galaxies, although Romeo (see Section~\ref{subsec:case studies}) has the earliest time of disk onset, $\approx 10.2 - 11$ Gyr ago ($z \approx 2$).

Figure~\ref{fig:disktime} shows that galaxies that experienced earlier disk onset have young stars (age < 250 Myr) with larger $v_{\rm \phi} / \sigma_{\rm Z}$ today, that is, currently form stars on dynamically cooler orbits. While we do not show it, we find similarly strong correlation between the time of disk onset and the average $v_{\rm \phi} / \sigma_{\rm Z}$ and $v_{\rm \phi} / \sigma_{\rm tot}$ for both all and young stars. Table~\ref{table:table2} lists the Pearson correlation coefficients, which are close to 0.9, and their corresponding p-values, which are $\sim 10^{-5} - 10^{-6}$, indicating a strong correlation. We find similar results using instead the Spearman rank coefficients. 

We also examined correlations with other galaxy properties.
Surprisingly, the time of disk onset does not correlate much with the galaxy's current stellar mass, within our sample.
It also does not correlate with any component of $\sigma$, for either all or young stars.
However, the time of disk onset \textit{does} correlate with the median $v_{\rm \phi}$ of all or young stars, with Pearson coefficients ranging between 0.75 and 0.85 for all stars and 0.73 and 0.80 for young stars. Despite the lack of correlation with $\sigma$, each threshold has its highest correlations with $v_{\rm \phi} / \sigma$, \textit{not} with $v_{\rm \phi}$ alone, further supporting that $(v_{\rm \phi} / \sigma)_{\rm now}$ is a more fundamental (and dimensionless) metric than $\sigma_{\rm now}$ to compare different galaxies' formation histories.

The strong correlation with the $v_{\rm \phi} / \sigma$ of young stars today is particularly striking, considering it only moderately correlates with $v_{\rm \phi}$ of young stars. This correlation implies that, after disk onset, these disks settled at relatively similar rates, in order to preserve rank ordering. In particular, after disk onset, mergers and internal processes tend not to significantly alter the overall Hubble-time-averaged rate of disk settling.

In general, galaxies in LG-like pairs experienced earlier disk onset than isolated galaxies. This follows from Section~\ref{subsec:galaxy mass trends}, where LG-like galaxies had lower disk-wide $\sigma$ and higher disk-wide $v_{\rm \phi} / \sigma$ than isolated galaxies of similar (current) stellar mass, consistent with their larger radii \citep{Bellardini22}. The earlier settling times for galaxies in LG-like environments follows from the fact that such galaxies assemble (form stars) earlier than those in isolated environments, which in turn follows from the fact that their dark-matter halos formed earlier \citep{Garrison-Kimmel19, Santistevan20}.
Thus, galaxies in LG-like environments, despite ending up at similar stellar mass today as galaxies in isolated environments, formed stars earlier, their disks started to settle earlier, and their disks are radially larger and dynamically cooler today.

This dependence on LG environment provides important insight into the assembly history of the MW (and M31) compared with other galaxies of similar mass today. As discussed in Section~\ref{subsec:observation}, young stars in the MW are dynamically colder than those in other nearby galaxies. If we naively extrapolate and place the MW's (local) $v_{\rm \phi} / \sigma_{\rm tot} \approx 8$ along the relation in Figure~\ref{fig:disktime}, the trends within our simulated sample indicate that the MW disk began to settle earlier than our galaxies' disks ($\gtrsim 11 \Gyr$ ago). Remarkably, this simple extrapolation agrees with recent estimates of the age of the MW's thick disk: recent works argue that its coherent rotation may have begun $\approx 12 - 13 \Gyr$ ago \citep{Conroy22, Xiang22}. We postulate that this dependence on LG environment provides critical insight into this seemingly early assembly of the MW.

\begin{table*}
\centering
\begin{tabular}{|c|c|c|c|c|} 
\hline
metric & $v_{\phi} / \sigma_{Z} \ \mathrm{young}$ & $v_{\rm \phi} / \sigma_{\rm tot} \ \rm{young}$ & $v_{\rm \phi} / \sigma_{Z} \ \rm{all}$ & $v_{\rm \phi} / \sigma_{\rm tot} \ \rm{all}$ \\ 
& corr. coeff.~(p-value) & corr. coeff.~(p-value) & corr. coeff.~(p-value) & corr. coeff.~(p-value) \\
\hline
$v_{\phi} / \sigma_{\rm Z, now} > 1.8 $ & $0.90~(1.3 \times 10^{-6}$) & $0.91~(8.1 \times 10^{-6}$) & $0.86~(6.7 \times 10^{-5}$) & $0.92~(3.3 \times 10^{-6}$) \\ 
$v_{\phi} / \sigma_{\rm tot, now} > 1$ & $0.91~(6.1 \times 10^{-6}$) & $0.86~(8.8 \times 10^{-5}$) & $0.87~(5.1 \times 10^{-5}$) & $0.93~(2.4 \times 10^{-6}$) \\
$v_{\phi} / \sigma_{\rm tot, form} > 1$ & $0.89~(1.7 \times 10^{-5}$) & $0.89~(2.4 \times 10^{-5}$) & $0.81~(4.0 \times 10^{-4}$)& $0.88~(3.00 \times 10^{-5}$) \\
$t_{\rm burst}$ & $0.78~(3.0 \times 10^{-3}$) & $0.70~(0.011)$ & $0.55~(0.067)$ & $0.63~(0.028)$ \\
\hline
\end{tabular}
\vspace{-2 mm}
\caption{
\textbf{Pearson correlations and p-values between the degree of rotational support of stars and different lookback times of galaxy transitions}. The first three rows are the lookback times of the onset of disk settling, using the three different metrics discussed in Section~\ref{subsec:disk settling time}, while the fourth row is the lookback time corresponding to the galaxy's transition from bursty to steady star formation from \citet{Yu2021}. \textit{All three times of disk onset show strong correlations with the degree of rotational support: earlier-forming disks are kinematically colder, both for young stars and for all stars, than later-forming disks.}
%\vspace{-2 mm}
}
\label{table:table2}
\end{table*}

\section{Summary and Discussion}
\label{section:discussion}

\subsection{Summary of Results}

We summarize the main results from our analysis of stellar disks across the formation histories of 14 MW-mass galaxies from the FIRE-2 simulations.
Most importantly, \textit{the kinematics of stars today do not simply reflect their kinematics at formation, but neither do they simply reflect post-formation dynamical heating; both processes are important}.
In particular, although the present-day dispersion of stars, $\sigma_{\rm now}$, increases monotonically with age, $\sigma_{\rm form}$ does not.

We identified \textbf{Three Eras of Disk Formation}, when stars formed with different kinematics and experienced different degrees of post-formation dynamical heating:

\textbf{1) Pre-Disk Era} (typically $\gtrsim 8 \Gyr$ ago): stars formed on dispersion-dominated orbits, with $(v_{\rm \phi} / \sigma_{\rm tot})_{\rm form} < 1$ and $v_{\rm \phi} \lesssim 20 \kms$.
$\sigma_{\rm form}$ increased over time, reflecting the deepening of the gravitational potential. Thus, the oldest stars formed with the \textit{lowest} dispersion. However, present-day dynamics versus stellar age do not reflect this trend: $\sigma_{\rm now}$ increases monotonically with age, meaning that the oldest stars now have the \textit{highest} dispersion.
Thus, post-formation dynamical heating is most important during this era.
Furthermore, the typical $v_{\rm \phi, now} \approx 60 \kms$, so the oldest stars often have been `spun up' since formation.
As a result of both of these trends,  $(v_{\rm \phi} / \sigma_{\rm tot})_{\rm now}$ remains below 1, similar to formation.

\textbf{2) Early-Disk Era} (typically $\approx 4 - 8 \Gyr$ ago): began at the time of disk onset, when $v_{\rm \phi} / \sigma_{\rm tot} > 1$ and grew rapidly, and it ended when the growth of $v_{\rm \phi} / \sigma$ slowed significantly. Similarly, $\sigma_{\rm form}$ peaked at the start of this era, and it decreased steadily throughout, reaching a minimum by the era's end. The amount of dispersion added by post-formation heating was constant with age throughout this era, so the slope of $\sigma_{\rm now}$ with age is unchanged from that at formation. The majority of $\sigma_{\rm now}$ for these stars was in place at formation, not added via post-formation dynamical heating.

\textbf{3) Late-Disk Era} (typically $\lesssim 4 \Gyr$ ago): began when the growth of $(v_{\rm \phi} / \sigma_{\rm tot})_{\rm form}$ slowed. Stars formed with nearly constant $\sigma_{\rm tot, form}$. $\sigma_{\rm tot, now}$ increases monotonically with age, because post-formation dynamical heating increases with age, and the oldest stars in this era have $\sigma_{\rm now}$ with an equal contribution from post-formation heating and formation.

\textbf{We examined the evolution of the three different components of velocity}.
\textit{At formation}, $\sigma_{\rm R,form} > \sigma_{\rm \phi, form} > \sigma_{\rm Z, form}$ for all ages. Remarkably, across all eras and ages, the relative amount in each component is relatively unchanged.
\textit{At present}, $\sigma_{\rm R, now}$ continues to dominate at all ages, but $\sigma_{\rm Z, now} \approx \sigma_{\rm \phi, now}$ for stars that formed in the Early- and Pre-Disk Eras (ages $\gtrsim 4 \Gyr$).
\textit{Post-formation dynamical heating} primarily increased $\sigma_{\rm R}$ at ages $\lesssim 4 \Gyr$ but acted nearly isotropically for older stars.

\textbf{We examined the dependence of stellar kinematics on galaxy stellar mass}.
$\sigma_{\rm now}$ of all stars increases with galaxy stellar mass, but $(v_{\rm \phi} / \sigma)_{\rm now}$ of all stars is nearly constant. Young stars (age < 250 Myr) exhibit opposite trends: $\sigma_{\rm now}$ is independent of mass, while $(v_{\rm \phi} / \sigma)_{\rm now}$ increases with mass.

\textbf{We compared disk-wide and local ($\lesssim 250 \pc$) velocity dispersions for cold gas and young stars at formation}: local dispersions are $\approx$ 3 times lower than disk-wide dispersions for cold gas and $\approx$ 2 times lower for stars at formation. Thus, the velocity dispersion strongly depends on the spatial scale measured over, which is critical for observational comparisons.

\textbf{We compared locally-measured kinematics in our 14 simulated galaxies to those observed in 3 Local Group galaxies and nearby star-forming galaxies}, compiling measurements of $\sigma$ and $v_{\rm \phi} / \sigma$ for young stars (age $\lesssim 100 \Myr$) in the MW, M31, M33, and 19 galaxies from PHANGS-MUSE.
\textit{The MW is significantly kinematically colder than both M31 and M33, and all but one of the PHANGS galaxies.}
For young stars, all of our simulations show broad agreement with observations, with 3 of our galaxies even exhibiting lower locally-measured $\sigma_{\rm tot}$ than the MW. 

\textbf{We quantified the time of disk onset, the lookback time that each galaxy's disk began to settle}, using thresholds in $v_{\rm \phi} / \sigma$, measured both today and at formation. 
\textit{The degree of rotational support of young stars today correlates strongly with the time of disk onset: earlier-settling galaxies currently form colder disks.}
Most of our galaxies began to settle $\approx 4 - 8 \Gyr$ ago, with galaxies in LG-like environments generally starting to settle earlier than isolated galaxies. Romeo, a galaxy in a LG-like pair, began to settle the earliest, $\approx 10 - 11 \Gyr$ ago, likely the closest to the MW.

In Appendix~\ref{a:cr}, \textbf{we examine the impact of including self-consistent cosmic-ray injection/transport from stars}, assuming a constant diffusion coefficient, which in the FIRE-2 simulations leads to galaxies with lower stellar masses, lower rotational velocities, and lower dispersions.
\textit{Overall, the inclusion of cosmic rays does not significantly change $v_{\rm \phi} / \sigma$ at a given stellar mass}.

\subsection{Discussion}

\subsubsection{Caveats}

We first discuss key limitations and caveats to our analysis of 14 MW-mass galaxies from the FIRE-2 simulations.

First, we reiterate the selection function of our simulated sample.
We selected isolated or LG-like paired dark-matter halos from dark-matter-only simulations at $z = 0$ with $M_{\rm 200m} \approx 1 - 2 \times 10^{12} \Msun$, with no additional selection based on galaxy properties or merger/formation history, etc.
Thus, we did not choose our galaxies specifically to be analogues of the MW (or M31 or M33).
As such, our galaxy-averaged results represent randomly sampled formation histories of MW-mass galaxies, which should fairly sample the cosmological scatter in formation histories.

Second, these FIRE-2 simulations do not model all of the physics that may be relevant for disk formation, in particular, AGN feedback from black holes. The degree to which AGN may impact stellar dynamics in the solar annuli of MW-mass galaxies (which in this work we define as $R = 6 - 10 \kpc$ and $|Z| < 3 \kpc$) is unclear. While AGN feedback is likely most crucial for more massive galaxies and most significantly alters stellar dynamics in the galaxy's innermost regions \citep{Dubois13, Weinberger17, Mercedes-Feliz23, Wellons23}, growing evidence from simulations indicates that AGN can both directly and indirectly effect the kinematic and structural properties of MW-mass and low-mass galaxies \citep{Dashyan18, Koudmani21, Irodotou22}. For example, \citet{Grand17} found that the primary effect of AGN in MW-mass galaxies was to suppress central star formation, which in turn prevents the formation of overly massive bulges. Indeed, \citet{Grand17} found that a simulation of a MW-mass galaxy reran from $z = 1$ without AGN feedback had a smaller fraction of stars on disc-like orbits (defined as having orbital circularities > 0.7), more dominant bulges, and enhanced star formation in the outer disk leading to a more extended disk component compared with versions ran with AGN feedback.
%given stellar radial redistribution and AGN feedback's probable impact on quenching and global galaxy properties such as stellar mass and SFR, its effects may impact our results.

Additionally, our analysis of birth kinematics has limitations. We determined a star particle's position and velocity relative to the galaxy's major axes, which are stable and well-defined at present, but not necessarily at earlier times, when accretion and mergers were more significant and even the notion of a single main galaxy progenitor can be ill-defined \citep{Wechsler02, Giocoli07, Santistevan20}. Thus, cleanly separating the $R$, $\phi$, and $Z$ components is difficult at early times. However, all components of the dispersion at formation display similar trends as $\sigma_{\rm tot}$, which is independent of the axes' determination, implying that this limitation's impact on our overall qualitative results is likely minor.

Furthermore, as we discussed in Section~\ref{subsec:selection}, our simulations store snapshots every $20 - 25 \Myr$, so our star `formation' properties include this additional $\approx 12 \Myr$ of post-formation dynamical evolution, on average. This early heating can be significant: using an idealized simulation of an isolated MW-like galaxy with sub-parsec resolution (0.05 pc), \citet{Renaud13} found that stars that formed with $\sigma_{\rm tot, form} \approx 10 \kms$ were heated to $15 \kms$ in just 10 Myr. Similarly, our Figure~\ref{fig:heating_rate} shows that $\sigma_{\rm tot}$ most rapidly increased post-formation by $\approx 30 \kms$ in $\approx 125 \Myr$.
We tested if this early evolution could change our conclusions about the relative importance of `formation' dispersion versus post-formation heating. For stars younger than 10 Gyr, $\sigma_{\rm tot, now} / \sigma_{\rm tot, form} < 1.95$ for all but $\approx 3.5 \Gyr$ old stars, meaning that post-formation heating is less than 95\% of their $\sigma_{\rm tot, form}$. If we assume that stars were heated at the same effective rate as the youngest stars in Figure~\ref{fig:heating_rate} for $\approx 12 \Myr$, our revised $\sigma_{\rm tot, now} / \sigma_{\rm tot, form}$ still remains below 1.95 for the same ages. Even if we assume an initial effective heating rate that is $4 \times$ larger, $\sigma_{\rm tot, now} / \sigma_{\rm tot, form}$ remains less than 2 for the same ages. Thus, our results about the relative importance of post-formation heating remain unchanged.

Lastly, our analysis did not account for uncertainties in stellar age.
Stellar ages are difficult to measure accurately, and uncertainties on stellar ages for large populations are generally $\gtrsim 20 - 30\%$ \citep[for example][]{Soderblom10}, though the synergy of astrometric and spectroscopic surveys with asteroseismological data has driven tremendous progress in the past decade \citep[for example][]{Silva18, Mackereth19}.
Previous works showed that the inclusion of observationally-motivated age uncertainties of 20 - 30\% in simulations can (1) obscure features from mergers in the age-velocity dispersion relationship \citep{Martig14, Buck20}, and (2) decrease inferred exponential heating rates via the `smearing' of age bins \citep{Aumer16}. In McCluskey et al. ( in prep.) we will examine the kinematic trends of our simulations versus age at $z = 0$, incorporating realistic age uncertainties, which will allow us to compare to observational trends in much more detail.
%We preformed initial tests that included 20\% age uncertainties and similarly found that merger-driven features were weakened. However, we defer further analysis of age uncertainties to our next paper that will focus on more rigorous comparisons to observation.

\subsubsection{Disk galaxy formation in FIRE simulations}

As we discussed in the Introduction, our work builds upon a series of analyses of FIRE simulations that studied the formation histories of MW-mass galaxies, in particular, the co-evolution of their stellar and gaseous dynamics, morphology, star formation burstiness, metallicity spatial variations, and virialization of the inner CGM \citep{Ma-17, Garrison-Kimmel18, Yu2021, Stern21, Bellardini22, Gurvich22, Hafen22, Trapp22, Yu22, Hopkins23}.
% In this picture, a dynamically-cold disk is able to form because of a qualitative change in the nature of feedback, in particular, the inability of feedback to drive gaseous outflows that fuel burst cycles.
We further quantified the notion that MW-mass disk galaxy formation occurred in distinct phases,
%largely defined by the co-evolution and shared transition times of the CGM, ISM, star-formation time variability, and metallicity spatial variations. Our work furthers this picture by
showing that stellar kinematics, both at formation and at present, follow broadly contemporaneous evolutionary phases.

In particular, what we define as the transition from the Early-Disk to the Late-Disk Era coincides with the virialization of the inner CGM \citep{Stern21}, the transition from bursty to steady star formation \citep{Yu2021}, and a transformation in the energetics and angular momentum of both accreted gas \citep{Hafen22} and the ISM \citep{Gurvich22}. During this stage of steady star formation in the Late-Disk Era, $\sigma_{\rm form}$ was low and constant, with the gas turbulence reaching a floor maintained by the thermal and turbulent pressure of the warm neutral/ionized medium \citep[for example][]{Leroy08}, stellar feedback, and spiral arms.

%As discussed in Section~\ref{subsec:disk settling time}, we defined the transition from the Pre-Disk to the Early-Disk Era, the `Time of Disk Onset', as when the stellar $v_{\rm \phi} / \sigma$ permanently exceeded some threshold value. Similarly, 
\citet{Yu2021} quantified the lookback time that star-formation transitioned from bursty to steady -- the `burst time' or $t_{\rm b}$ -- for 12 of the 14 galaxies that we analyze.
Table~\ref{table:table1} lists each galaxy's $t_b$ from \citet{Yu2021}.
%\citet{Yu2021} noted that although their times shifted based on the exact threshold value used to define steady star-formation, this shift did not qualitatively impact their results, just as we found for our dynamical threshold value. 
%Thus, the relative ordering of our galaxies and their qualitative trends are of greater interest than the exact times themselves.
The transition to steady star formation typically occurred $\approx 2 \Gyr$ after our `time of disk onset', when stellar $v_{\rm \phi} / \sigma_{\rm tot} > 1$. Nominally, this implies that star formation remained bursty for a while during the Early-Disk Era, in agreement with our finding that $\sigma_{\rm form}$ remained high (but declining) throughout this era.
However, our times of disk onset are based on stellar dynamics within the solar annulus today ($R = 6 - 10 \kpc$ and $|Z| < 3 \kpc$) and only moderately correlate with $t_{\rm b}$, having Spearman coefficients between $0.75 - 0.78$. By contrast, \citet{Yu2021} included all stars that formed within 20 kpc physical of the galactic center. If instead we extend our radial range to $R < 12 \kpc$, this yields times that (1) are on average 1.2 Gyr later than our original times, because of the hotter kinematics of the inner region, and (2) more strongly correlate with $t_{\rm b}$, with Spearman coefficients between 0.94 and 0.96 for our three dynamical thresholds. Thus, these estimates for when a galaxy's disk began to form/settle can depend on the selection of stars today, with stars in the solar neighborhood yielding earlier settling times.
We will investigate radial trends in these simulations in future work.

%Given that this transition occurred only after the disk began to settle, some aspect of rotationally-dominated motion may facilitate the transition to steady star-formation.

%The high correlation between the time of disk onset (when the galaxy transitioned into the Early-Disk Era) and the burst time (when the galaxy transitioned into the Late-Disk/steady star-formation Era) indicates that the duration of the Early-Disk Era is similar for most of our galaxies. If the Early-Disk Era's length significantly varied between galaxies, the order that our galaxies entered the era would not agree well with the order they left. Thus, disks, once formed, evolved similarly within our sample.

%Two of our three methods used to determine the time of disk onset depend only on the present-day kinematics of stars. Specifically, these times correspond to the age of the oldest rotationally-dominated stars at present(with the requirement that all younger populations also currently exhibit rotationally-dominated motion). Thus, the age of the oldest, rotationally-dominated stars at present correlates to both the time of disk onset and the time when the SFR stabilized, the inner-CGM virialized, and gas accretion became dominated by hot mode accretion.

Using the same FIRE-2 simulations, \citet{Yu22} argued that they can subdivide their bursty star-formation era into two eras: an early bursty and chaotic phase, when stars formed with irregular morphology, and a subsequent bursty phase when stars formed with disky morphology. These bursty star-formation eras generally correspond to our Pre-Disk and Early-Disk eras. \citet{Yu22} also studied how both the birth and present-day orbits of stars reflect the era when they formed, classifying stars as belonging to the thin-disc, thick-disc, and isotropic spheroid based on their orbital circularity at both formation and at $z = 0$. They found that, in general, a star's orbital classification at formation reflects the era in which it formed: stars primarily formed on isotropic spheroid orbits in the early bursty star-formation phase, thick-disc orbits in the later bursty star-formation phase, and thin-disc orbits in the steady star-formation phase. However, \citet{Yu22} noted that this is not a one-to-one mapping: $\approx 34\%$ of stars that formed with thin-disc orbits now have thick-disc orbits, while a similar fraction of stars that formed with thick-disc orbits now have isotropic spheroid orbits.
Furthermore, the fraction of stars that were born `thin' but are now `thick' increases with age.
Our results agree with theirs, and we further quantify the degree of post-formation heating/perturbations that stars experienced.

%Recently, \citet{Hopkins23} used an extensive set of numerical experiments in cosmological simulations to identify the physical mechanisms that cause the formation of disks and the end of bursty star formation. That is, \citet{Hopkins23} investigated what causes galaxies to transition from (what we call) the Pre-Disk Era to the Early-Disk Era and from the Early-Disk Era to the Late-Disk Era. \citet{Hopkins23} found that (two different aspects of) the gravitational potential drives these transitions. The shape of the potential, specifically the of development of a sufficiently centrally-concentrated mass profile, drives the initial formation of the disk ($\approx$ Pre-Disk to Early-Disk transition). A sufficiently centralized potential provides the galaxy with a dynamic center, allowing for the conservation of angular momentum,  stabilizes the disk against strong non-axisymmetric modes, such as spiral arms or bars, and promotes orbit-mixing, facilitating the coherence of angular momentum.  
%Similarly, the depth of the potential (specifically, the depth of the potential becoming sufficiently large at the radii of star formation) drives the transition to smooth star formation ($\approx$ Early-Disk to Late-Disk transition). A sufficiently deep potential `traps' and/or confines outflows near the disk, enabling short recycling times such that star formation bursts `blur' together in time.  

\subsubsection{Comparison to other cosmological zoom-in simulations}

Various works have used different cosmological zoom-in simulations to study the formation histories of MW-mass galaxies \citep[for example][]{Bird12, Brook12, Martig14, Grand16, Buck20, Agertz21, Bird-21}.
Across a wide range of different numerical implementations of hydrodynamics, star formation, and feedback, cosmological simulations consistently agree that the dynamical history of a MW-mass disk arises from \textit{both} cosmological disk settling and post-formation dynamical heating; one cannot neglect one or the other.
As we discuss below, most current-generation zoom-in simulations of MW-mass galaxies also broadly agree in both the \textit{typical} disk-wide velocity dispersion of young stars at $z \approx 0$ and the shape/magnitude of the stellar age dependence, as measured today. 
%We first note simulations that we show strong agreement with, and then discuss works which find slightly different results than we do and where these differences may stem from. 

%Earlier zoom-in simulations struggled to recreate realistic disk galaxies 

Our results agree with the VINTERGARTEN simulation \citep{Agertz21}, which used the same initial conditions as m12i but simulated with the adaptive mesh refinement code RAMSES \citep{Teyssier02}. Their $\sigma_{\rm Z, now}$ versus age \citep[Figure 14 of][]{Agertz21} is nearly identical to ours for m12i, both in value and shape: both of our simulations show near exponential increases in $\sigma_{\rm Z, now}$ with age for stars younger than 6 Gyr and a jump at 6 Gyr. Our results also agree well with the NIHAO-UHD suite \citep{Buck20}: their 5 galaxies, which span $\Mstar = 1.5 - 16 \times 10^{10} \Msun$, have $\sigma_{\rm Z, now} \approx 20 - 35 \kms$ for young stars, and $\approx 60 - 120 \kms$ for stars 12 Gyr old, similar to ours. Any difference in the exact age dependence of each galaxy's dispersion follows from its unique formation history.

\citet{Grand16} analyzed 16 MW-mass galaxies from the Auriga simulations, also finding broadly similar disk-wide $\sigma_{\rm Z, now}$ and $\sigma_{\rm Z, form}$ as we do. Similarly to FIRE, NIHAO-UHD, and VINTERGARTEN, Auriga simulations have typical disk-wide $\sigma_{\rm Z, now} \approx 20 - 25 \kms$. 
They also concluded that cosmological disk settling was generally the primary effect that set the relation between the velocity dispersion and the age of stars today.
However, \citet{Grand16} found that bars were the strongest contributor to vertical dynamical heating. This seemingly contradicts our results, because stars in \textit{all} of our galaxies had significant post-formation increases to their disk-wide $\sigma_{\rm Z}$, including stars in unbarred or weakly-barred galaxies. Furthermore, even our galaxies that do house significant bars near $z = 0$ (such as m12m) did not at early times, and our results indicate that heating is most significant for stars that formed in the Pre-Disk Era. Although we do not study the effect of bars in this work, Ansar et al., (in prep.) quantifies the incidence of bars in the FIRE-2 simulations, and in future work we will address the drivers of dynamical heating in our simulations, including bars. While it is possible that the bar heated these old stars more recently, bars most likely do not heat \textit{only} old stars.

Our results may differ from those of \citet{Grand16} because of numerical differences. 
First, Auriga uses the subgrid model from \citet{Springel03}, treating the ISM as a two-phase medium with an effective equation of state that inhibits the formation of dense gas, so that star formation occurs at much lower density, $n > 0.13$ cm$^{-3}$, compared with $n > 1000$ cm$^{-3}$ in FIRE-2. \citet{House11} found that low thresholds in density impose a dispersion `floor' in $\sigma_{\rm form}$, even in high-resolution simulations (50 pc), because stars form from higher-temperature gas. \citet{House11} find that this floor effectively goes away if one increases the star-formation threshold from $n = 0.1$ to 1000 cm$^{-3}$  \citep[see also][]{Martig14, Kumamoto17}. That said, given that \citet{Grand16} find that $\sigma_{\rm form}$ decreases over time, this dispersion `floor' may have little effect on their results.

Second, certain mechanisms that induce dynamical heating in our simulations may be different (and less significant) in the Auriga simulations. The simulations in \citet{Grand16} had baryonic mass resolution of $4 \times 10^4 \Msun$, $8 - 10$ times larger than ours, and they used lower spatial resolution, with gravitational force softenings of 375 pc physical for stars and gas (which is comparable to the thin-disk scale height of the MW), while our FIRE-2 simulations use $2.7 - 4$ pc for star particles and $\approx 1$ pc minimum and $\sim 4$ pc ISM-averaged for gas cells. Thus, Auriga simulations do not resolve the dense ISM, including important dynamical structures like GMCs that likely heat stars particles after their formation in FIRE. 
Furthermore, the modeling of stellar feedback in Auriga is markedly different: supernovae launch `wind particles' that are temporarily decoupled from the hydrodynamics until the particle encounters sufficiently low-density gas,  which does not couple that feedback directly to dense gas in a star-forming region. In turn, our simulations may exhibit more gas turbulence. 
That said, the Auriga simulations do model certain physics that these FIRE-2 simulations do not include, most importantly, AGN feedback.
%\citep[Implementations of magnetic fields in FIRE-2 show that they do not significantly change galaxy-wide properties;][]{Su17, Hopkins2020b}.

\citet{Grand16} also found that spiral arms did not considerably heat the disk, while we proposed that spiral arms are likely responsible for the significant increase in $\sigma_{\rm tot, now}$ of young stars, in part because $\sigma_{\rm R}$ dominates the overall dispersion \citep[see also][]{Orr22}. However, \citet{Grand16} only considered \textit{vertical} heating. 
Because the Auriga simulations do not resolve GMCs, spiral arms likely have negligible effects on the vertical dispersion, but may have more significant impacts on the in-plane dispersions. For example, while the FIRE-2 simulations resolve GMCs, young stars in FIRE-2 have vertical effective heating rates that are $\approx 4 \times$ lower than their radial rates, as Figure~\ref{fig:heating_rate} showed.
Many previous works -- both observational and theoretical -- only examined the vertical velocity dispersion. However, vertical post-formation dynamical heating operates differently than total and in-plane heating. For example, Figure~\ref{subsec:formation vs post-formation} (top) shows that the amount of vertical dispersion added post-formation increases approximately linearly across all ages, while in-plane heating plateaus for Early-Disk stars. Similarly, Figure~\ref{subsec:formation vs post-formation} (bottom) shows that the present-day vertical dispersion has the largest relative impact from post-formation heating: the dispersion added from heating typically begins to equal (or surpass) the dispersion from formation at $\approx$ 4 Gyr for the vertical component, but does not rival the in-plane dispersions from formation until $\approx$ 11 Gyr. Furthermore, the vertical dispersion typically exhibits the smallest values. For these reasons, we do not recommend using vertical heating to draw inferences regarding total heating. 

As we discussed above, nearly all current cosmological zoom-in simulations form MW-mass galaxies with typical \textit{disk-wide} $\sigma_{\rm Z, now} \gtrsim 20 \kms$, not as dynamically cold as the (solar-neighborhood) measurements of MW today. However, \citet{Bird-21} analyzed a single MW-mass galaxy, \texttt{h277}, selected from the \textit{g14} suite of simulations \citep{Christensen12}, generated using the parallel $N$–body+SPH code GASOLINE \citep{Wadsley04}.
That simulation had \textit{disk-wide} $\sigma_{\rm Z, now} < 10 \kms$, measured within an annulus of $R = 8 - 10 \kpc$, for stars younger than 1 Gyr, similar to (primarily locally-measured) values for the MW \citep[for example][]{Nordstrom2004, Mackereth19}.
%, while our simulations have $\sigma_{\rm Z, now} = 13 - 25$ km/s for young stars (see Figure~\ref{fig:stellar_mass}).

The cold disk-wide kinematics of \texttt{h277} likely reflect its formation history: its disk began to settle early, $\approx 9 \Gyr$ ago, and it had no major mergers since $z \approx 2$.
%, similar to estimates for the MW's history \citep[for example][]{Belokurov18, Helmi18, Conroy22, Xiang22}.
%Both of these factors likely contribute to colder kinematics.
%, as we found a strong correlation between the time of disk onset and how kinematically cold the stellar disk is at $z = 0$ for our 14 galaxy sample.
That said, one of our galaxies, Romeo, has a similar time of disk onset and merger history as \texttt{h277} but has a larger (disk-wide) $\sigma_{\rm Z} \approx 16 \kms$ for stars 1 Gyr old today. However, we note that Romeo has a local $\sigma_{\rm Z} \approx 12 \kms$, which is closer, albeit still higher, than measurements of the MW.
Therefore, differences in our simulations' physical and numerical modeling likely contribute to our unequal dispersions. Although the baryonic mass resolution, modeling of dense ISM, and molecular-based star-formation criteria are broadly similar between our FIRE-2 simulations and \texttt{h277}, key numerical differences exist between our simulations: \texttt{h277} used SPH for the hydrodynamics solver (instead of MFM); used larger gravitational force softening of 173 pc for all particle species; used the `blastwave' feedback model for core-collapse supernovae in which cooling is turned off for gas within the blast wave radius to mimic the adiabatic expansion of a remnant \citep{Stinson06}; used only thermal energy injection for white-dwarf (Type Ia) supernovae; and did not include stellar winds or radiative feedback. While it is unclear how any of these individual aspects do or do not affect their results, differences in feedback modeling may contribute to their lower dispersions to some degree. FIRE-2 accounts for energy and momentum injection for both core-collapse and white-dwarf supernovae, and includes stellar winds, radiation pressure, and photo-ionization. This feedback launches starburst-induced, star-forming outflows at early times (see the top-right panel of our Figure~\ref{fig:fig1} and \citet{el-badry18}), which \citet{Bird-21} notes are absent in their simulation. As such, it is possible that our additional feedback channels contribute to larger gas velocity dispersions and subsequently larger $\sigma_{\rm form}$ values in our simulations, though it is unclear if this persists to later cosmic times. Similarly, \citet{Hopkins12} and \citet{Agertz13} both show that the structure of the ISM can non-trivially vary depending on the feedback model used.
Despite any differences, we again emphasize that both the disk-wide $\sigma_{\rm Z, form}$ and $\sigma_{\rm Z, now}$ exhibited similar trends with age in \texttt{h277} and our FIRE-2 simulations.

\subsubsection{Connections to the early formation of the MW}

As we discussed in Sections~\ref{subsec:galaxy mass trends} and \ref{subsec:disk settling time}, the cold kinematics of MW stars may indicate that the MW's disk began to settle unusually early in its history. For example, using our predicted result that the disk-wide kinematics of young stars today correlates with when the disk began to settle, and placing the MW's value of $v_{\rm \phi} / \sigma_{\rm tot} \approx 8$ along the relation from our simulations in Figure~\ref{fig:disktime}, this implies that the MW's time of disk onset was $\gtrsim 11 \Gyr$ ago, which agrees with recent estimates from observations \citep[for example][]{Belokurov22, Conroy22, Xiang22}. Thus, our analysis of disk settling times in cosmological simulations and its relation to present-day kinematics complements and augments these works.

That said, measurements of the MW's kinematics are largely restricted to one local patch: the solar neighborhood. As we show in Figure~\ref{fig:local_vs_global}, locally-measured velocity dispersions in our simulations (at formation) are typically $\approx 2$ times smaller than disk-wide measurements. In turn, while all of our disk-wide dispersions are larger than the MW's, 3 of our galaxies have present-day local dispersions lower than the MW's. Furthermore, the galaxy-to-galaxy scatter in the relation between disk-wide and local kinematics is significant. For example, stars today younger than 100 Myr in one of our most MW-like galaxies, Romeo, have disk-wide $\sigma$ that is $1.1 - 1.2$ times higher than its local value, while m12i, which is otherwise similar to Romeo, has disk-wide values that are $1.5 - 1.7$ times higher than local values.
More work is needed to determine how the MW's disk-wide dispersion compares to its local dispersion, and if these differences affect its placement on our disk onset - young stellar kinematics relation.
In McCluskey et al. (in prep.) we will more rigorously compare our simulations to the MW as well as other observed galaxies.
%This also suggests that the FIRE-2 simulations do not produce a galaxy as kinematically cold as the MW today primarily because no galaxy in our sample started to settle as early as the MW.

We now discuss connections with some of these recent observational works. 
\citet{Conroy22} combined Gaia astrometry and H3 Survey spectroscopy to posit that the MW disk formed in three dynamical eras, broadly consistent with our Pre-Disk, Early-Disk, and Late-Disk Eras. \citet{Conroy22} proposed that the MW began in a `simmering phase', during which the star formation efficiency was low and stars formed kinematically hot, but that $\approx 12 - 13 \Gyr$ ago the MW transitioned into a `boiling phase', during which the SFE strongly increased and stars formed with increasingly `disky' dynamics, marking the `birth of the Galactic disk'. This thick disk continued to grow and settle in a `boiling phase' for $\approx 3 - 4 \Gyr$ (from $z \approx 4$ to $z \approx 1$) until the Gaia-Sausage-Enceladus (GSE) merger, after which the star formation efficiency decreased and stars formed in a dynamically-colder, thin disk until today.

Similarly, \citet{Belokurov22} combined Gaia astrometry with spectroscopy from the APOGEE Data Release 17. Although \citet{Belokurov22} lacked empirical age estimates, their general picture agrees with ours: old, metal-poor, in-situ stars were born in a `messy' phase, then the MW  `spun-up' as stars rapidly became metal-rich and rotational earlier than $\approx 8$ Gyr ago. \citet{Belokurov22} posited that this rapid formation of the MW's disk took place over 1 - 2 Gyr, with the GSE merger occurring after the initial formation of a thick-disk, such that the merger heated disk stars onto halo-like orbits.

On the other hand, \citet{Xiang22} conclude that the formation of the in-situ halo and thick-disk overlapped. In this picture, the GSE merger then occurred $\approx 11 \Gyr$ ago, $1 - 2 \Gyr$ earlier than previous estimates, and did not strongly heat pre-existing disk stars, or cause a transition to a thin-disk, but instead enhanced thick-disk formation.

Our results help connect these present-day observations to understanding the MW's formation history.
In particular, \textit{the presence of coherent rotation in a stellar population today does not necessarily indicate the presence of coherent rotation at the time of formation}.
Observations show that older, more metal-poor stars in the MW rotate slower than younger stars, but they still display some coherent rotation. In our simulations, this is not primarily the result of these Pre-Disk stars being born with coherent rotation, but rather, this tends to arise from post-formation torquing (see Figure~\ref{fig:fig1}), for example, caused by mergers (see Bellardini et al. in preparation).
m12m provides an illustrative example: it experienced several major mergers at early times, broadly similar to estimates of when the GSE merger occurred.
As Figure~\ref{fig:casestudies} shows, m12m has old stars with $v_{\rm \phi, now} \approx 200 \kms$, despite having $v_{\rm \phi, form} \approx 0 - 50 \kms$.
Thus, mergers like the GSE merger could have torqued old stars in the MW onto more coherently-rotating orbits.

Ultimately, the MW's kinematics reflect its individual formation history. While the MW provides a useful benchmark, the fact that most current cosmological simulations form galaxies with hotter kinematics than the MW may speak to the MW's early settling time rather than to limitations of the simulations.
Indeed, the comparison of the MW to M31 and M33 in Figure~\ref{fig:stellar_mass} affirms that the MW may be an outlier among galaxies at its mass.
Therefore, future works that study the kinematics of stars and their relation to stellar ages in external galaxies will provide excellent opportunities to place the MW in a statistical context, to study the combination of cosmological disk settling and dynamical heating across a range of formation histories \citep{Dorman2015, Leaman17, Quirk22}.

%%%%%%%%%%%%%%%%%%%%%%%%%%%%%%%%%%%%%%%%%%%%%%%%%%
\vspace{-3 mm}
\section*{Acknowledgements}

We thank Jonathan Stern for helpful comments and the participants of the Disk Formation Workshop at UC Irvine for their enlightening discussions. FM and AW received support from: NSF via CAREER award AST-2045928 and grant AST-2107772; NASA ATP grant 80NSSC20K0513; HST grants AR-15809, GO-15902, GO-16273 from STScI; and a Scialog Award from the Heising-Simons Foundation. CAFG was supported by NSF through grants AST-2108230 and CAREER award AST-1652522; by NASA through grants 17-ATP17-0067 and 21-ATP21-0036; by STScI through grant HST-GO-16730.016-A; by CXO through grant TM2-23005X; and by the Research Corporation for Science Advancement through a Cottrell Scholar Award. PFH recieved supported by NSF Research Grants 1911233 and 20009234, NSF CAREER grant 1455342, and NASA grants 80NSSC18K0562 and HST-AR-15800.001-A. We performed some of this work at the Aspen Center for Physics, supported by NSF grant PHY-1607611. We ran simulations using: XSEDE, supported by NSF grant ACI-1548562; Blue Waters, supported by the NSF; Frontera allocations AST21010 and AST20016, supported by the NSF and TACC; Pleiades, via the NASA HEC program through the NAS Division at Ames Research Center.

\vspace{-3 mm}
\section*{Data Availability} 
All of the Python code that we used to generate these figures are available at \url{https://fmccluskey.github.io}, which uses the publicly available package \url{https://bitbucket.org/awetzel/gizmo_analysis} \citep{Wetzel2020a}. The FIRE-2 simulations are publicly available \citep{Wetzel22} at \url{http://flathub.flatironinstitute.org/fire}.
Additional FIRE simulation data is available at \url{https://fire.northwestern.edu/data}.
A public version of the \textsc{Gizmo} code is available at \url{http://www.tapir.caltech.edu/~phopkins/Site/GIZMO.html}.
\bibliographystyle{mnras}
\bibliography{AVR}
% if your bibtex file is called example.bib
%%%%%%%%%%%%%%%%% APPENDICES %%%%%%%%%%%%%%%%%%%%%
\appendix
\vspace{-6 mm}
\section{Effect of Vertical Selection}
\label{a:vertical_selection}
\begin{figure}
\includegraphics[width = \columnwidth]{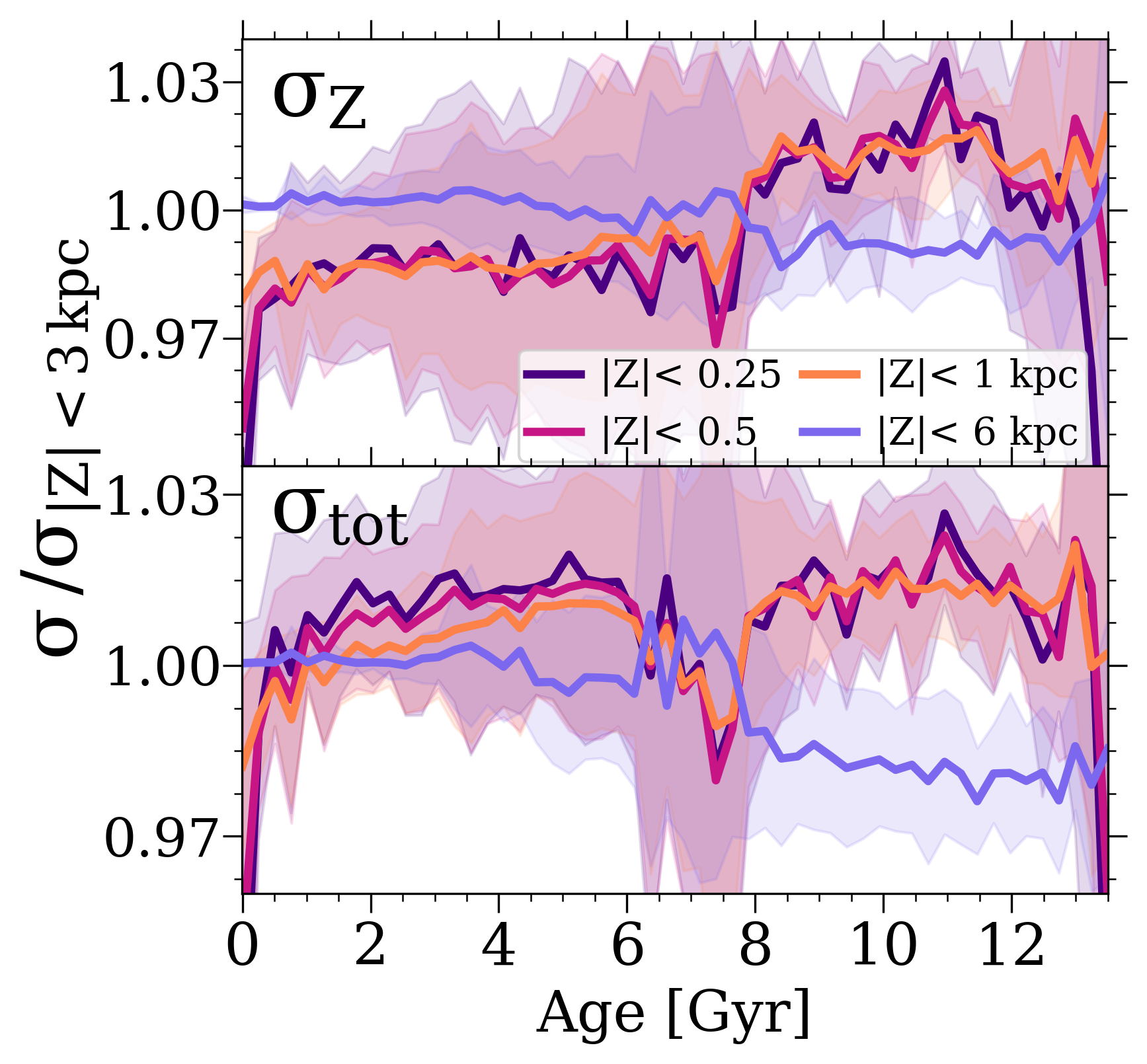}
\vspace{-7 mm}
\caption{
\textbf{Effect of vertical selection on $\sigma_{\rm Z}$ (top) and $\sigma_{\rm tot}$ (bottom)}:
present-day velocity dispersion for stars selected within 4 vertical ranges, divided by the dispersion for our fiducial range of $|Z| < 3 \kpc$, versus stellar age. We select all stars at $R = 6 - 10 \kpc$. Thinner vertical selections lead to slightly higher $\sigma_{\rm Z}$ and $\sigma_{\rm tot}$ for stars older than $\approx 8 \Gyr$, but slightly \textit{lower} $\sigma_{\rm Z}$ and \textit{higher} $\sigma_{\rm tot}$ for stars younger. However, different vertical selections change the dispersion by $\lesssim 3\%$, with the exception of the oldest and youngest stars, which have $\approx 5\%$ lower dispersions for thinner vertical ranges.
}
\label{fig:z_cut}
\end{figure}

By default, we select stars currently at $|Z| < 3 \kpc$, which includes stars in both the thin and thick disk, to sample disk stars at all ages in an unbiased manner and avoid assumptions about thin- versus thick-disk dynamics.
Figure~\ref{fig:z_cut} shows the effect of the vertical selection on $\sigma_{\rm Z, now}$ and $\sigma_{\rm tot, now}$ versus stellar age, specifically, the fractional change relative to our fiducial selection of $|Z| < 3 \kpc$. We also examined the effect on $\sigma_{\rm R, now}$ and $\sigma_{\rm \phi, now}$ and they show trends nearly identical to $\sigma_{\rm tot, now}$ so we do not show them separately.

At most ages, our vertical selection has minimal impact on any component of the dispersion, $\lesssim 3\%$. This is because most disk stars lie well within $|Z| < 1 \kpc$, so they dominate regardless of vertical selection.
Thinner vertical selection does yield slightly higher $\sigma_{\rm tot, now}$, although with significant scatter. Interestingly, thinner selection only yields higher $\sigma_{\rm Z, now}$ for stars older than $\approx 8 \Gyr$, and instead yield slightly \textit{lower} values of $\sigma_{\rm Z, now}$ for younger stars.
The youngest population formed near the mid-plane and experienced $\approx 125 \Myr$ of post-formation heating, on average, so most are still near the mid-plane and have low dispersions.
Any that are above $|Z| = 0.25 - 1 \kpc$ likely experienced an atypical amount of heating.
Again, all of these effects are at the level of $\lesssim 3\%$, where non-equilibrium effects like disk warping and asymmetries may become important.
%On the other hand, few old stars are currently within $|Z| = 0.25 - 0.5 \kpc$, so these thinner vertical selections may be biased towards kinematically-colder stars.

%These results can be simply explained using basic orbital dynamics. The vertical motion of stars resembles that of a simple harmonic oscillator: $v_{\rm Z}$ is greatest near the disk plane, but approaches 0 as the star reaches its maximum/minimum height. Greater vertical selections select more stars that are closer to their maximum/minimum heights, and as the mean $v_{\rm Z, now}$ is near 0 in the solar annulus, selecting these stars decreases the dispersion. 
%As shown in Figure~\ref{fig:fig1}, all age groups of stars in the solar annulus have mean $v_{\rm R, now} \approx v_{\rm Z, now} \approx$ 0. 

%However, we find that the impact of our vertical selection is minor only within individual age bins. 
%The vertical selection \textit{does} strongly impact our determinations of a galaxy's overall dispersion, as 
If instead we measure the dispersion for the galaxy overall (not in age bins) as in Section~\ref{fig:stellar_mass}, larger vertical selections non-trivially increase the dispersion.
Vertical dispersions within $|Z| < 3 \kpc$ are 10 - 35\% larger than within $|Z| < 1 \kpc$, because larger vertical selections include a higher fraction of older, higher-dispersion stars.
\vspace{-6 mm}
\section{Effect of In-Situ Selection}
\label{a:insitu_selection}
\begin{figure}
\includegraphics[width = \columnwidth]{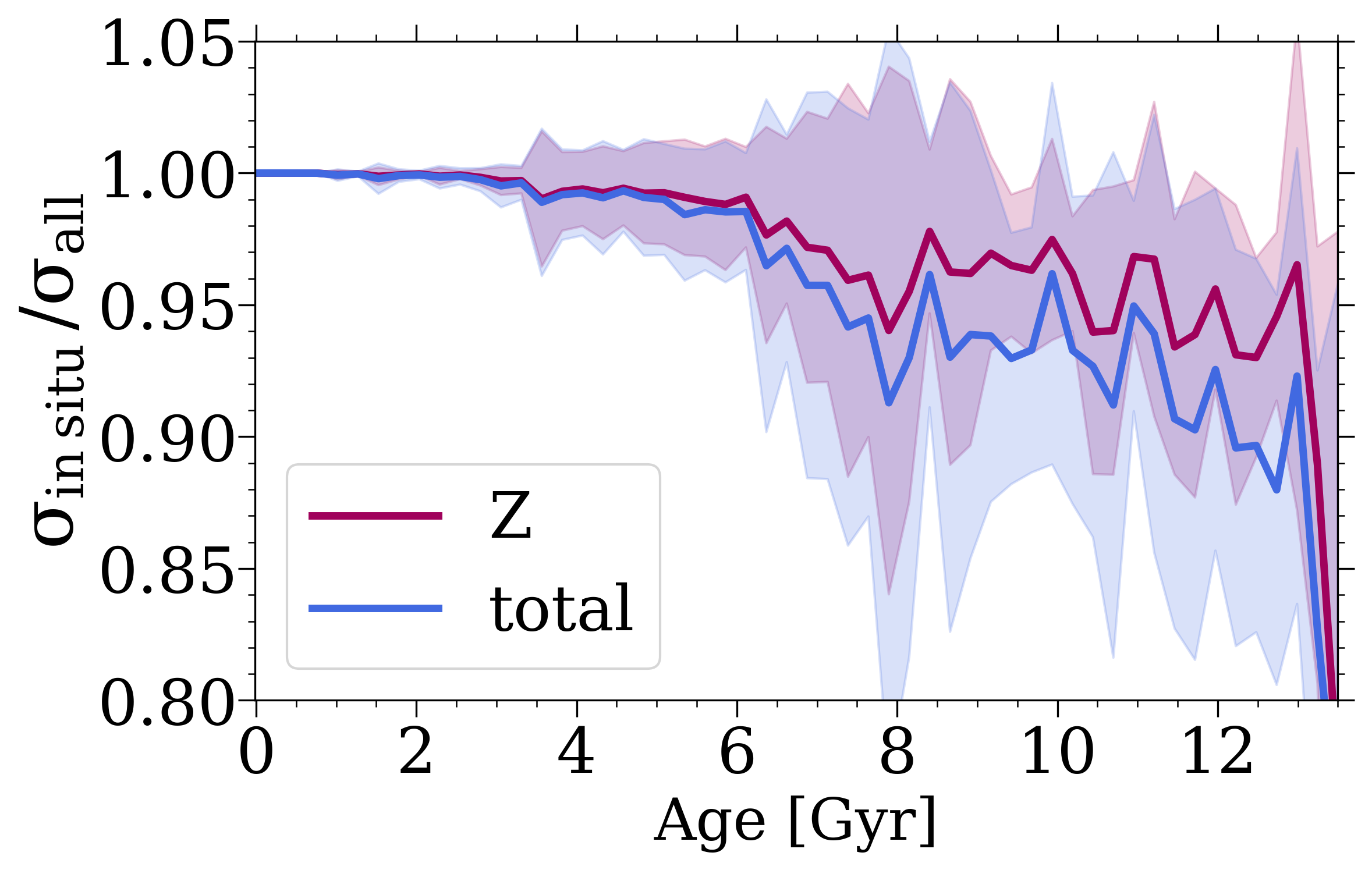}
\vspace{-7 mm}
\caption{
\textbf{Effect of excluding ex-situ stars on the present-day velocity dispersion:} $\sigma_{\rm Z, now}$ (top) and $\sigma_{\rm tot, now}$ (bottom) of only stars that formed in situ divided by that of all (in-situ + ex-situ) stars, versus age, averaged across 11 galaxies. Stars with ages $\lesssim 6 \Gyr$ show no significant difference, because almost all young disk stars formed in situ. Older than 6 Gyr, the dispersion of in-situ stars relative to all stars decreases with age, because the fraction of ex-situ stars increases with age, and ex-situ stars have larger present-day dispersions.
However, at almost all ages that we investigate, the difference in velocity dispersion is $\lesssim 10\%$.
}
\label{fig:form_cut}
\end{figure}

By default, we select only stars that are at $R = 6 - 10 \kpc$ and $|Z| < 3 \kpc$ at present. We include all of these stars when measuring present-day kinematics, but when measuring kinematics at formation, we select only in-situ stars that formed within 30 kpc comoving of the galactic center.
Figure~\ref{fig:form_cut} shows the effect of also including this in-situ selection on the \textit{current} velocity dispersion versus stellar age. As above, we only show $\sigma_{\rm Z, now}$ and $\sigma_{\rm tot, now}$, because $\sigma_{\rm R, now}$ and $\sigma_{\rm \phi, now}$ exhibit the same trends as $\sigma_{\rm tot, now}$. Overall, our in-situ selection has no effect on stars younger than $\approx 6 \Gyr$, because almost all of them formed in situ. However, the effects of in-situ selection become increasingly significant for older stars, when a more significant fraction formed ex situ. Not surprisingly, including those older ex-situ stars increases the dispersion. That said, this effect is $\lesssim 10\%$ for all but the oldest stars \citep[see][for analysis of these ancient stars]{El-Badry18-anc}.
\vspace{-6 mm}
\section{Effect of Cosmic-Ray Feedback}
\label{a:cr}

\begin{figure}
\includegraphics[width = \columnwidth]{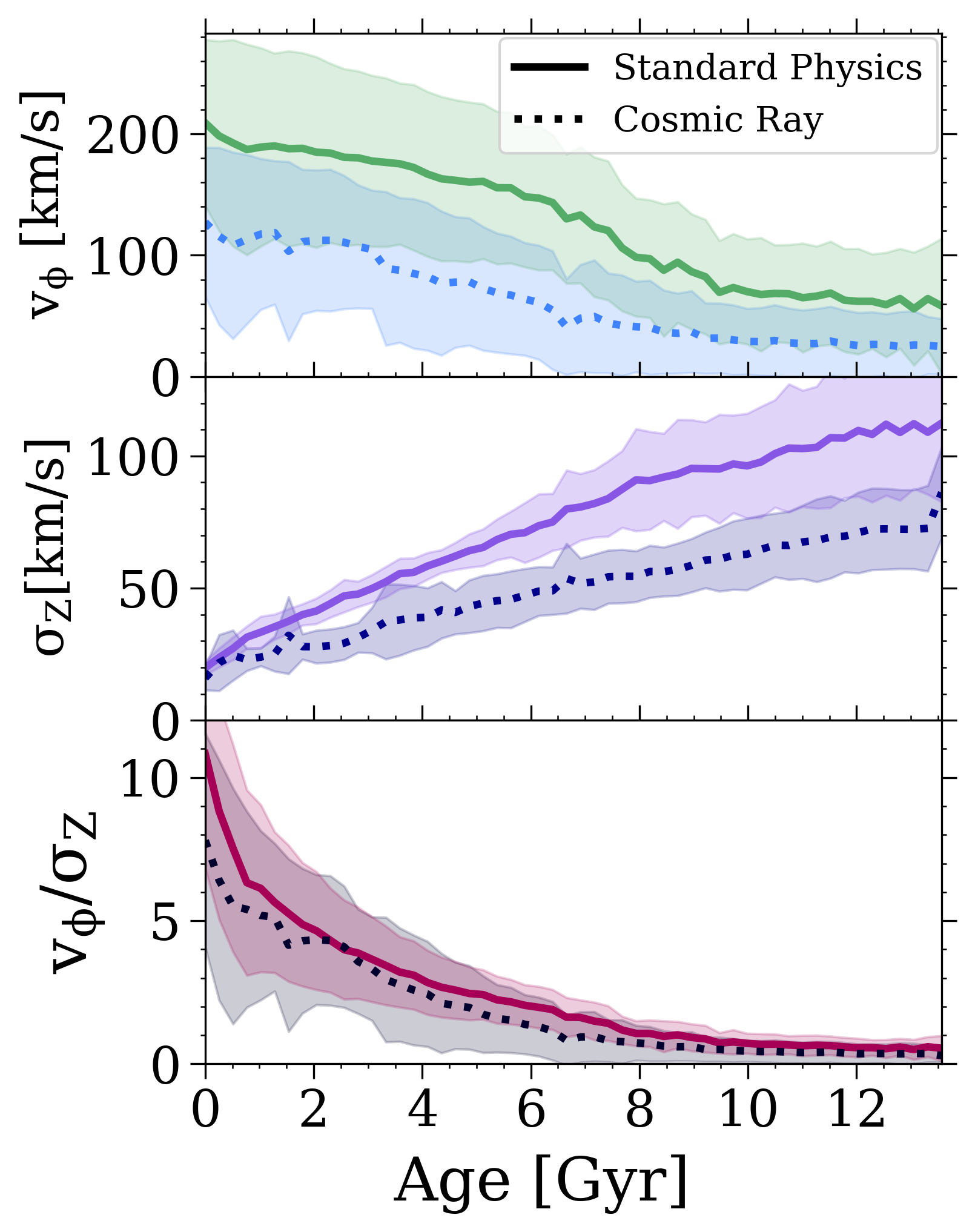}
\vspace{-6 mm}
\caption{
\textbf{Effects of cosmic-ray feedback on disk dynamical history}:
current stellar kinematics versus age, averaged across 8 isolated MW-mass galaxies simulated with fiducial FIRE-2 physics (solid) and additionally including cosmic ray (CR) injection and transport from stars (dotted), assuming a constant effective diffusion coefficient $\kappa = 3 \times 10^{29} \rm cm^2 s^{-1}$.
The shaded regions show the galaxy-to-galaxy standard deviation.
\textbf{Top}: median azimuthal velocity, $v_{\rm \phi}$. The addition of CR feedback reduces overall star formation, leading to lower-mass galaxies with lower velocities at all ages.
\textbf{Middle}: vertical velocity dispersion, $\sigma_{\rm Z}$. Given their lower stellar mass, the CR simulations have lower velocity dispersions, though the CR simulations converge to the fiducial simulations for the youngest stars.
\textbf{Bottom}: median $v_{\rm \phi} / \sigma_{\rm Z}$, as a measure of rotational support and `diskiness'.
Given the reduction of both $v_{\rm \phi}$ and $\sigma_{\rm Z}$ in the CR simulations, $v_{\rm \phi} / \sigma_{\rm Z}$ is similar at nearly all ages for the fiducial and CR simulations.
For the youngest stars, the CR simulations are fractionally dynamically hotter and less rotationally supported.
We find similar trends for the other velocity components and the total dispersion.
\vspace{-1 mm}
}
\label{fig:cr_phi}
\end{figure}
The fiducial FIRE-2 simulations that we examine do not include self-consistent cosmic ray (CR) injection and transport. However, CRs may be important in regulating the dynamical state of the ISM, and therefore the resultant stellar populations, because in the solar neighborhood the CR energy density is comparable to the gas thermal and turbulent energy densities \citep[for example][]{Boulares90}.

Here, we explore the effects of CR feedback from stars on the dynamical state of our MW-mass disks. One proposed way to form kinematically colder disks in cosmological simulations is the inclusion of previously neglected feedback channels, including CRs, because they can provide a spatially and temporally smoother form of feedback and potentially launch cooler galactic winds.
Previous work using FIRE-2 simulations found that the inclusion of magnetic fields, anisotropic conduction, and viscosity has no appreciable effects on any bulk galaxy properties, but that the inclusion of CRs, within reasonable uncertainty for the treatment of the CR diffusion coefficient, significantly can alter galaxy properties \citep{Su17, Hopkins20cr}. FIRE-2 simulations with CRs also include anisotropic CR transport with stream and advection / diffusion (with constant parallel diffusivity, $\kappa = 3 \times 10^{29} \rm cm^2 s^{-1}$), CR cooling (hadronic and Compton, adiabatic, and streaming losses), CR injection in supernova shocks, and CR–gas coupling \citep[see][for details of the numerical implementation of CRs in these simulations]{Chan19}.

\citet{Hopkins20cr} showed in the FIRE-2 simulations that the primary effect of CR feedback is to prevent the accretion of halo gas into the galaxy, leading to lower star formation and stellar mass by $\approx 2 - 4 \times$ in MW-mass galaxies at $z \lesssim 2$.
Furthermore, \citet{Chan21} found that the inclusion of CRs in FIRE-2 leads to lower velocity dispersions in gas, which might suggest dynamically cooler disks, although they did not examine $v_{\rm \phi} / \sigma$.
We show here that the \textit{opposite} is true for young stars when considering  $v_{\rm \phi} / \sigma$, that is, the inclusion of CRs leads to dynamically hotter stellar disks.

Figure~\ref{fig:cr_phi} shows the effect of CRs on present-day stellar kinematics versus age.
We include only the isolated galaxies from the \textit{Latte} suite, which have versions with CR feedback. We also include the lower-mass galaxies m12r and m12z, because our goal is only to examine the differential effects of including CRs. Figure~\ref{fig:cr_phi} (top) shows the current median azimuthal velocity, $v_{\rm \phi, now}$, versus stellar age, averaged across the same galaxies in the CR simulations and the fiducial simulations.
At all ages, galaxies with CR feedback have lower $v_{\rm \phi}$, with the largest absolute difference for the youngest stars, which arises primarily because of their lower stellar masses.

Figure~\ref{fig:cr_phi} (middle) shows the present vertical velocity dispersion, $\sigma_{\rm Z}$, versus age. The CR simulations have lower $\sigma_{\rm Z}$ at all ages, except for the youngest stars, which may arise from the effects of CRs on the ISM and halo gas occurring over long timescales. We also examined $\sigma_{\rm R}$, $\sigma_{\rm \phi}$, and $\sigma_{\rm tot}$ and found qualitatively similar results. 
Overall, the inclusion of CR feedback leads to lower velocity dispersions in stars, consistent with analyses of gas in these simulations \citep{Hopkins20cr, Chan21}. 

\begin{figure}
\includegraphics[width = \columnwidth]{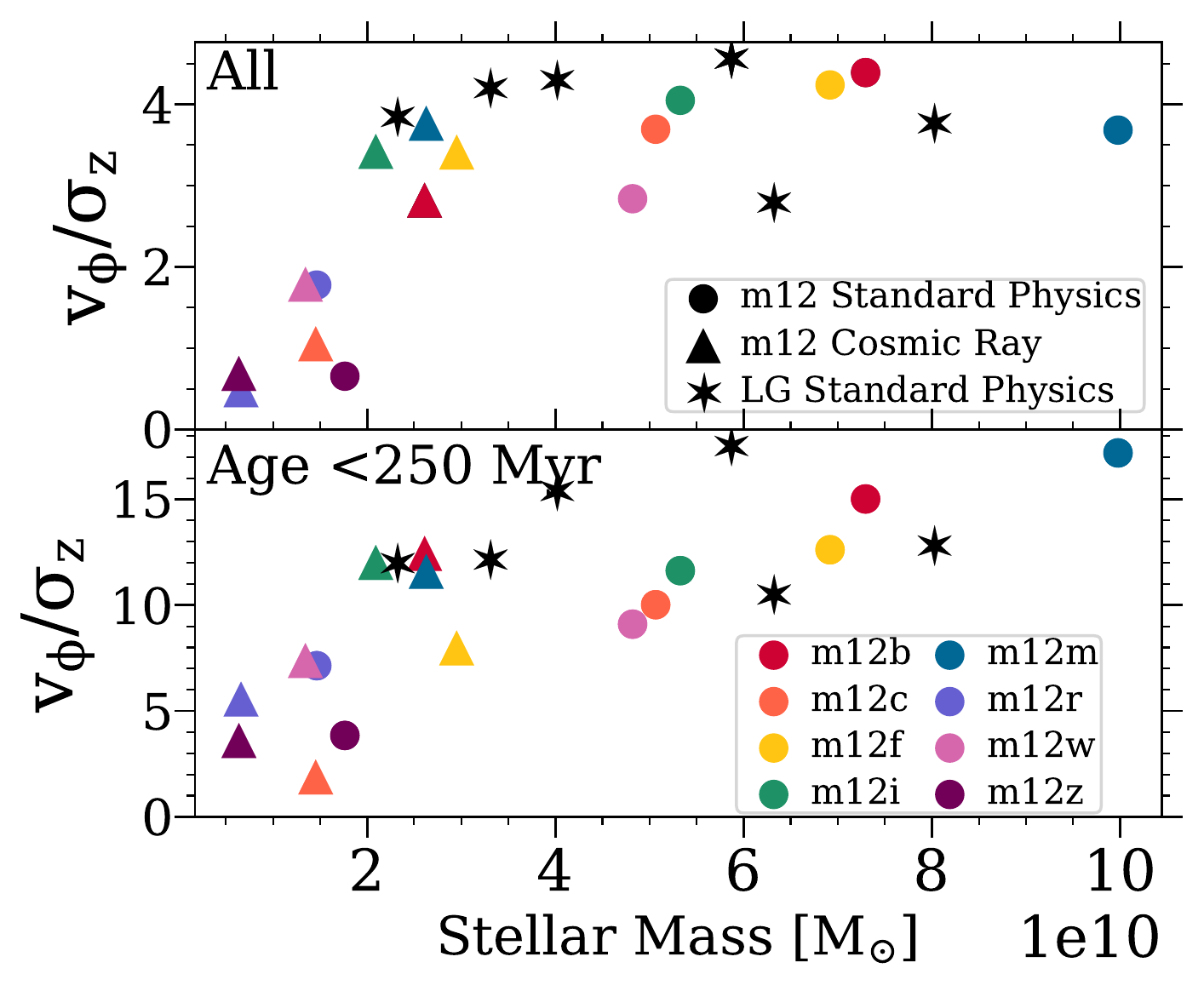}
\vspace{-6 mm}
\caption{
\textbf{Effects of cosmic-ray feedback on current stellar $v_{\rm \phi} / \sigma_{\rm Z}$ versus stellar mass}, across 8 isolated MW-mass galaxies simulated with fiducial FIRE-2 physics (circles) and additionally with CR injection and transport (triangles). Black stars show LG-like galaxies simulated with fiducial physics; although these galaxies lack CR counterparts, we include them in order to illustrate our fiducial sample's $v_{\phi} / \sigma_{\rm Z}$ - stellar mass relation. 
CR feedback leads to galaxies with $\approx 3$ times lower stellar mass.
\textbf{Top row}: $v_{\phi} / \sigma_{\rm Z}$ for stars of all ages.
The CR simulations have lower $\sigma_{\rm Z}$ but also lower stellar masses, leading them to land on a similar relation in $v_{\phi} / \sigma_{\rm Z}$.
\textbf{Bottom row}: same but for young stars with ages $< 250 \Myr$.
CR simulations have slightly lower $v_{\phi} / \sigma_{\rm Z}$ for 7 of the 8 galaxies, though again, they remain on a similar scaling relation.
\textit{Adding CR feedback does not lead to dynamically colder disks; it primarily reduces the stellar mass formed and moves galaxies along a similar relation.}
\vspace{-1 mm}
} 
\label{fig:cr_mass}
\end{figure}

Figure~\ref{fig:cr_phi} (bottom) shows the sample-averaged $v_{\rm \phi} / \sigma_{\rm Z}$ versus age. For most ages, stars in the CR simulations have slightly \textit{less} rotational support than in the simulations without CR feedback; the youngest stars show the most significant reduction at $\approx 30\%$. Thus, including CR feedback leads to fractionally dynamically hotter galaxies, despite their lower absolute velocity dispersions. However, Figure~\ref{fig:cr_mass} shows that CR feedback leads to galaxies with $\approx 3 \times$ lower stellar mass at present, as \citet{Hopkins20cr} showed.

Figure~\ref{fig:cr_mass} (top row) shows $v_{\rm \phi} / \sigma_{\rm Z}$ for all stars versus stellar mass, while Figure~\ref{fig:cr_mass} (bottom row) shows the same but for young stars (age < 250 Myr). We show our 6 fiducial LG-like galaxies (as black stars) to clarify our fiducial sample's mass-relation across this entire mass range, given that none of our fiducial isolated simulations have masses within $2 - 4 \times 10^10 \Msun$, the range containing half of their CR counterparts. That said, we advise against making direct comparisons between any individual fiducial LG-like simulations and CR isolated simulations, because each galaxy has its own unique history and our fiducial LG-like simulations generally have slightly colder kinematics than our isolated simulations. We emphasize that Figure~\ref{fig:cr_mass} shows the same trends with stellar mass as for our fiducial simulations without CR feedback in Figure~\ref{fig:stellar_mass}.
Thus, the inclusion of CR feedback primarily acts to reduce a galaxy's stellar mass and correspondingly shifts its $v_{\rm \phi}$ and $v_{\rm \phi} / \sigma_{\rm Z}$ along the existing relation. This leads to galaxies that are slightly less disky, with slightly lower $v_{\rm \phi} / \sigma_{\rm Z}$.
We conclude that the inclusion of CR feedback, at least as implemented in FIRE-2 and assuming a constant diffusion coefficient, does not lead to dynamically colder disks.

%%%%%%%%%%%%%%%%%%%%%%%%%%%%%%%%%%%%%%%%%%%%%%%%%%

% Don't change these lines
\bsp	% typesetting comment
\label{lastpage}
\end{document}